\documentclass[11pt,a4paper]{article}
\usepackage{jheppub}

\usepackage[english]{babel}
\usepackage[colorlinks=true,linkcolor=blue,citecolor=blue,urlcolor=blue]{hyperref}

\usepackage{amsmath,graphicx,verbatim,epsfig}
\usepackage{amsmath,graphicx}
\usepackage{color}
\usepackage{datetime}
\usepackage{slashed}

\usepackage{mathtools}

\usepackage{listings}

\usepackage{amssymb}
\usepackage{emptypage}
\usepackage{graphicx}
\usepackage{bm}
\usepackage{placeins}
\usepackage{float}
\usepackage{subfigure}
\usepackage{caption}
\usepackage{epsfig}
\usepackage{array}
\usepackage[usenames,dvipsnames]{xcolor}

\usepackage{lipsum}
\usepackage{wrapfig}
\usepackage{etoolbox}
\usepackage[normalem]{ulem} 
\usepackage{bbm}
\usepackage{braket}
\usepackage{verbatim}
\usepackage{booktabs}
\usepackage{multirow}
\usepackage{xfrac}

\usepackage[toc,page]{appendix}

\usepackage{afterpage}
\usepackage[normalem]{ulem} % \sout{old text} for strikeout

\newcommand{\st}{{\scriptscriptstyle T}}

\newcommand{\vek}[1] {\boldsymbol{#1}}

\newcommand{\y}{\vek{y}}

\newcommand{\nn}{\nonumber}

\allowdisplaybreaks[3]

\newcommand{\htb}[1]{{\color{blue} #1}}

\newcommand{\blue}[1]{{\color{blue} {\bf #1}}}

\begin{document}
\abstract{The future runs of LHC offer a unique opportunity to measure correlations between two partons inside the proton, which have never been experimentally detected. The process of interest is the production of two positively charged W-bosons decaying in the muon channel. We present a detailed analysis of proton-proton collisions at $\sqrt{s}$ = 13 TeV, where we combine Monte Carlo event generators with our calculations of parton correlations. We carefully compare double parton scattering to relevant background processes and trace a path towards a clean signal sample. Several observables are constructed to demonstrate the effect of parton correlations with respect to clear benchmark values for uncorrelated scatterings. We find that especially spin correlations can be responsible for large effects in the variables we study, because of their direct relation with the parton angular momentum and, therefore, the directions of the muon momenta. We estimate the significance of the measurements as a function of the integrated luminosity and conclude that the LHC has the potential to detect, or put strong limits on, parton correlations in the near future. \\\\

\centerline{\textbf{\Large DRAFT: \today}}}

\preprint{\vbox{
\hbox{MITP/19-044}}}

%%%%%%%%%%%%%%%%%%%%%%%%%%%%%%%%%
%%%%%%%%%%%%%%%%%%%%%%%%%%%%%%%%%

\title{Confronting same-sign W-boson production with parton correlations}

\author[a]{Sabrina Cotogno,}
\emailAdd{sabrina.cotogno@polytechnique.edu}
\affiliation[a]{CPHT, CNRS, Ecole Polytechnique, Institut Polytechnique de Paris, Route de Saclay, 91128 Palaiseau, France}

\author[b]{Tomas Kasemets,}
\emailAdd{kasemets@uni-mainz.de}
\affiliation[b]{PRISMA Cluster of Excellence \& Mainz Institute for Theoretical Physics\\ Johannes Gutenberg University, 55099 Mainz, Germany}

\author[c]{and Miroslav Myska}
\emailAdd{miroslav.myska@fjfi.cvut.cz}
\affiliation[c]{FNSPE, Czech Technical University in Prague, Brehova 7, 115 19 Prague, Czech Republic}
\date{\bf{Draft:} \blue{ \today, \currenttime}}

%%%%%%%%%%%%%%%%%%%%%%%%%%%%%
%%%%%%%%%%%%%%%%%%%%%%%%%%%%%

\maketitle

\frenchspacing
\allowdisplaybreaks

% !TEX root = WWLongPaper.tex
%%%%%%%%%%%%%%%%%%%%%%%%%%%%%%%%%%%%%%%%
\section{Introduction}
%%%%%%%%%%%%%%%%%%%%%%%%%%%%%%%%%%%%%%%%
Double parton scattering is the simultaneous collision of two pairs of mutually correlated partons in two independent hard interactions. 
Double parton scattering (DPS) carries resemblance with, but differs from, the much more common single parton scattering (SPS). 
Since two partons inside one proton are related, they cannot be treated as independent free partons. The amount of inter-parton correlations in DPS is unknown and, to a large extent, 
so are its consequences. {\it It is the purpose of this article  to demonstrate the experimental implications that parton correlations can have on experimental 
observables and pave the way to explicit measurements of the degree to which two partons in a proton are interconnected}.

In order to reach this goal, we delve into the details of proton collisions at center-of-mass energy $\sqrt{s}=13$ TeV producing equally (positively) charged $W$-bosons, referred to as same-sign W-boson (SSW) production. 
Once enough statistics is collected at the LHC, this process will be one of the best probes of partonic correlations and of double parton scattering in 
general \cite{Cao:2017bcb,Ceccopieri:2017oqe,Luszczak:2014mta,Golec-Biernat:2014nsa,dEnterria:2012jam,Myska:2013duq,Myska:2012dj,Gaunt:2010pi,Kulesza:1999zh}. 

The theory of DPS has seen rapid developments in the last decade, and it is  fair to say that factorization into hard scatterings and parton distributions is 
now on a similar footing as in SPS \cite{Diehl:2018wfy,Buffing:2017mqm,Diehl:2015bca}. A formalism has been developed to simultaneously treat the cross section contribution from DPS and SPS without double counting \cite{Diehl:2017kgu}.
The theoretical developments have been accompanied by a large increase in the number of measurements of DPS, see e.g.~\cite{Akesson:1986iv,Alitti:1991rd,Abe:1993rv,Abe:1997xk,Abe:1997bp,
Abazov:2009gc,Aaij:2012dz,Aad:2013bjm,Chatrchyan:2013xxa,Abazov:2014fha,Aad:2014kba,Aaboud:2016dea,Aaij:2015wpa,Abazov:2015nnn,Aaij:2016bqq,Sirunyan:2017hlu,Abazov:2014qba,Abazov:2015fbl,
Aaboud:2016fzt,Khachatryan:2016ydm,Aaboud:2018tiq,Sirunyan:2019zox}. 
Advancements in these measurements, which are possible due to the increased statistical power, will enable the study of correlations in DPS.  In particular, the LHC is now reaching integrated luminosities large enough to probe the SSW process and recently first 
experimental observations of DPS in the SSW final state have been made \cite{Sirunyan:2019zox,Sirunyan:2017hlu}.

Since the early stages of the DPS theory, the presence of kinematical and quantum correlations between two partons has been acknowledged as an intrinsic consequence of 
the composite structure of hadrons~\cite{Paver:1983hi,Mekhfi:1985dv,Mekhfi:1983az}. However, until recently parton correlations have largely been ignored, either 
because they were considered to be quantitatively unimportant, or, more likely, the contact with experiments was out of reach.  
Thanks to the opportunities given by the LHC, a renewed interest towards correlations in DPS  flourished, and substantial work has been put in their theoretical formulation 
and modeling, see e.g.~\cite{Diehl:2017wew,Kasemets:2017vyh,Blok:2017alw,Treleani:2018dbg,Blok:2013bpa} and references therein. 
Quark model calculations  show strong correlations in the valence region \cite{Rinaldi:2018bsf,Chang:2012nw,Rinaldi:2013vpa,Rinaldi:2014ddl,Rinaldi:2016jvu,Rinaldi:2016mlk,
Broniowski:2013xba,Broniowski:2016trx,Kasemets:2016nio} but give limited information about the region of small momentum fractions, where DPS predominantly occurs. 
Such strong correlations were found also in a calculation of electron-positron double parton distributions (DPDs) \cite{Mondal:2019rhs}. Sum rule improved DPDs also induce correlations between the kinematical variables \cite{Diehl:2020xyg,Diehl:2018kgr,Gaunt:2009re}. Kinematical correlations in the production of 
SSW were studied in~\cite{Gaunt:2010pi,Ceccopieri:2017oqe}. The generation of correlations by single to double parton splitting has been investigated by several groups 
\cite{Diehl:2011yj,Gaunt:2012dd,Blok:2011bu,Blok:2013bpa,Ryskin:2011kk,Ryskin:2012qx}. Despite being suppressed in SSW production, the splitting can give significant contributions to the cross section \cite{Cabouat:2019gtm}. The SSW cross section, 
including correlations, was derived in \cite{Kasemets:2012pr}. 

Regarding quantum correlations, while for instance color effects are Sudakov-suppressed at high energy~\cite{Mekhfi:1988kj,Manohar:2012jr}, spin correlations can remain 
sizable after evolution from smaller to larger scales~\cite{Diehl:2014vaa}. The polarized contributions can be constrained by positivity bounds \cite{Diehl:2013mla,Kasemets:2014yna}, 
which have similar theoretical status as positivity constraints on single parton distributions (PDFs). In \cite{Echevarria:2015ufa} the correlations between the 
spin of the two partons were first quantitatively connected to an observable cross section, but no clear observable for their detection was found.

In the previous letter \cite{Cotogno:2018mfv}, we demonstrated that the effect of spin correlations can be measured in some observables of SSW production. With the present paper we extend and complement that analysis. 
Our work quantifies the impact of different types of correlations in DPS, identifies observables which are particularly  suited for their measurement, and provides extended discussions on how to, in a practical way, treat the backgrounds. 

The structure of the paper is as follows: Section~\ref{SecDoubleW} briefly presents the framework employed to study the production of SSW, 
such as the factorized cross-section, while Section~\ref{Sec:4} contains a description of the various models used to include kinematical and spin correlations. 
In Section~\ref{Sec:parton}, the different correlation scenarios are explored at the level of the partonic cross section. Since we want to reach a realistic description 
of the results at the LHC, we devote Section~\ref{Sec:FSI} to an extended discussion on how to suppress the background and obtain a clean signal.
We get to the heart of correlation measurements in Section~\ref{Sec:Corr}, where we show the effects of correlations on  several variables and estimate the feasibility 
for the detection of correlations at the LHC. In particular, the same observables studied in Section~\ref{Sec:parton} are calculated after the successful 
suppression of the background and the identification of a suitable phase-space region. Finally, Section~\ref{template} underlines  potential problems deriving 
from neglecting correlations while measuring DPS cross sections, and in Section~\ref{Conclusions} we discuss what conclusions can be drawn from the results.

% !TEX root = WWLongPaper.tex
%%%%%%%%%%%%%%%%%%%%%%%%%%%%%%%%%%%%%%%%
\section{DPS production of same-sign W-boson}
\label{SecDoubleW}
%%%%%%%%%%%%%%%%%%%%%%%%%%%%%%%%%%%%%%%%

We analyze the effect that different types of inter-parton correlations have in the production of two $W^+$ bosons through DPS at the LHC. 
Among the various kinds of correlations accessible in DPS, we focus on the quantum correlation between the spin of the partons and the 
kinematic correlations between their momentum fractions. We neglect other sources of correlations, such as color correlations 
(shown to be Sudakov suppressed \cite{Buffing:2017mqm,Manohar:2012jr,Mekhfi:1985dv}), and the effect of flavor and fermion number interference, see e.g. \cite{Kasemets:2017vyh}.

The signature of the process is the detection of two positively charged muons (or electrons) $\mu^+$ in the final state as the result of the leptonic decay of each $W^+$, 
and missing energy due to the invisibility of the neutrinos. We study the tree-level results from quark-antiquark annihilation for the flavors $u,d,c,s$.  Each hard process is then of the kind: 
\begin{equation}
q\bar q\rightarrow W^+ \rightarrow\mu^+\nu_\mu.
\end{equation}
The active quarks can be unpolarized ($q$) or in a definite polarization state (longitudinal polarization $\Delta q$).
The cross section of the SSW process in DPS, in presence of spin correlations, reads \cite{Kasemets:2013nma}:
 \begin{equation}
 \begin{aligned}
\frac{d\sigma}{\prod_{i=1}^2 d\eta_{i}dk_{\st_i}{}^{2}d\eta_{\nu_i}}&=\left(\frac{4\pi}{s}\right)^2
\frac{1}{C}\sum_{q_1 q_2 q_3 q_4}K_{q_1\bar q_3}K_{q_2\bar q_4} \\ 
&\times \left\{\left(\omega_1^{-}\omega_2^{-}\right)^2 \int d^2 \boldsymbol {y}(f_{q_1q_2}+f_{\Delta q_1 \Delta q_2})(\bar f_{\bar q_3\bar q_4}+\bar {f}_{\Delta \bar q_3 \Delta \bar q_4})\right. \\ 
&\quad+\left(\omega_1^{-}\omega_2^{+}\right)^2 \int d^2 \boldsymbol {y}(f_{q_1\bar q_4}-f_{\Delta q_1 \Delta \bar q_4})(\bar f_{\bar q_3 q_2}-\bar {f}_{\Delta \bar q_3 \Delta q_2})\\ 
&\quad+\left(\omega_1^{+}\omega_2^{-}\right)^2 \int d^2 \boldsymbol {y}(f_{\bar q_3 q_2}-f_{\Delta \bar q_3 \Delta q_2})(\bar f_{ q_1\bar q_4}-\bar {f}_{\Delta  q_1 \Delta \bar q_4}) \\ 
& \left.\quad+\left(\omega_2^{+}\omega_2^{+}\right)^2 \int d^2 \boldsymbol {y}(f_{\bar q_3 \bar q_4}+f_{\Delta \bar q_3 \Delta \bar q_4})(\bar f_{q_1 q_2}+\bar {f}_{\Delta  q_1 \Delta  q_2})\right\}  , 
\end{aligned}
 \label{CrosSecSab}
\end{equation}
where $\omega_i^{\pm}=1\pm\tanh\left(\frac{1}{2}(\eta_{i} - \eta_{\nu_i})\right)$. The quantities $\eta_{i}$, $\eta_{\nu_i}$, and $k_{\st_i}$ are the rapidity of the produced muon, rapidity of the neutrino and transverse momentum of the muon from 
hard interaction $i$. $C$ is a symmetry factor which is set to 2 because of the indistinguishability of 
the final states from the two hard interactions. $K_{q_i \bar q_j}$ encodes the dependence on coupling factors, the width of the $W$-boson etc. and is given in Appendix~\ref{AppendixA}.
 
The expression involves two different DPDs, the unpolarized ($f_{qq}$) and longitudinally polarized ($f_{\Delta q\Delta q}$) distributions for 
quarks and antiquarks. Transverse quark polarization does not contribute due to the left-handed (right-handed) nature of the coupling between the $W$-boson with quarks (antiquarks). The mixed spin configurations 
(e.g. $f_{q\Delta q}$) are not allowed because of the parity conserving nature of QCD.
The inclusion of the longitudinally polarized distributions contributes to a change in both the magnitude and shape of the final-state distributions. 
The direct effect of polarization on the distributions of final-state particles through the hard cross section is a feature of polarization in DPS not shared with the other correlations. It originates in the difference in angular momentum between particles 
in different spin states. 
The arguments of the distributions read $f_{ab}(x_1,x_2, \boldsymbol y;\mu_1,\mu_2)$ and $\bar f_{ab}(\bar x_1,\bar x_2, \boldsymbol y;\mu_1,\mu_2)$, where $x_i,\bar x_i$ are the longitudinal fractions 
of momentum of the partons $a$ and $b$. $\bm y$ is the separation in the transverse plane between the two hard scatterings, and $\mu_i$ is the renormalization scale related to parton $i$\footnote{When not 
needed, the explicit dependence on the renormalization scale is omitted.}. For the SSW process, the natural choice for the renormalization scales is the hard scale of the individual interactions, i.e. the mass of 
the W-boson, $\mu_1=\mu_2=Q=m_W$. 

In order to focus on the spin and longitudinal momentum correlations, we will assume factorization between the longitudinal and transverse dependences, with the $\bm y$-dependent profile $G(\bm y)$ 
independent of flavor, parton type, and longitudinal momenta, such that:
\begin{equation}
f_{ab}(x_1,x_2,\bm y;Q_0)=g_{ab}(x_1,x_2;Q_0) G(\bm y),
\label{Factor1}
\end{equation}
where 
\begin{equation}
 \int d^2\bm y G^2(\bm y)=\sigma_{\text{eff}}^{-1}.
 \label{sigmaeffective}
 \end{equation} 
The factorization~\eqref{Factor1} eliminates all possible correlations between 
momentum fractions and transverse separation. If one further assumes that the $x_i$-dependent function $g$ is equal to the product of two single parton distributions (PDFs), one arrives at the definition of 
$\sigma_{\text{eff}}$ often used in DPS phenomenology and extracted experimentally. In spite of the fact that the DPDs entering the proper factorization theorems for DPS cannot be formally factorized as 
in~\eqref{Factor1}, see~\cite{Buffing:2017mqm}, and the limitations from ignoring the transverse-longitudinal kinematical correlations, 
see~\cite{Gaunt:2009re,Chang:2012nw,Rinaldi:2013vpa,Rinaldi:2014ddl,Rinaldi:2015cya,Diehl:2014vaa}, the effective cross section $\sigma_{\text{eff}}$ can be useful. 
In particular, eq.~\eqref{Factor1} allows us to single out the different effects of longitudinal kinematical correlations in the various observables.

Given an expression for the DPDs at an initial low energy scale $Q_0$ we implement (unpolarized and polarized) double DGLAP  evolution equations (dDGLAP) up to a maximum mass scale given by the mass of 
the produced particle, i.e.\ $Q=m_W$~\cite{Kirschner:1979im,Shelest:1982dg}.
We implement dDGLAP as two independent 
evolutions, one for each parton, with the kinematical constraints $x_1+x_2\le 1$ and $\bar x_1+\bar x_2\le 1$, see e.g.~\cite{Diehl:2014vaa}.  These constraints alone already introduce kinematical 
longitudinal correlations, which are investigated in the following sections.
We neglect the contribution from $1\rightarrow 2$ splitting~\cite{Diehl:2017kgu}, as this is suppressed for SSW production. However, a significant effect  on the total cross section has been found in~\cite{Cabouat:2019gtm}, prior to imposing phase-space cuts to eliminate the SPS contribution. We expect that these types of cuts largely reduce also the splitting contribution of DPS, but it would be interesting to further quantify this statement through a dedicated investigation. 

% !TEX root = WWLongPaper.tex
%%%%%%%%%%%%%%%%%%%%%%%%%%%%%%%%%%%%%%%%
\section{Models of double parton distributions}
\label{Sec:4}
%%%%%%%%%%%%%%%%%%%%%%%%%%%%%%%%%%%%%%%%

To investigate parton correlations in the DPS cross section, we implement different models for the DPDs at the initial scale $Q_0$. The different DPDs are then evolved up to the hard scale of the $W$-boson mass and used as input to the cross section formula~\eqref{CrosSecSab}. 
While presenting the four main DPD models we  use for studying correlations, we include, for completeness, the case where correlations are entirely absent.

\subsubsection*{No correlation}
If all correlations are removed, at all scales, one can factorize the dependence on $x_1$ and $x_2$  in eq.~\eqref{Factor1}.  In this case, the DPDs are given by:
\begin{equation}
f_{ab}(x_1,x_2, \bm y;Q)=f_a(x_1;Q)f_b(x_2;Q)G(\bm y),
\label{nocorr}
\end{equation}
where $f_a(x,Q)$ is the PDF from single parton scattering for the parton $a$. 
In the absence of correlations, the single PDFs evolve separately with the unpolarized single DGLAP evolution equations.   
The factorized form~\eqref{nocorr} is then valid across all energy scales, and the (separate) evolutions of the two single PDFs do not create correlations.  One should bear in mind that imposing separately $x_1<1$ and $x_2<1$, as on the 
right-hand side of eq.~\eqref{extremeFact}, does not ensure $x_1+x_2<1$, as required by momentum conservation when the two partons come from the same parent hadron. 
We will not include the ``no correlation" scenario in our numerical results.

\subsubsection*{Minimal correlation}
The DPDs are modeled as the product of single PDFs at an initial scale, that is:
\begin{equation}
f_{ab}(x_1,x_2, \bm y;Q_0)=f_a(x_1;Q_0)f_b(x_2;Q_0)G(\bm y),
\label{extremeFact}
\end{equation}
meaning that all kinds of correlations are set to zero at the initial scale $Q_0$. Eq.~\eqref{extremeFact} is not valid at any scale different from $Q_0$. Correlations between the longitudinal momenta of the partons arise 
as the result of the double DGLAP evolution equations. 
We use this scenario as a baseline for our analysis and call it ``minimally correlated", to point out that correlations cannot in principle be erased in DPDs.  
However, the correction introduced by the unpolarized double DGLAP evolution compared to two DGLAP evolution kernels is minimal in the kinematical region we are interested in, and the minimally 
correlated scenario is quantitatively equivalent to the uncorrelated one for our level of accuracy. The cross section $\boldmath{\sigma}_{\text{\bf min-corr}}$ is given by eq.~\eqref{CrosSecSab} where the polarized DPDs are set to zero.

\subsubsection*{Positive polarization} 
 In this scenario we include parton polarization by using polarized distributions which individually saturate the positivity bounds. Setting all the other polarized DPDs to zero, 
 the positivity bound on the longitudinally polarized DPDs is $| f_{\Delta q\Delta q}|\le f_{qq}$. Saturating this bound leads to polarized distributions equal to the unpolarized ones at the initial scale~\cite{Diehl:2013mla,Diehl:2014vaa}, i.e.:
\begin{equation}
f_{\Delta a \Delta b}(x_1,x_2, \bm y;Q_0)=f_{ab}(x_1,x_2, \bm y;Q_0)=f(x_1;Q_0)f(x_2;Q_0)G(\bm y).
\label{LongiFact}
\end{equation}
The factorized form~\eqref{LongiFact} is valid only at the initial scale, while at higher scales the polarized double DGLAP evolution equation introduces the correlations as previously described. 
The kernels that govern the evolution of the polarized DPDs are responsible for a relative decrease of the polarized DPDs for values of $Q$ larger than the initial scale $Q_0$. Therefore, the amount of polarization 
is maximal at $Q=Q_0$ and decreases towards higher values of $Q$.  The cross section $\boldmath{\sigma}_{\text{\bf pos-pol}}$ is given by the full expression in \eqref{CrosSecSab}.

\subsubsection*{Mixed polarization} 
The positivity bound only limits the modulus of the polarized distributions, allowing for any combination of signs separately for each set of partons. 
To explore the sensitivity to these signs, we consider polarized distributions which include a mixture of positive and negative distributions. 
Specifically, in this scenario we individually saturate the positivity bound at the initial scale with a negative sign when the two selected partons 
are both quarks or antiquarks and with a positive sign when the pair is composed by one quark and one antiquark~\cite{Cotogno:2018mfv}. Namely:
\begin{equation}
f_{\Delta a \Delta b}(x_1,x_2, \bm y;Q_0)=(-1)^nf_{ab}(x_1,x_2, \bm y;Q_0)
\label{MixFact}
\end{equation}
where $n=1$ if $ab=q q, \bar q\bar q$, and $n=2$ otherwise. The cross section $\boldmath{\sigma}_{\text{\bf mix-pol}}$ is given by the full expression in eq.~\eqref{CrosSecSab}. 

\subsubsection*{Longitudinal correlations}
 Longitudinal kinematical correlations are explicitly introduced in this scenario. The product of single PDFs used as initial ansatz is corrected by an $x_i$-dependent factor, to account for the kinematical constraint of double parton scattering as explained in~\cite{Gaunt:2009re}. The factorized form~\eqref{extremeFact} is no longer valid at the initial  scale because of a  $x_i$-dependent global factor:
\begin{equation}
f_{ab}(x_1,x_2, \bm y;Q_0)=f_a(x_1;Q_0)f_b(x_2;Q_0)X_{\text{corr}}(x_1,x_2)G(\bm y),
\label{mildFact}
\end{equation}
where $X_{\text{corr}}(x_1,x_2)=(1-x_1-x_2)(1-x_1)^{-2}(1-x_2)^{-2}$. In this case, the longitudinal correlations are present at the initial scale thanks to the explicit factor, and they travel 
towards smaller momentum fractions during evolution. The cross section $\boldmath{\sigma}_{\text{\bf long-corr}}$ is given by~\eqref{CrosSecSab} with the polarized distributions set to zero.

% !TEX root = WWLongPaper.tex
%%%%%%%%%%%%%%%%%%%%%%%%%%%%%%%%%%%%%%%%
\section{DPS parton level results}
\label{Sec:parton}
%%%%%%%%%%%%%%%%%%%%%%%%%%%%%%%%%%%%%%%%

Here, we present the results derived from a calculation of the interaction of point-like partons in pQCD convoluted with the DPDs, according to the factorization theorem for DPS~\cite{Diehl:2011yj,Diehl:2015bca}. 
We refer to this part of the analysis as parton level (PL).

We set the initial scale for the models of the DPDs to $Q_0=1\,\text{GeV}$ and implement (unpolarized and polarized) double DGLAP evolution (dDGLAP)  
to a final scale $Q=m_W$. $Q_0$ should be a low scale, chosen around the scale where perturbative calculations start to be valid. The reason for this is that, 
once the positivity bounds are saturated at  $Q_0$, they will be satisfied at all larger scales, but will typically be violated if perturbation 
theory is used to evolve the DPDs down to lower scales.
The single parton PDFs we use are the leading-order MSTW2008lo distributions~\cite{Martin:2009iq}. A change of the single PDF-set used does impact 
the DPDs, in particular the gluon distributions \cite{Diehl:2014vaa}. However, as we point out below, while discussing the correlation sensitive variables,
the qualitative results of our analysis are fairly stable to this change.
We set $\sqrt{s}=13$ TeV and  $\sigma_{\text{eff}}=15$ mb, 
which is in the range extracted by the CMS collaboration in SSW production~\cite{Sirunyan:2019zox,Sirunyan:2017hlu}.  One should be careful not to over-interpret this quantity, as it has several shortcomings which will be discussed further in Section~\ref{template}. 

The  parton level results of this section are calculated using the following phase-space cuts (``Baseline Selection"): 
\begin{equation}
4\,\text{GeV}\leq k_{\st_i}\leq 45.5\, \text{GeV},\quad |\eta_{i}|\leq3.3.
\label{selection0}
\end{equation}
The constraint $k_{\st_i}\leq 45.5\, \text{GeV}$ does not affect the results, as the amount of the cross section at larger transverse muon momentum is negligible.
The range of $\eta_{i}$ is chosen wider than the typical experimental acceptance at the LHC, and the selection serves as a starting point for the study of final-state distributions in Section~\ref{Sec:FSI}, 
where the cuts are tightened to match the current detector ranges. 
The cross section is calculated with numerical integration 
using the Vegas algorithm within the Cuba Library~\cite{Hahn:2004fe} and the four scenarios are normalized to the value of $\sigma_{\text{min-corr}}$ (1.74 fb) at the parton level. 
The reason for this normalization is that we want to stay as close as possible to the available experimental information on $\sigma_\text{eff}$, which is typically based on the assumption of uncorrelated or weakly correlated DPS events.

\begin{figure}[t]
\begin{center}
\subfigure[]{\includegraphics[width=0.42\textwidth]{%
    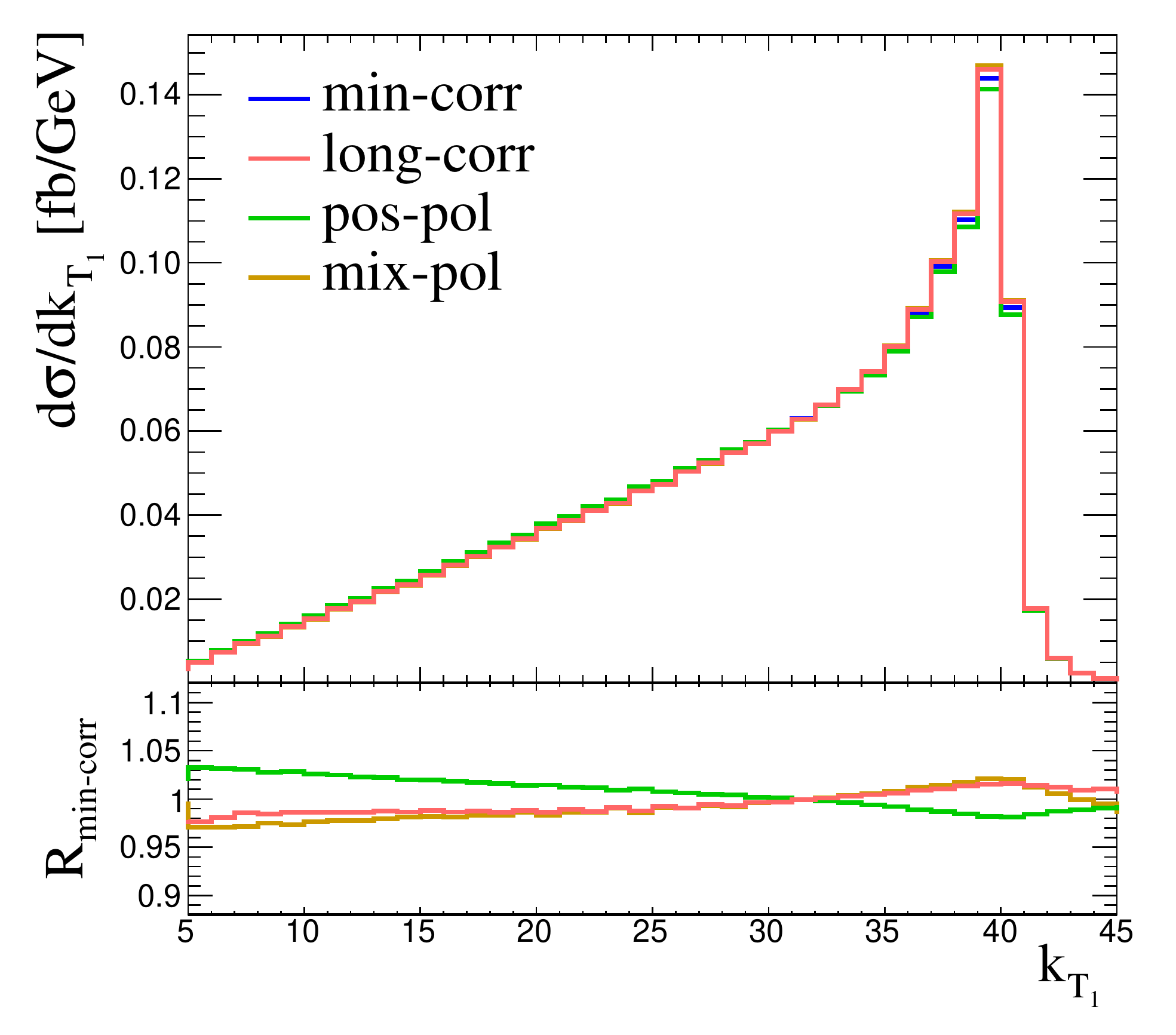}}
\subfigure[]{\includegraphics[width=0.42\textwidth]{%
    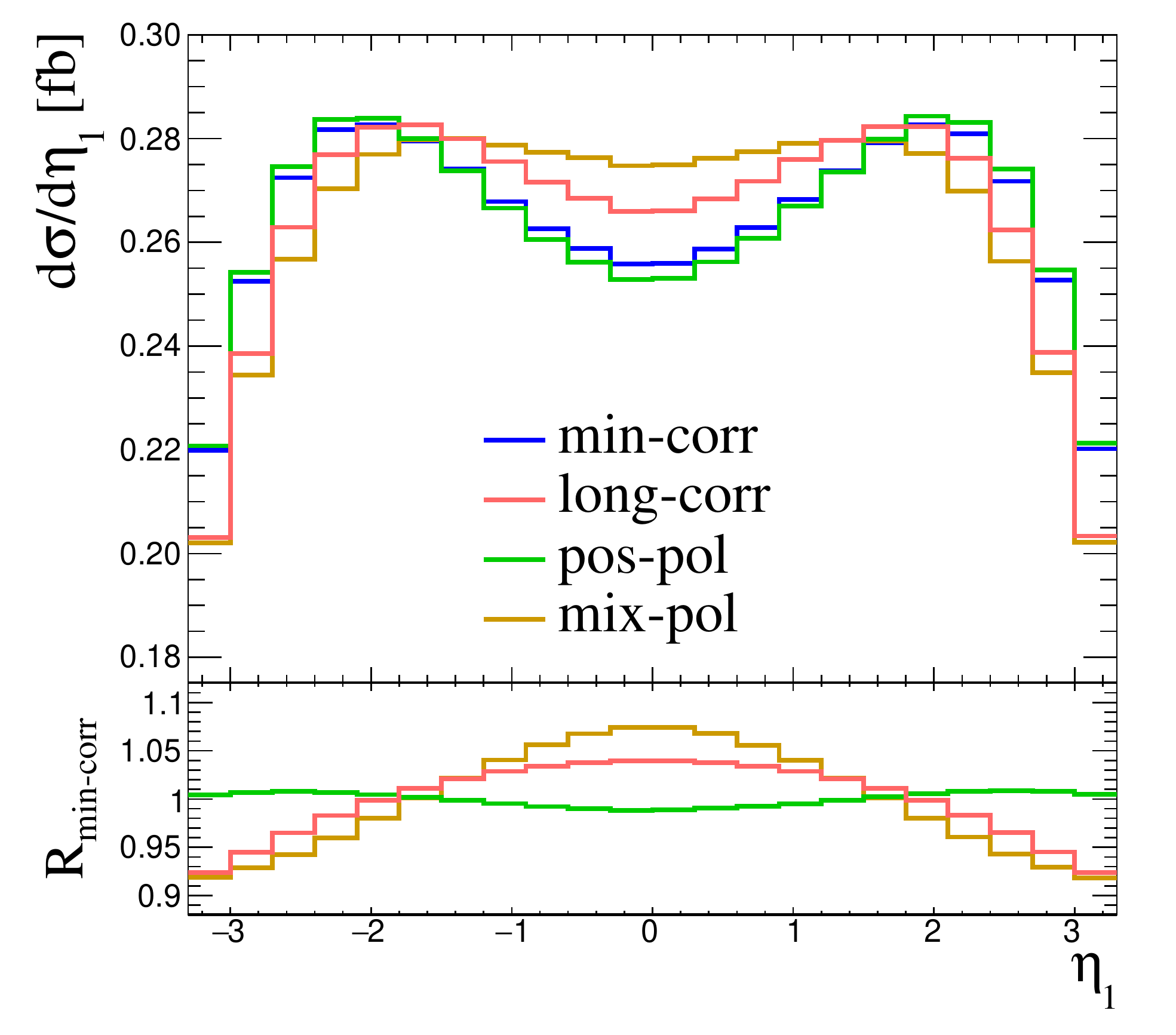}} 
\caption{\label{fig:0_muOne_eta_pt} The distributions of the muon transverse momentum and rapidity for the different models. The distributions for $\mu_{2}$ are identical and are not shown. 
The ratio $R_{\text{min-corr}}$ shows the comparison of different models to the min-corr scenario.}
\label{kTandEta}
\end{center}
\end{figure}

The cross section differential in the rapidity and transverse momentum of one of the muons is shown in  Fig.~\ref{kTandEta}.  
The distributions of $\mu_1$ (muon from hard interaction 1) and $\mu_2$ (muon from hard interaction 2) are equivalent, as we do not impose any hierarchy 
in magnitude between the hard scales. The labeling is of course purely theoretical, as no experimental distinction between the two detected muons can be made.

The $k_{\st_1}$-distributions in Fig.~\ref{kTandEta}(a) are all peaked around the value $k_{\st_1}=m_W/2$. 
The cross section value, for all the curves, decreases by nearly two orders of magnitude from the value at the peak and $k_{\st\text{max}}=45$ GeV. 
The situation is different for the rapidity distribution in  Fig.~\ref{kTandEta}(b), for which the rapidity range selected in~\eqref{selection0} 
leaves out a non negligible portion of the cross section. The rapidity distribution is symmetric under $\eta_{i}\rightarrow-\eta_{i}$, as expected upon noticing that the 
cross section formula~\eqref{CrosSecSab} is invariant under this exchange. The maximum values of the cross section are reached at $\eta_{1}\in[1.5,2.5]$ with small 
differences between the different scenarios. For all curves, the cross section decreases both towards central and peripheral values of the rapidity interval. The lower panel of the figures shows the ratio 
of the three correlated scenarios to the min-corr result, i.e. $R_{\text{min-corr}}=d\sigma_{\text{corr}}/d\sigma_{\text{min-corr}}$. 
This ratio gives a clear demonstration of the extent to which the shapes of the distributions depend on the partonic correlations.

\begin{figure}[t]
\begin{center}

\subfigure[]{\includegraphics[width=0.42\textwidth]{%
    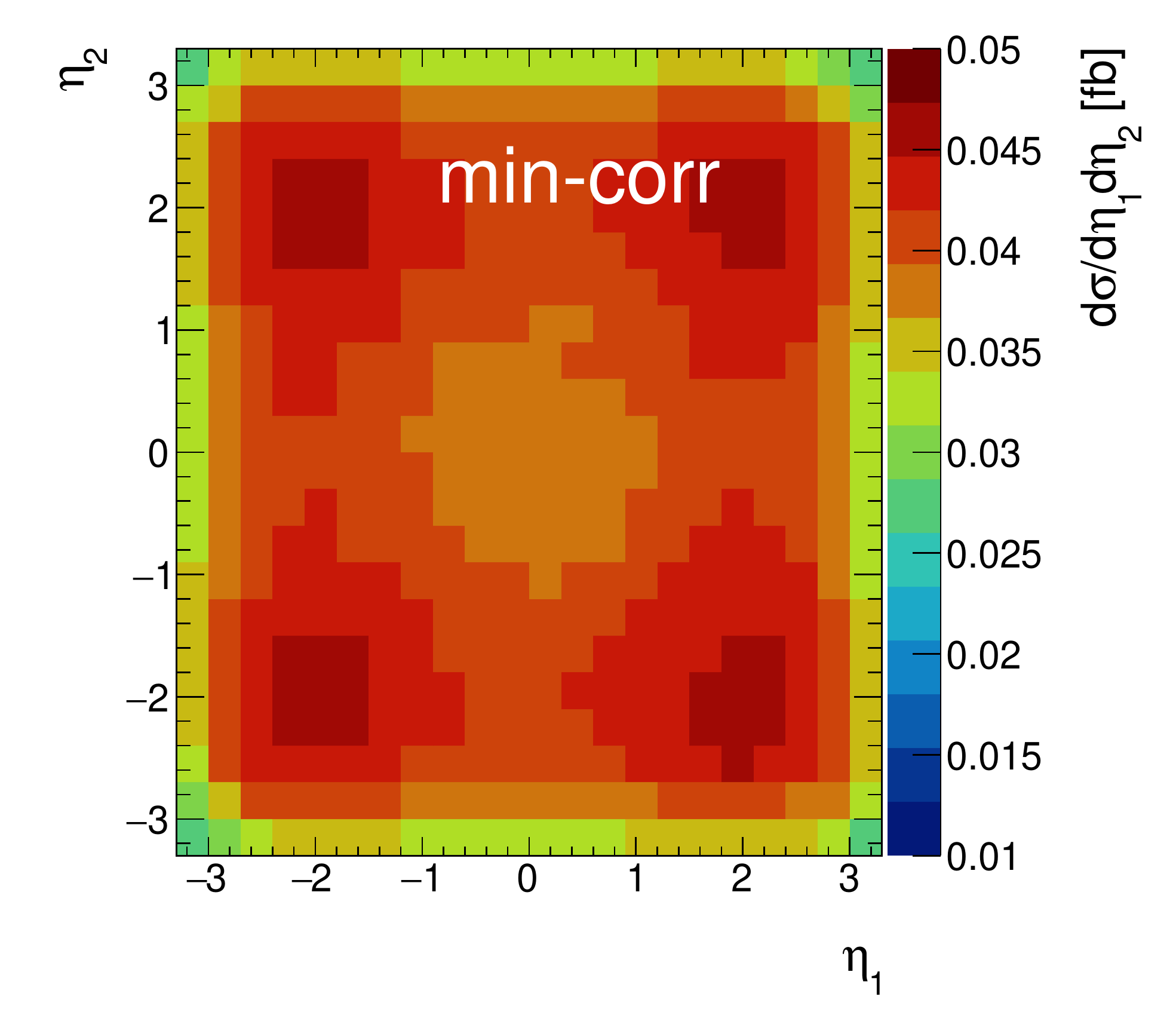}
    \label{2DPlotmincorr}} 
\subfigure[]{\includegraphics[width=0.42\textwidth]{%
   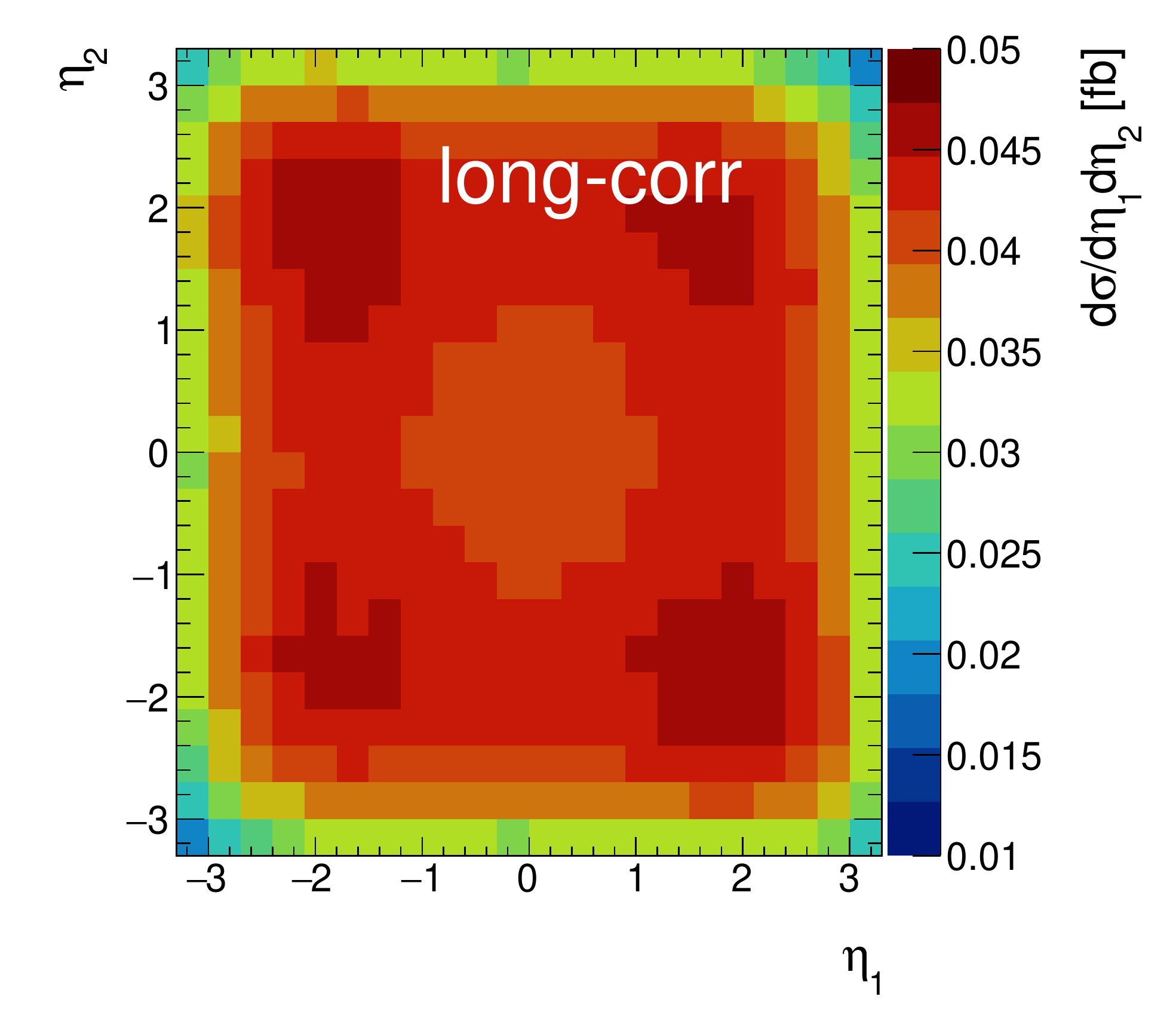}} 

\subfigure[]{\includegraphics[width=0.42\textwidth]{%
    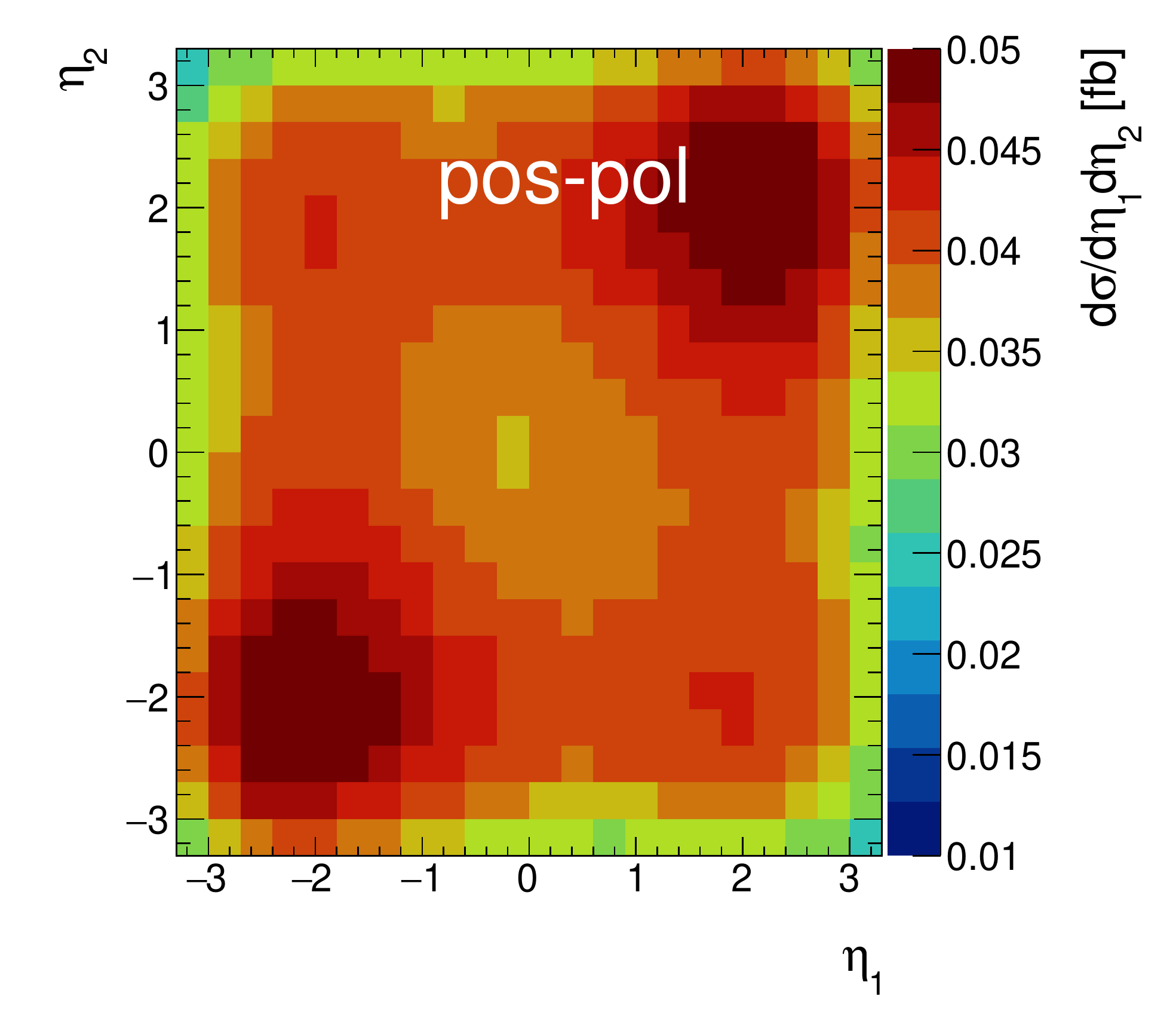}} 
\subfigure[]{\includegraphics[width=0.42\textwidth]{%
    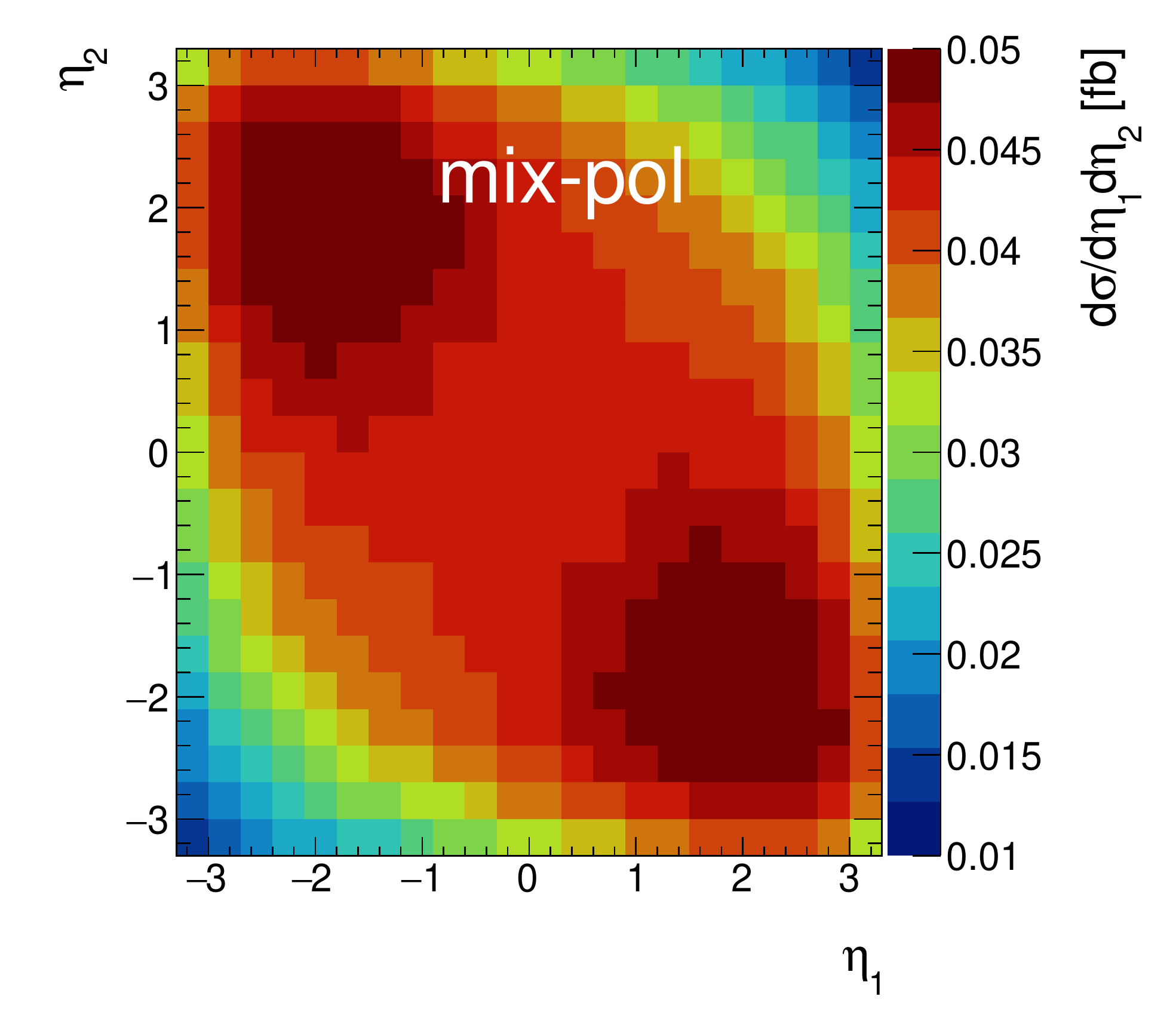}} 
    
    \caption{\label{fig:0_etaeta_2} Double-differential distributon for rapidity of the two muons, with cuts as in~\eqref{selection0}. 
    Top left: minimally correlated scenario. Top right: longitudinal correlation. Bottom left: positive polarization. Bottom right: mix polarization.}
    \label{2DPlots}
\end{center}
\end{figure}

We now turn into the analysis of variables which are particularly correlation sensitive. Fig.~\ref{2DPlots} shows the double differential cross section
with respect to $\eta_1$ and $\eta_2$, $d\sigma/d\eta_1 d\eta_2$, for the four scenarios.
 As expected, when correlations are absent  the distribution is symmetric in all four quadrants  (a). There are some slight deviations from this symmetry when longitudinal correlations are introduced (b), and the symmetry is largely distorted in the presence of polarization (c and d). 
The two polarized scenarios lead to opposite effects. The pos-pol (mix-pol) model increases the rate of muons produced in the same (opposite) hemisphere.
 
\begin{figure}[t]
\begin{center}
\subfigure[]{\includegraphics[width=0.42\textwidth]{%
    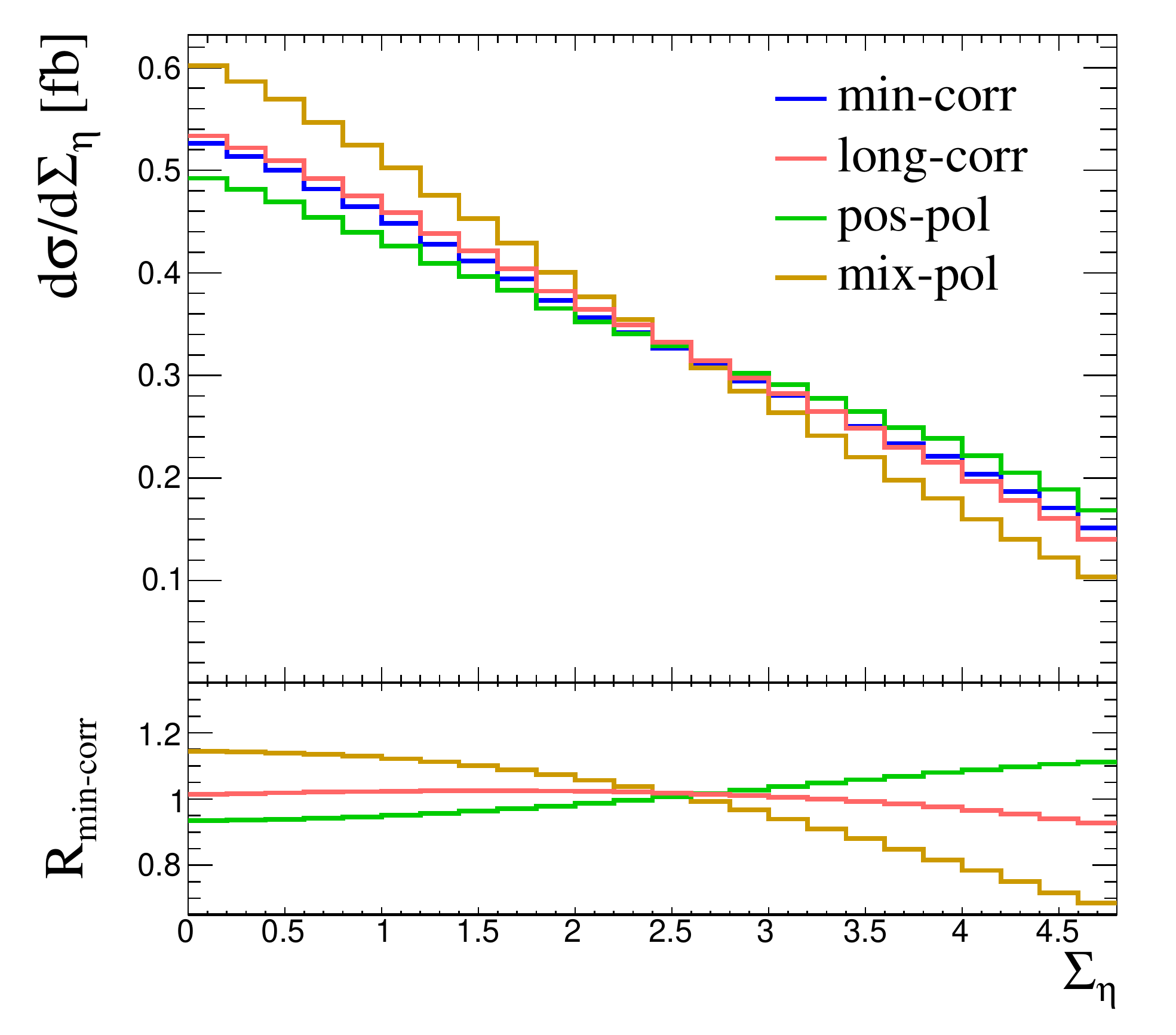}} 
\subfigure[]{\includegraphics[width=0.42\textwidth]{%
    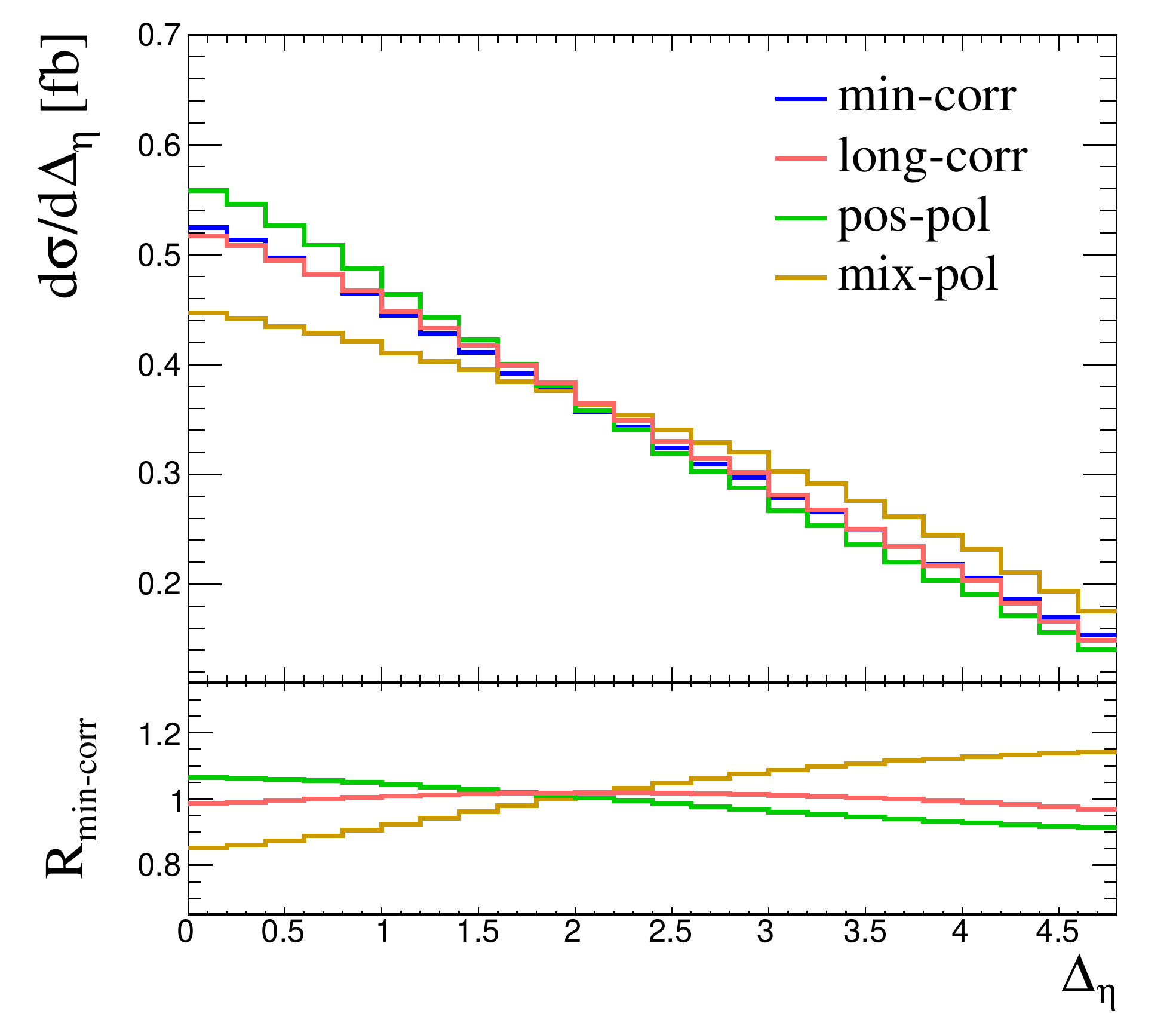}} 
    \subfigure[]{\includegraphics[width=0.42\textwidth]{%
    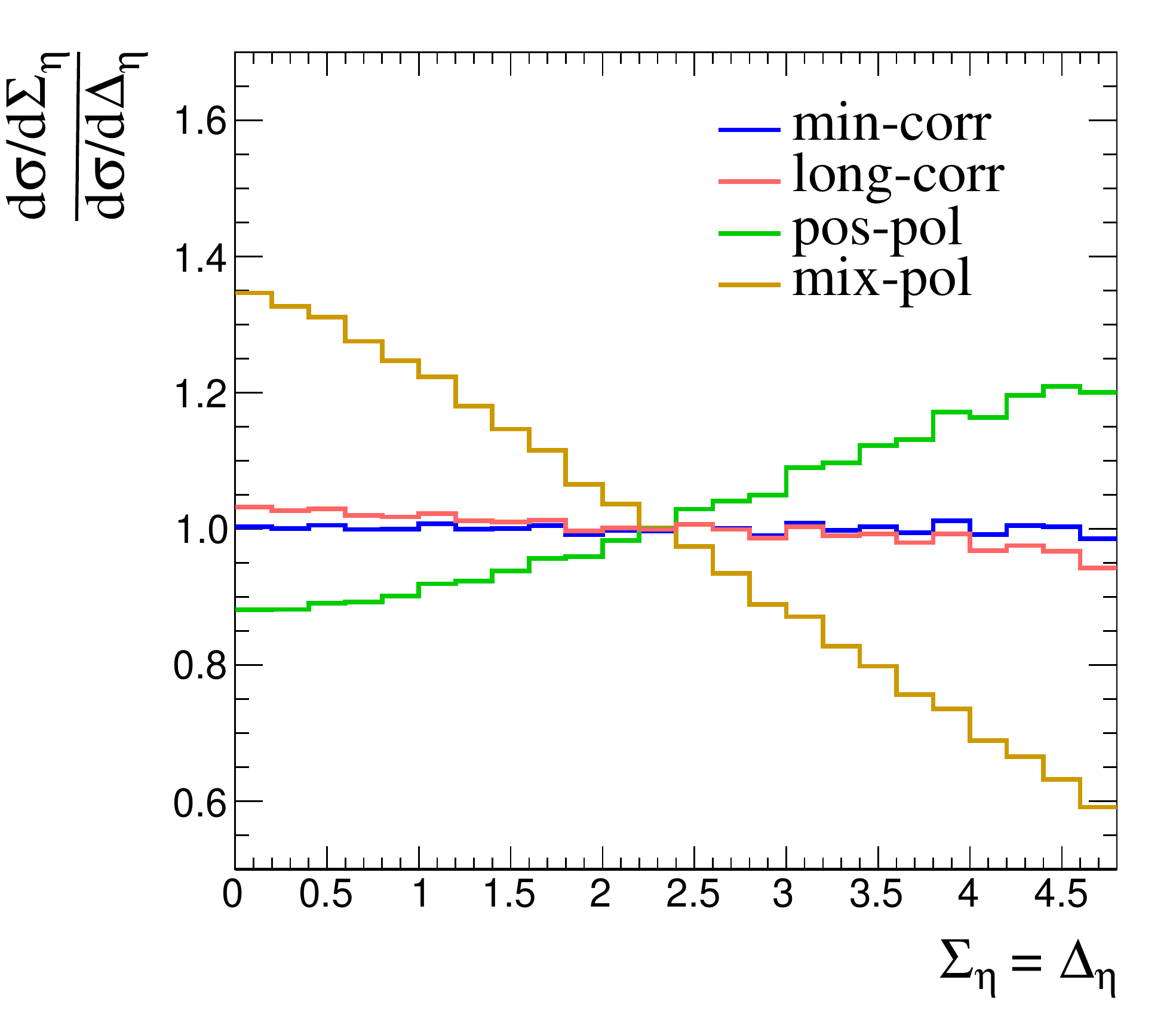}
    \label{RatioSum}} 
     \subfigure[]{\includegraphics[width=0.42\textwidth]{%
     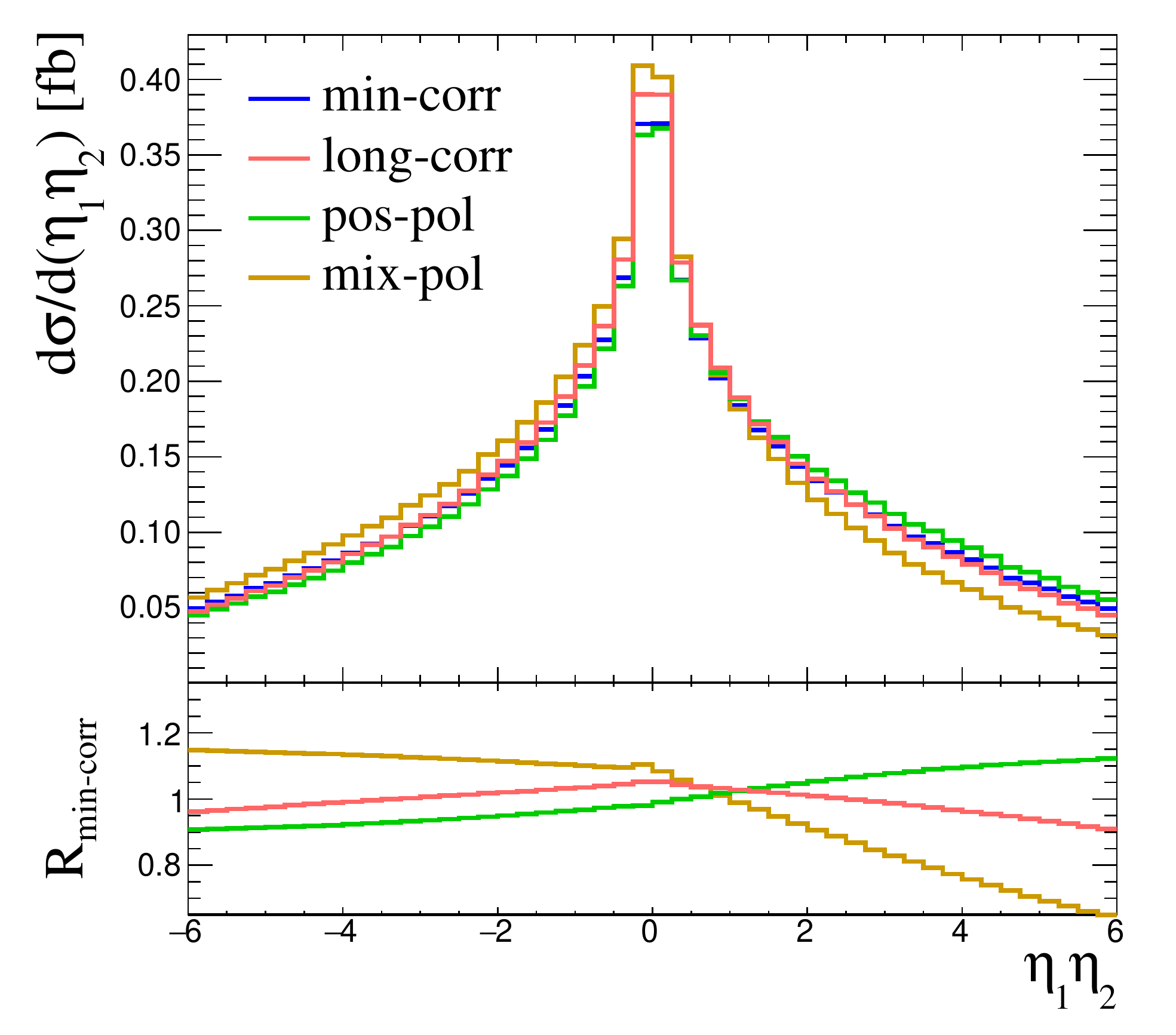}
     \label{EtaProduct}}   
    \caption{\label{fig:0_mupair_etaDifSum} Upper panels: distributions of the muon rapidity sum $\Sigma_\eta$ (top left) and difference $\Delta_\eta$ (top right) for the different models. 
    Lower panels: distribution of the ratio between the sum and difference of rapidities versus the 
    values of $\Sigma_\eta$ and $\Delta_\eta$ respectively in the relevant ranges (bottom left) and
    the product of the muon rapidity for the different models (bottom right). The ratio $R_{\text{min-corr}}$ shows the comparison of different models to the min-corr scenario.
}
\label{SumDiff}
\end{center}
\end{figure}

A less differential version of this observation is represented by Fig.~\ref{EtaProduct}, where we display the distribution in the rapidity product 
$d\sigma/d(\eta_1\eta_2)$. The curves are symmetric with respect to the axes $\eta_1\eta_2=0$  when correlations are absent and asymmetric in the correlated cases. 
All four scenarios have a similar shape when the two muons are produced in opposite hemispheres (i.e. $\eta_1 \eta_2$ is negative), and a shape dependence is 
turned on when the product approaches zero. Such a separation of the phase space into portions corresponding to different signs of the rapidity product allows us to identify the amount 
of the cross section corresponding to measurements of muons detected in the same hemisphere of the detector ($\sigma(\eta_1\eta_2\ge 0)$) or opposite ones 
($\sigma(\eta_1\eta_2\le 0)$). This separation is extremely convenient because it further translates into a number that measures this unbalance, namely:
\begin{equation}
A=\frac{\sigma(\eta_{1}\eta_{2}<0)-\sigma(\eta_{1}\eta_{2}>0)}{\sigma(\eta_{1}\eta_{2}<0)+\sigma(\eta_{1}\eta_{2}>0)}.
\label{asymmetry}
\end{equation}
The asymmetry must be exactly zero when  correlations are absent, as clear from eq.~\eqref{CrosSecSab}, whereas any deviation from zero will be a sign of correlations. 
The questions whether a significant deviation from zero can be detected and which kind of correlations are the best candidates for producing such a distortion 
will be discussed in Section~\ref{Sec:Corr}. Another advantage of this variable is that the values obtained for each scenario 
are stable under certain crucial modifications, for instance the change of the initial scale of the models $Q_0$, the PDF set used and the specific value of $\sigma_{\text{eff}}$. 
These are all sources of uncertainty that can, however, affect the magnitude of the cross section and, therefore, the significance of a measurement of $A$ (see Section~\ref{Sec:Corr}). 
The first row of Table~\ref{tab:allnumbers} shows the values of the variable $A$ for the Baseline Selection at PL. The largest asymmetry is obtained with the mix-pol model, 
which favors the production of muons in the opposite direction rather than in the same hemisphere, indicated by a positive value of $A$. 
A very small positive asymmetry is obtained in the longitudinal correlation case, while this quantity is almost zero for the min-corr scenario, confirming that it effectively 
represents the absence of correlations. A negative asymmetry is instead obtained for the positive polarization, where the cross section moves towards positive values of $\eta_1\eta_2$ 
(see Figures \ref{2DPlots} and \ref{EtaProduct}).

Correlations can also manifestly shape the distribution of the sum and difference of muon rapidities. After defining $\Sigma_\eta=|\eta_1+\eta_2|$ and $\Delta_\eta=|\eta_1-\eta_2|$, we show the differential cross sections $d\sigma/d\Sigma_\eta$ and $d\sigma/d\Delta_\eta$ in the upper part of Fig.~\ref{SumDiff}. 

The rapidity difference only shows some dependence on the model and, therewith, the different correlations. The two polarized scenarios have opposite effects on the shape, 
where mix-pol gives rise to a somewhat broader distribution while pos-pol results in a more steeply falling spectrum. For the rapidity sum, the pattern is inverted and mix-pol 
(pos-pol) results in a more narrow (broader) distribution. In both cases, the effect of the long-corr scenario is mild. 
Different scenarios lead to different values for the slopes of the linear fit to the curves. We denote $S_{lin}~(\Sigma_{\eta})$ and $S_{lin}~(\Delta_{\eta})$ the result of the linear slope fit
for the sum and difference of the rapidity distributions respectively and show the values in Table~\ref{tab:allnumbers}, see the rows corresponding to the Baseline Selection at PL.
Similarly to the asymmetry, the measurements of the slopes can  discriminate correlations in double parton scattering, although in a less clear way than the 
asymmetry $A$, since the baseline value for the uncorrelated scenario is uncertain. 

\begin{figure}[t]
\begin{center}

\subfigure[]{\includegraphics[width=0.42\textwidth]{%
    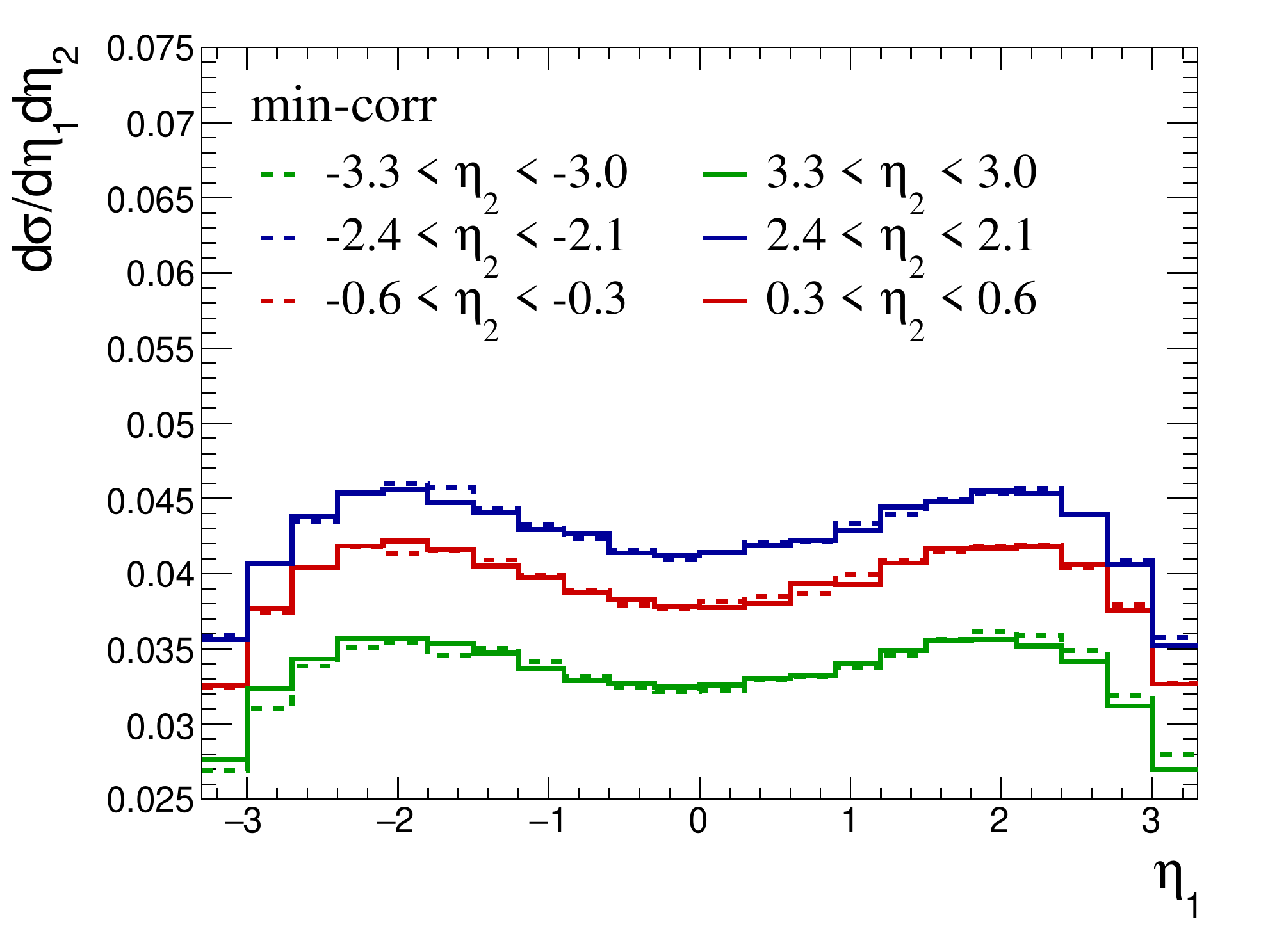}
    \label{slicesmincorr}} 
\subfigure[]{\includegraphics[width=0.42\textwidth]{%
    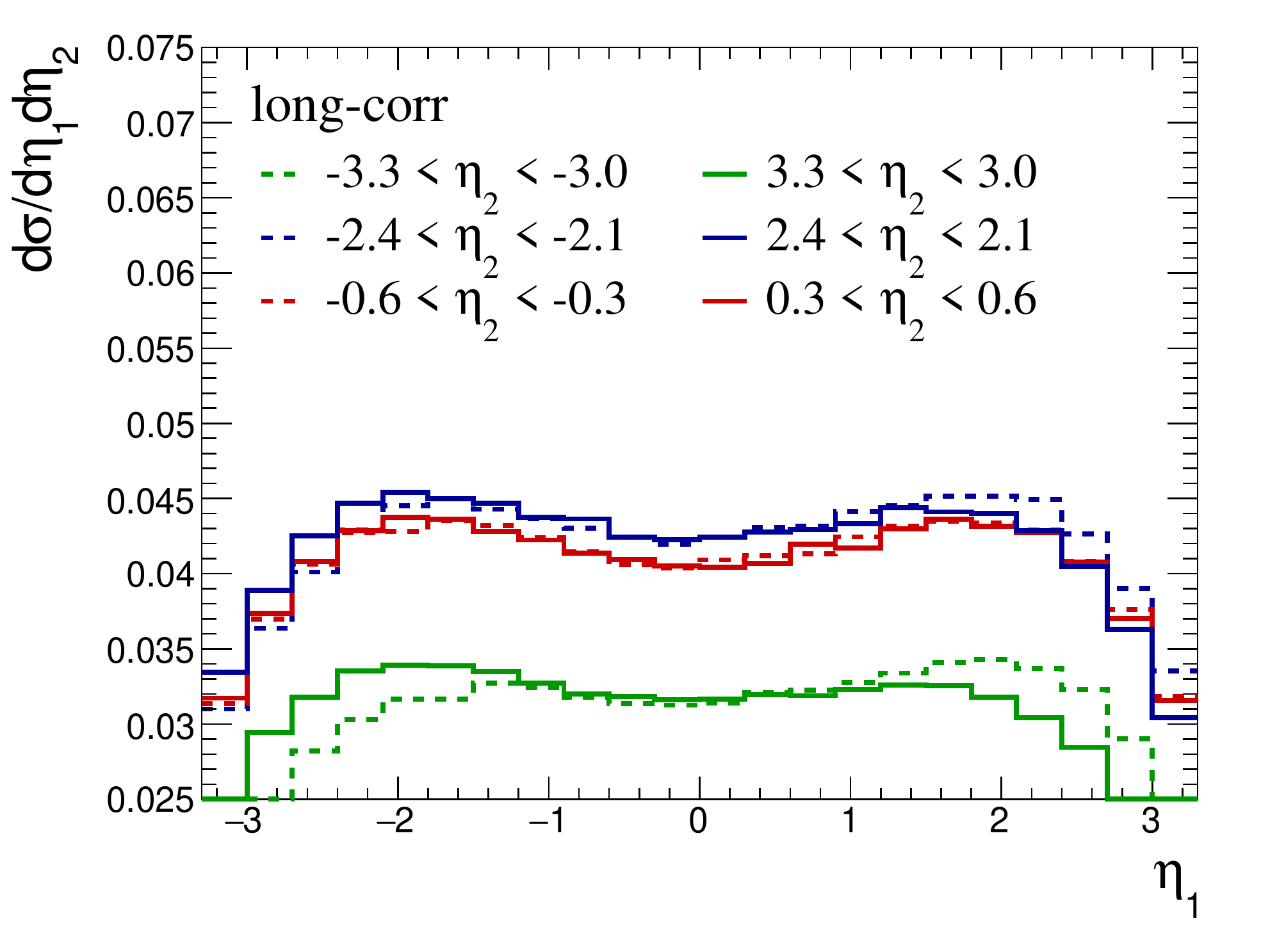}
    \label{slicesxcorr}} 
\subfigure[]{\includegraphics[width=0.42\textwidth]{%
    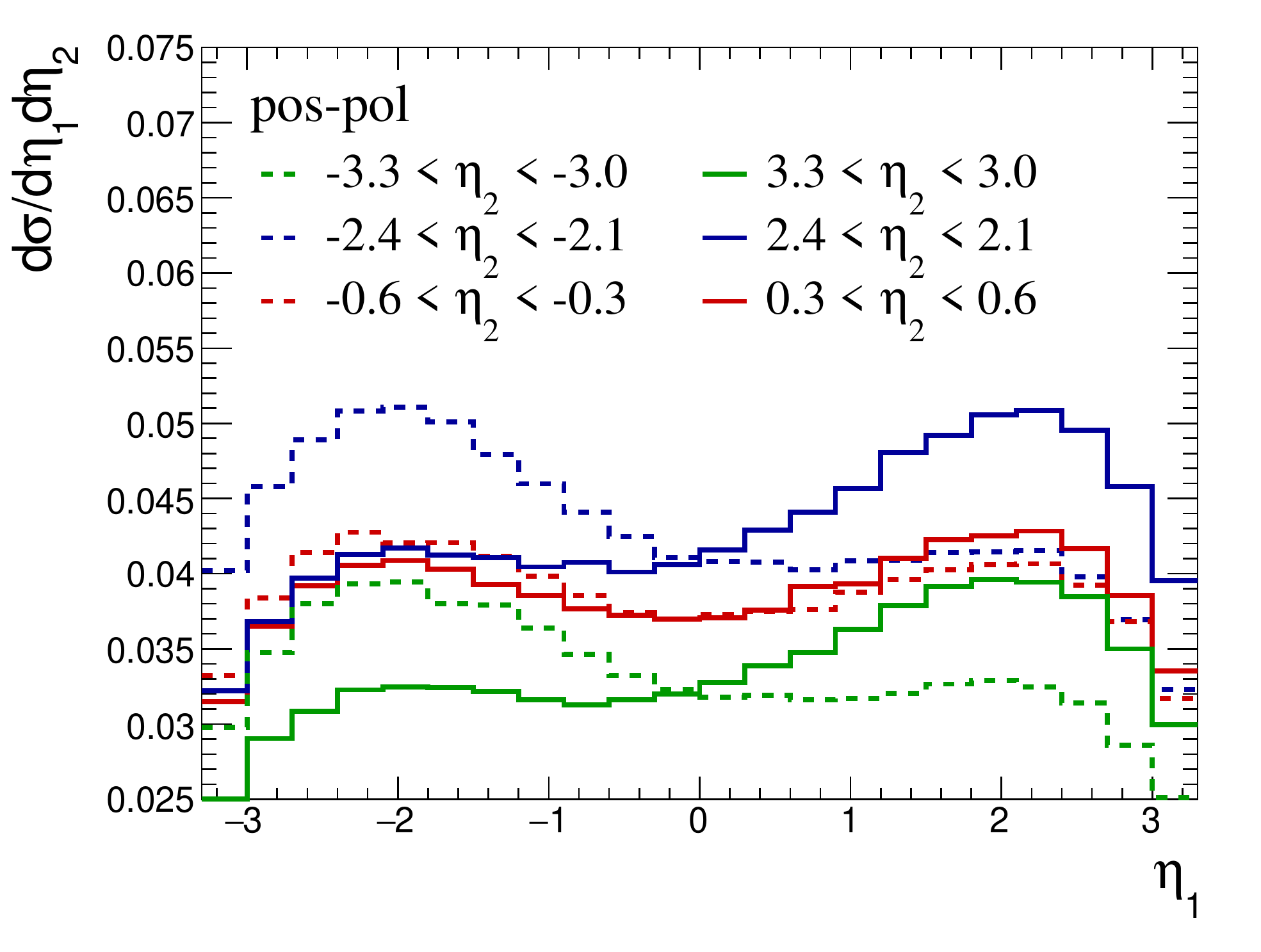}
    \label{slicespospol}} 
\subfigure[]{\includegraphics[width=0.42\textwidth]{%
    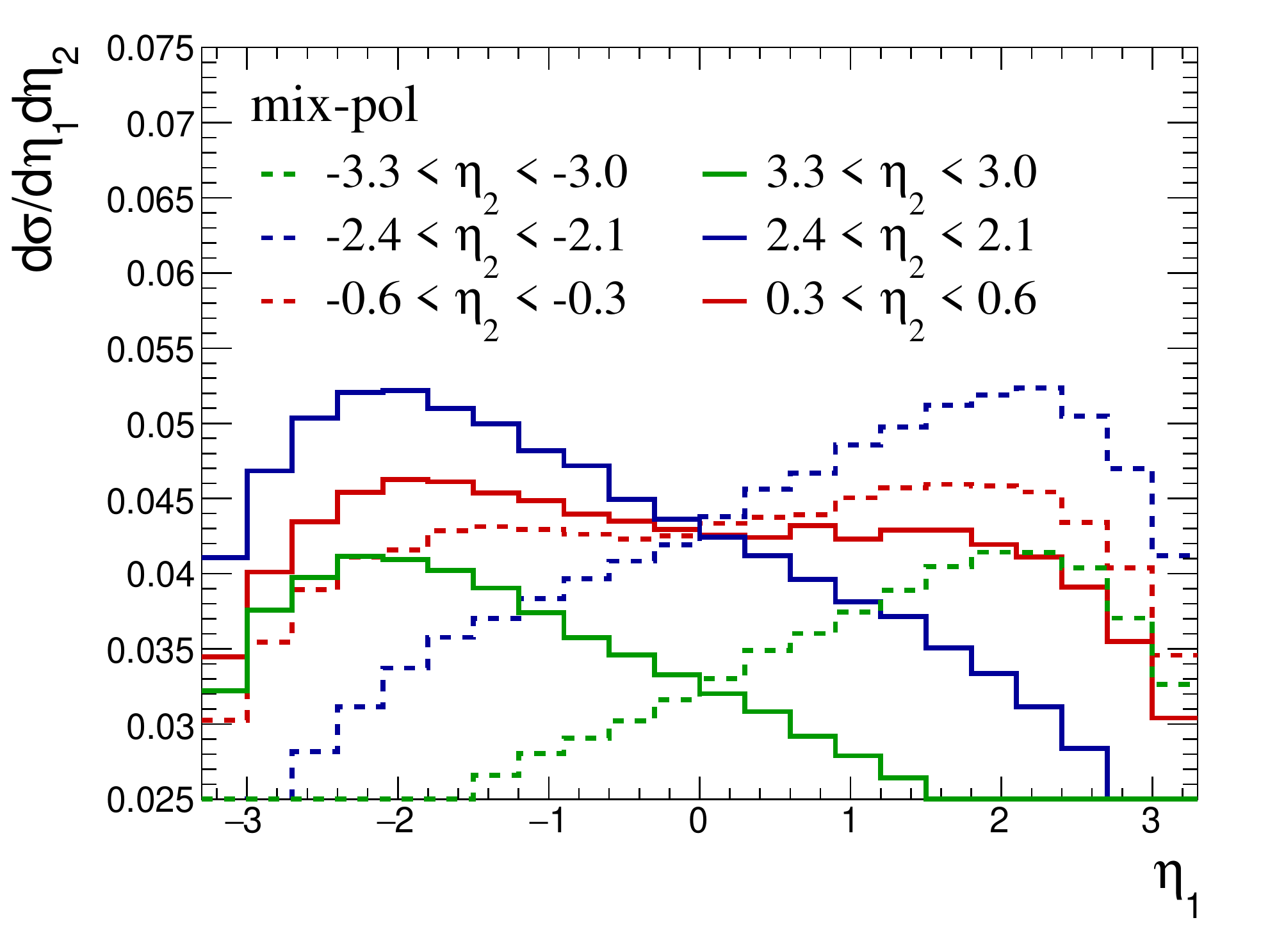}
    \label{slicesmixpol}} 
    
    \caption{\label{fig:0_etaslice_2} Double-differential distribution for rapidity of the two muons plotted for fixed value of $\eta_2$. Different scenarios are plotted in separate panels as indicated in the figures.}
    \label{slices}
\end{center}
\end{figure}

\begin{figure}[h!]
\begin{center}
\subfigure[]{\includegraphics[width=0.42\textwidth]{%
    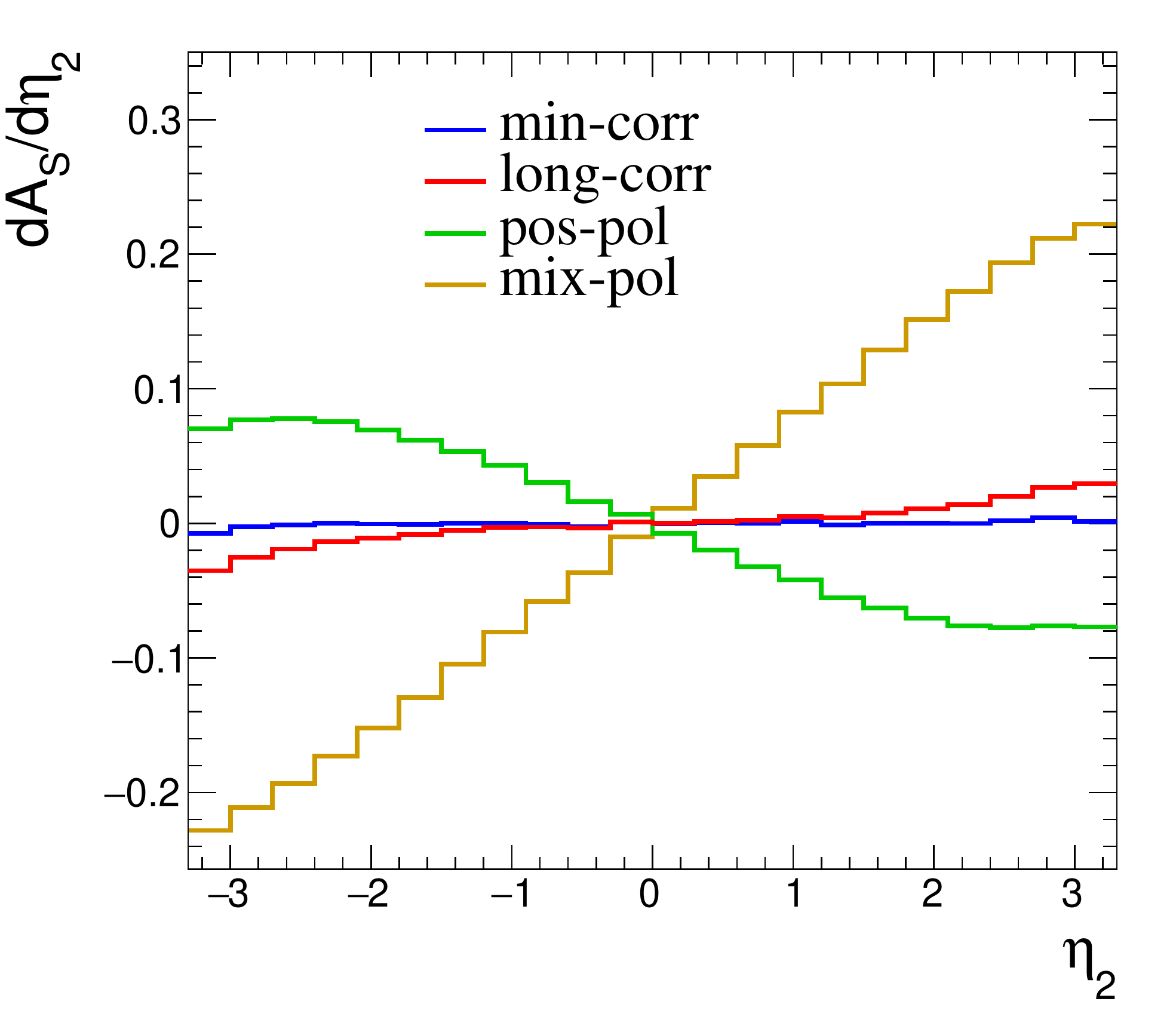}} 
\caption{\label{fig:0_sliceAsym} The distribution of the asymmetry \eqref{eta-asymm} as a function of the second muon rapidity.}
\label{AsymmSlice}
\end{center}
\end{figure}

In fact, the shapes of the cross section in these variables depends on the details of the models used for the double parton distributions even for the uncorrelated scenario. 
For example, changing the single PDFs at the starting scale, can impact the spectrum. Since we are hunting for correlations, this type of dependence on the uncorrelated cross 
section can be problematic. One way to circumvent this issue experimentally is to construct the DPS uncorrelated distribution directly from measured single $W$ production. 
However, this of course still induces some remaining uncertainties. Here we follow an alternative path, and construct variables where we can make exact predictions of 
the uncorrelated result and therewith a more direct access to the correlations. 

In absence of correlations, the cross section distribution in the sum and difference of the muon rapidities have to be exactly equal. This is due to the symmetry of the cross 
section in changing the sign of one of the rapidities (e.g. $\eta_2 \rightarrow -\eta_2$) while keeping the other one fixed. Keeping the decay of the $W$'s fixed, this would be 
induced by a change in the momentum factions probed in the two DPDs $f(x_1,x_2,\y) f(\bar{x}_1,\bar{x}_2,\y) \rightarrow f(x_1,\bar{x}_2,\y) f(\bar{x}_1,x_2,\y)$. 
For two uncorrelated events, where the DPDs can be expressed in terms of PDFs, the two sides are equal. When longitudinal correlations between the partons are induced, this symmetry can be broken. 
Moreover, this symmetry is directly violated in the partonic cross section in \eqref{CrosSecSab} when including the polarized contribution.

A useful variable to consider in order to exploit such information and simultaneously minimize the effect of unknown factors, is the bin-by-bin ratio between the sum and difference distributions. 
This variable is displayed in Fig.~\ref{RatioSum}, the linear slope of the curves is indicated as $S_{lin}(\Sigma_\eta/ \Delta_\eta)$ and its values are reported in the last row of Table~\ref{tab:allnumbers}. 
This ratio, in case of uncorrelated DPS, has to be constant and equal to one. Fig.~\ref{RatioSum} shows a very strong dependence on the different scenarios. The min-corr is, as expected, close to unity in the entire range. 
The long-corr scenario results only in tiny deviations from one. However, the two polarized scenarios differ from unity and from one another. 
The pos-pol scenario gives an increasing curve, while mix-pol leads to a relatively steeply falling result.  This is naturally to be expected, as the two changes induced by these 
scenarios in Figures \ref{SumDiff}(a) and (b) enhance one another in this cross section ratio. 

Naturally the complete information on the rapidity distributions is already contained in the double-differential distributions. 
If one had such a two-dimensional distribution available, the one-dimensional distributions  $d\sigma/d\Sigma_\eta$ and $d\sigma/d\Delta_\eta$ could be obtained by summing all the events 
on the slices $\Sigma_\eta= \Delta_\eta=$ const in the relevant ranges, i.e. along lines parallel to the bisector of the first and third quadrants ($\Delta_\eta=\text{const}$) or the 
second and fourth quadrants ($\Sigma_\eta=\text{const}$). Similarly, the asymmetry~\eqref{asymmetry} would simply be given by the sum of all the events in the second and fourth quadrants 
(opposite hemisphere) minus those in the first and third quadrants (same hemisphere), normalized to be a fraction of all events. 

To conclude the discussion of the piece of information that follows from the double-differential distribution of the muon rapidities, we now present another variable that involves 
changes in the rapidity distributions. We construct a one-dimensional distribution obtained by rapidity slicing, i.e. looking at the two-dimensional variable $d\sigma/d\eta_1d\eta_2$ for fixed interval of $\eta_2$. The situation is displayed in Fig.~\ref{slices}, where the different curves are obtained by varying the interval of 
$\eta_2$, and the different scenarios are represented in separate panels for clarity. Once again, correlations create an asymmetric pattern, leading to lines that change their shapes when 
the $\eta_2$ slices change. As previously explained, interchanging $\eta_2\rightarrow -\eta_2$ must leave the distribution unchanged if correlations are not present, a fact that is also 
visible from Fig.~\ref{2DPlotmincorr}, upon noticing that the variable we are interested in is obtained by slicing the two-dimensional distributions using lines $\eta_2=\text{const}$.  
In the min-corr scenario of Fig.~\ref{slicesmincorr} all the curves are equal, up to normalization, for different slices of $\eta_2$. The situation is slightly 
modified by the longitudinal correlations in Fig.~\ref{slicesxcorr}, whereas any symmetry is lost for the polarized scenarios~\ref{slicespospol} and~\ref{slicesmixpol}, with a shift 
in opposite direction, as expected from Fig.~\ref{2DPlots}. We can construct a variable that summarizes the previous information and allows us to visualize this unbalance:
\begin{equation}
\frac{dA_S}{d\eta_2}=\frac{d\sigma(\eta_1>0)}{d\eta_2}-\frac{d\sigma(\eta_1<0)}{d\eta_2},
\label{eta-asymm}
\end{equation}
displayed in Fig.~\ref{AsymmSlice}. The curve of the uncorrelated scenario is expected to be constantly zero over the entire range of $\eta_2$, by the definition~\eqref{eta-asymm}. This is true for 
the line of min-corr scenario, that is our proxy for uncorrelated physics, it slightly deviates from zero for long-corr, and it becomes outstandingly different from a constant zero curve in the presence of polarization.

% !TEX root = WWLongPaper.tex
%%%%%%%%%%%%%%%%%%%%%%%%%%%%%%%%%%%%%%%%
\section{DPS hadron level results with physics background}
\label{Sec:FSI}
%%%%%%%%%%%%%%%%%%%%%%%%%%%%%%%%%%%%%%%%

In this section, we embed the parton level results on parton correlations into the study 
of the hadron level (HL) distributions using general-purpose Monte Carlo generators. 
For the study of the effects of correlations in the final states, it is actually crucial to deal
directly with individual events rather than with calculated distributions.
%especially with respect to the tests of decision tree analysis.
This step allows us to include the full Underlying Event surrounding the actual process and to better model the measurable distributions of the leptons, taking into account initial\htb{-} and final-state radiations. 
Another reason for this procedure is to prepare a more 
realistic definition of the kinematic region where the signal process is enhanced and measurable with respect to the 
physics background processes.

\subsection{Signal process}

The double parton scattering process producing two positively charged W bosons decaying in the muon channel 
is obtained using the generator Herwig 7.1.2~\cite{Bahr2008,Bellm:2015jjp}, which is fully capable of generating double W events. 
We utilized the possibility of accessing the information about the outgoing leptons 
directly from the matrix element (i.e. at PL) before corrections are applied on their momenta. Our method for preparing the hadron level event datasets
is based on a re-weighting procedure, as shortly explained below. As a result, we obtain a perfect correspondence between the theoretical PL results of Section~\ref{Sec:parton} and the PL results generated by the 
reweighted Herwig for all the correlation scenarios.

\begin{figure}
\begin{center}
\subfigure[]{\includegraphics[width=0.42\textwidth]{%
    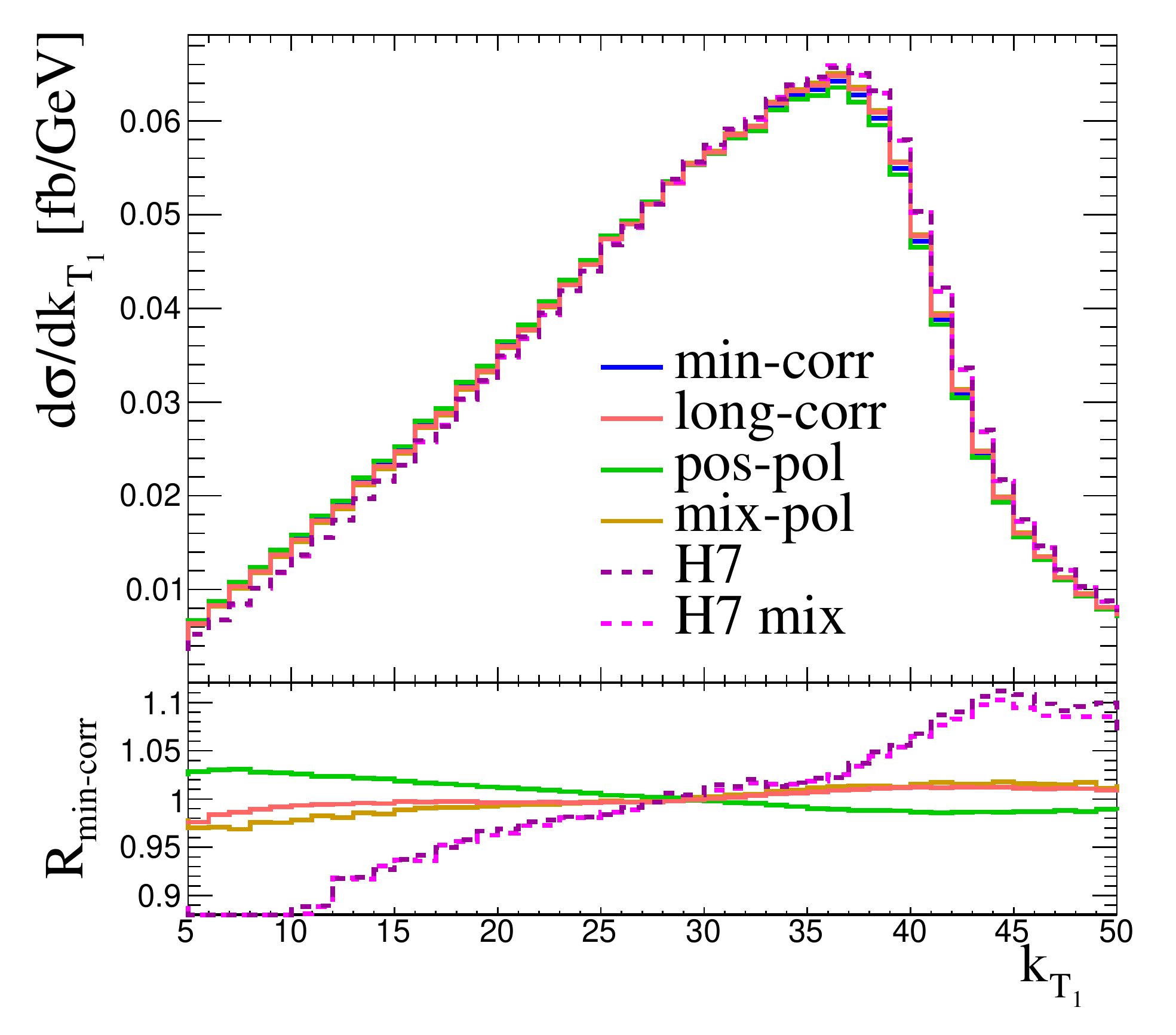}} 
\subfigure[]{\includegraphics[width=0.42\textwidth]{%
    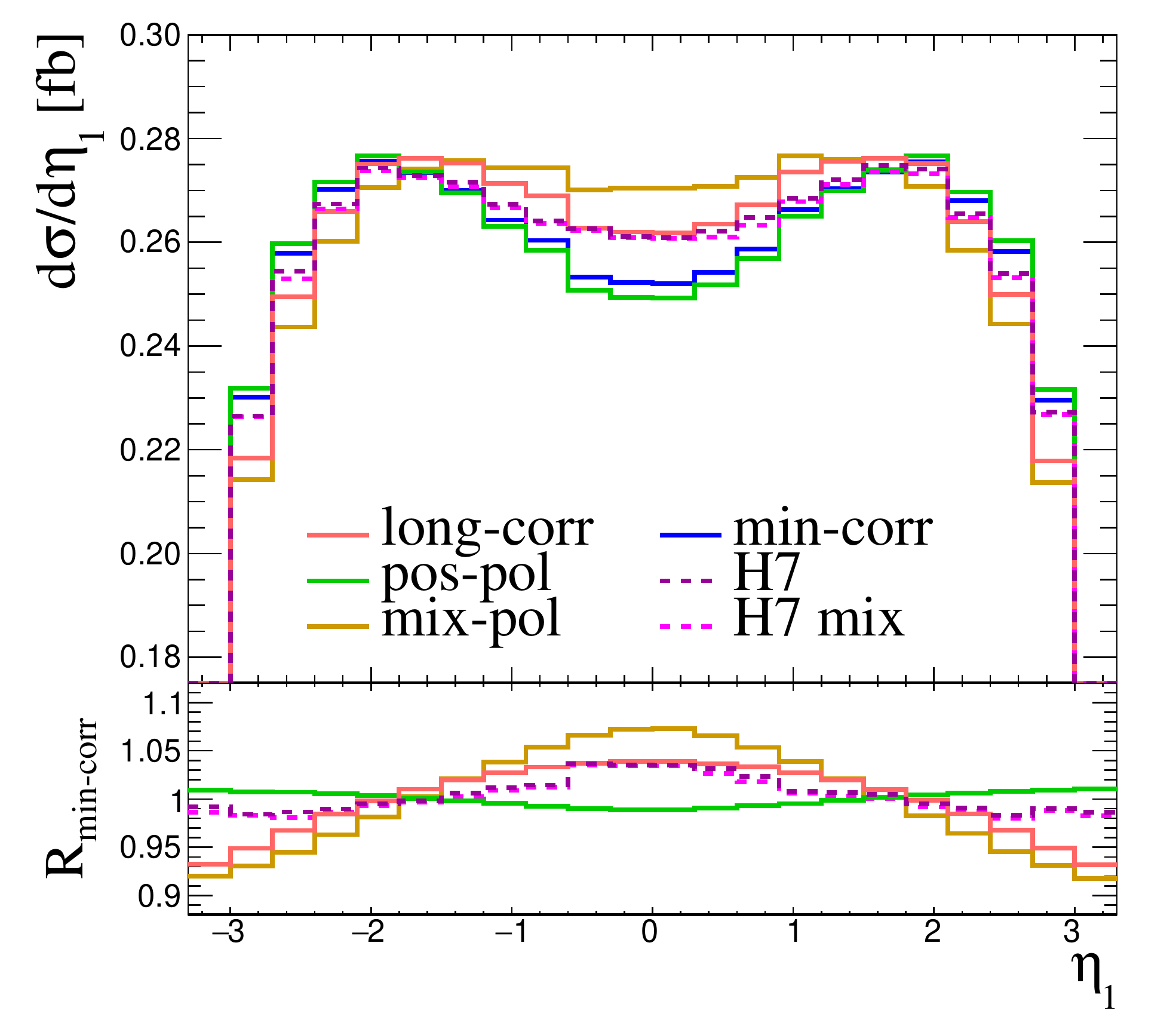}} 
\caption{\label{fig:0f_muOne_eta_pt} The distributions of the final-state muon transverse momentum and rapidity for the different models. The distributions for $\mu_{2}$ are identical (except the one for H7 model). 
The ratio $R_{\text{min-corr}}$ shows the comparison of different models to the min-corr scenario.}
\end{center}
\end{figure}

We initially calculate  the partonic cross section differential in four variables ($k_{T_{1}}$, $k_{T_{2}}$, $\eta_1$, and $\eta_2$), with the methods of Section~\ref{Sec:parton} using the different DPD models
(see Section \ref{Sec:4}). 
We divide the phase-space region (\ref{selection0}) into almost half a million subregions with 
unequal sizes. The same four-dimensional quantity is then obtained from the event generator and compared 
to the theoretical results at the parton level. According to this comparison, we appropriately change the default weights of the Herwig events.
%The normalization of the total cross section is the same for all the correlation scenarios (1.744 fb). 
After the event re-weighting, the distributions generated by Herwig at the 
PL  are identical to the ones we have calculated for each correlation scenario. This validates the procedure of re-weighting, which has 
been found sufficient and fully reliable. 
At the hadron level, the re-weighted Herwig events represent the same events as if they were generated according 
to our model of parton interactions and correlations. 

In order to demonstrate the quantitative effect of the Monte Carlo generator on the studied distributions, i.e. the differences between 
parton (PL) and hadron level (HL) distributions, we keep the events which satisfy the phase space (\ref{selection0}) 
at PL and apply similar criteria to HL muons too. Only the upper limit on muon transverse momentum is
removed for the HL selection. 
In essence, the main effect we observe is the smearing of the sharp 
$k_{\st}$ peak into a broader distribution, evident from the comparison of Fig.~\ref{fig:0_muOne_eta_pt} and~\ref{fig:0f_muOne_eta_pt}.
We have to point out here that we show this comparison despite the fact that 
there might be a few events missing in the HL distributions due to the blocked event migration in the way, where 
events that do not fit into PL cuts actually might satisfy HL cuts. We quantify this 
inconsistency to a level below 2 per cent of the total cross section.

At this point, we would like to mention more technical details about the event generation.
In its default setting, Herwig does not produce entirely independent hard scatterings, since there are 
several mechanisms that guarantee the validity of the conservation laws. As a result, the secondary process is statistically 
slightly softer than the primary process. Therefore, we have to perform the re-weighting procedure for 
all types of correlations, including the min-corr model, which is our reference for uncorrelated physics. 
For further comparison, two additional types of Herwig events have been prepared, labeled as H7 and H7 mix. The H7 events are 
obtained using a default settings, while H7 mix events are prepared through the merging of two random single W events. 
The numerical results related to the events H7 and H7 mix are shown in the third and fourth column of Table~\ref{tab:allnumbers}, 
including the values for the different variables and phase-space cuts. 
Taking the asymmetry $A$ as our principal indicator of correlations (any deviation from zero is a sign of correlation), we notice that the default 
setting H7 results into an asymmetry of 0.01, which is actually quite large (the same task was performed, for comparison,  using Pythia 8.235 \cite{Sjostrand:2006za,Sjostrand:2007gs}, and an asymmetry of 0.02 was obtained). On the contrary, even though Herwig produces muons with slightly different
distributions of transverse momentum and rapidity, see Fig. \ref{fig:0f_muOne_eta_pt}, the final value of the asymmetry $A$ is exactly zero for the H7 mix sample.
In a more global view,  the computational corrections of the full event generator to the PL muons do not affect the variables of interest more than 3$\%$
(see Table~\ref{tab:allnumbers} and compare selections ``Baseline PL" and ``Baseline HL") and, thus, the distortions introduced by correlations are not simply washed out. 

\subsection{Background processes}

\begin{figure}
\begin{center}
\subfigure[]{\includegraphics[width=0.42\textwidth]{%
    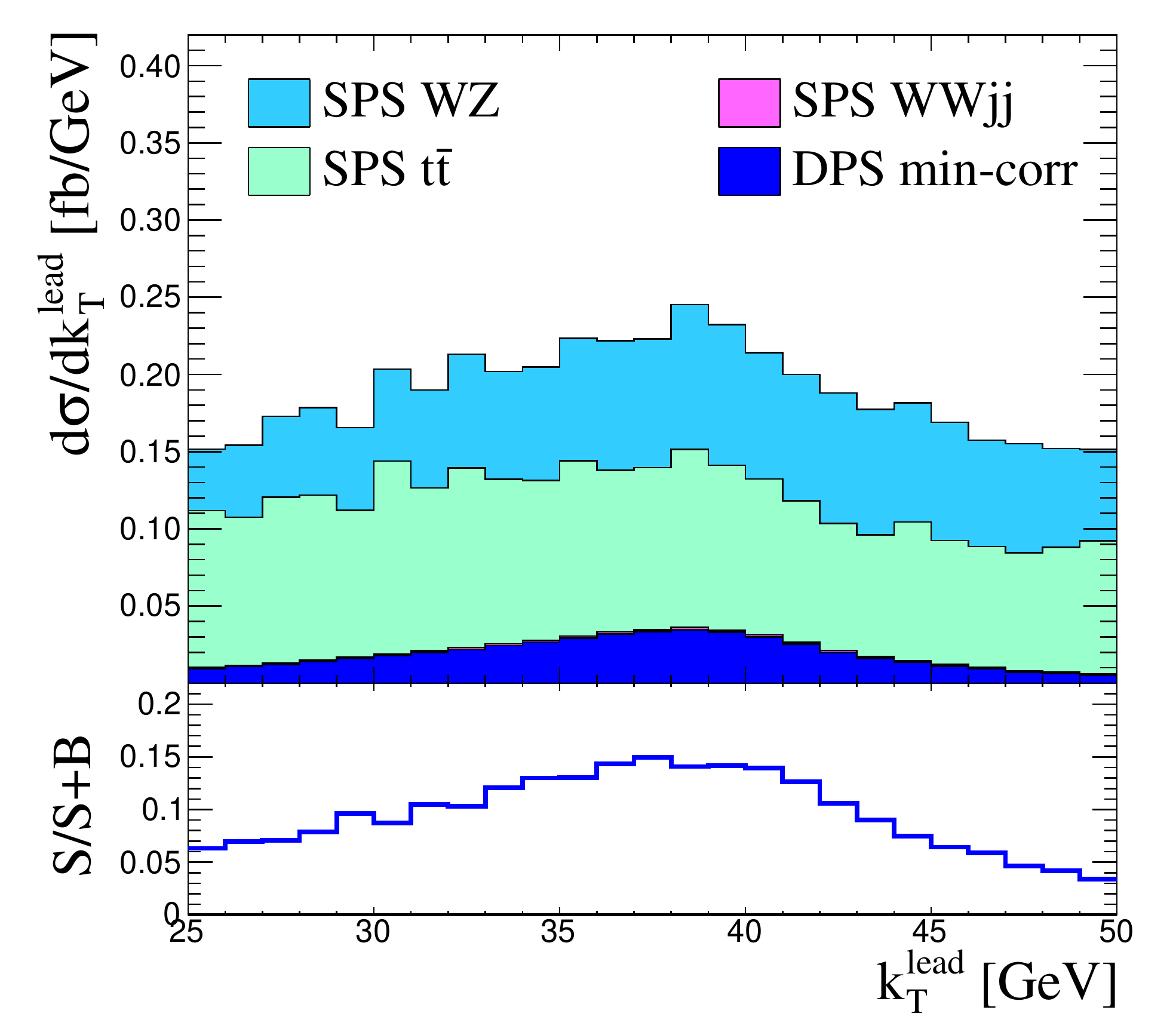}} 
\subfigure[]{\includegraphics[width=0.42\textwidth]{%
    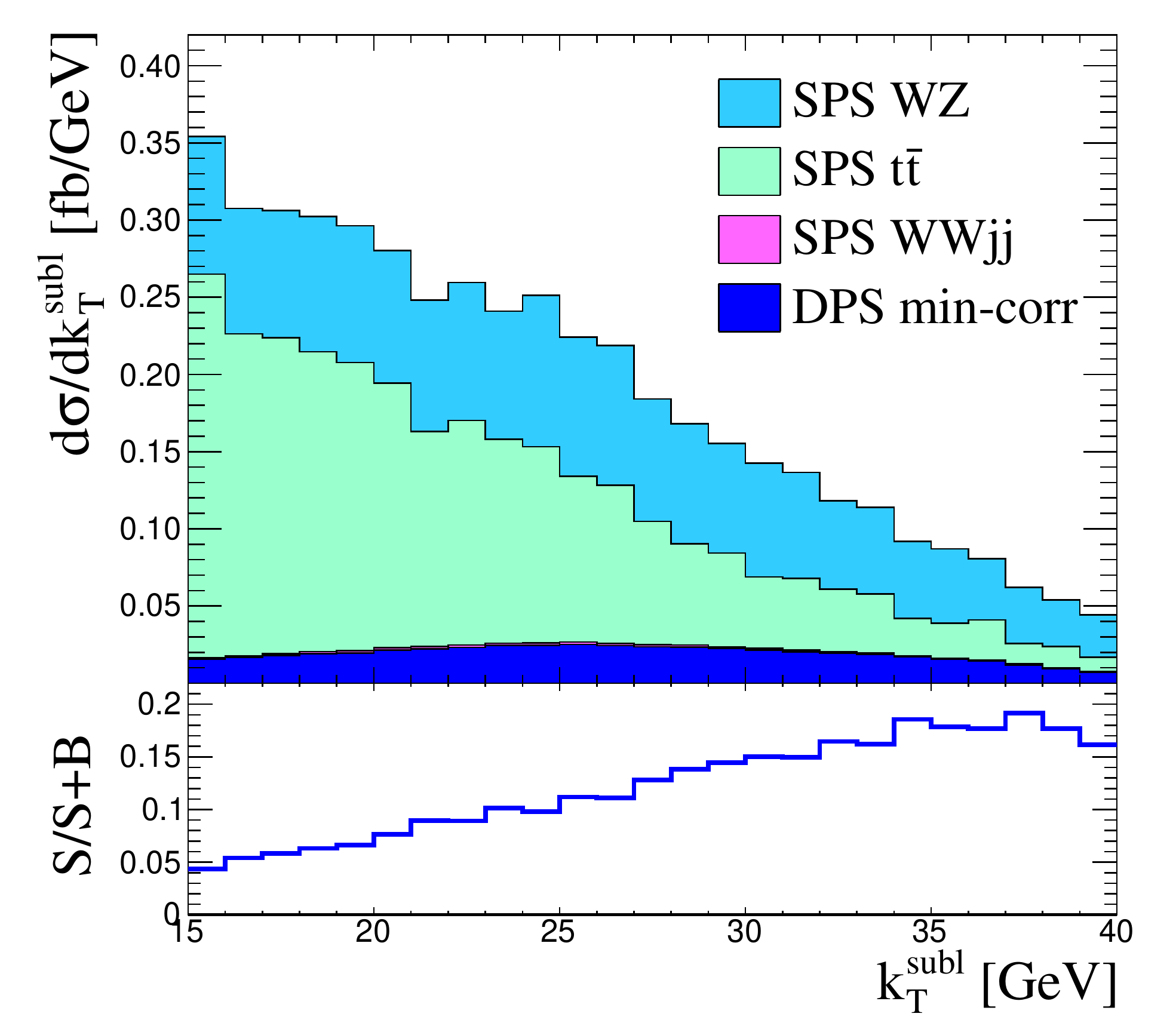}} 
\caption{\label{fig:4_muon_kt} The distributions of the transverse momentum of the leading (a) and sub-leading (b) muon. The signal contribution to the total yield is drawn at bottom 
to optically capture its shape. }
\end{center}
\end{figure}

\begin{figure}
\begin{center}
\subfigure[]{\includegraphics[width=0.42\textwidth]{%
    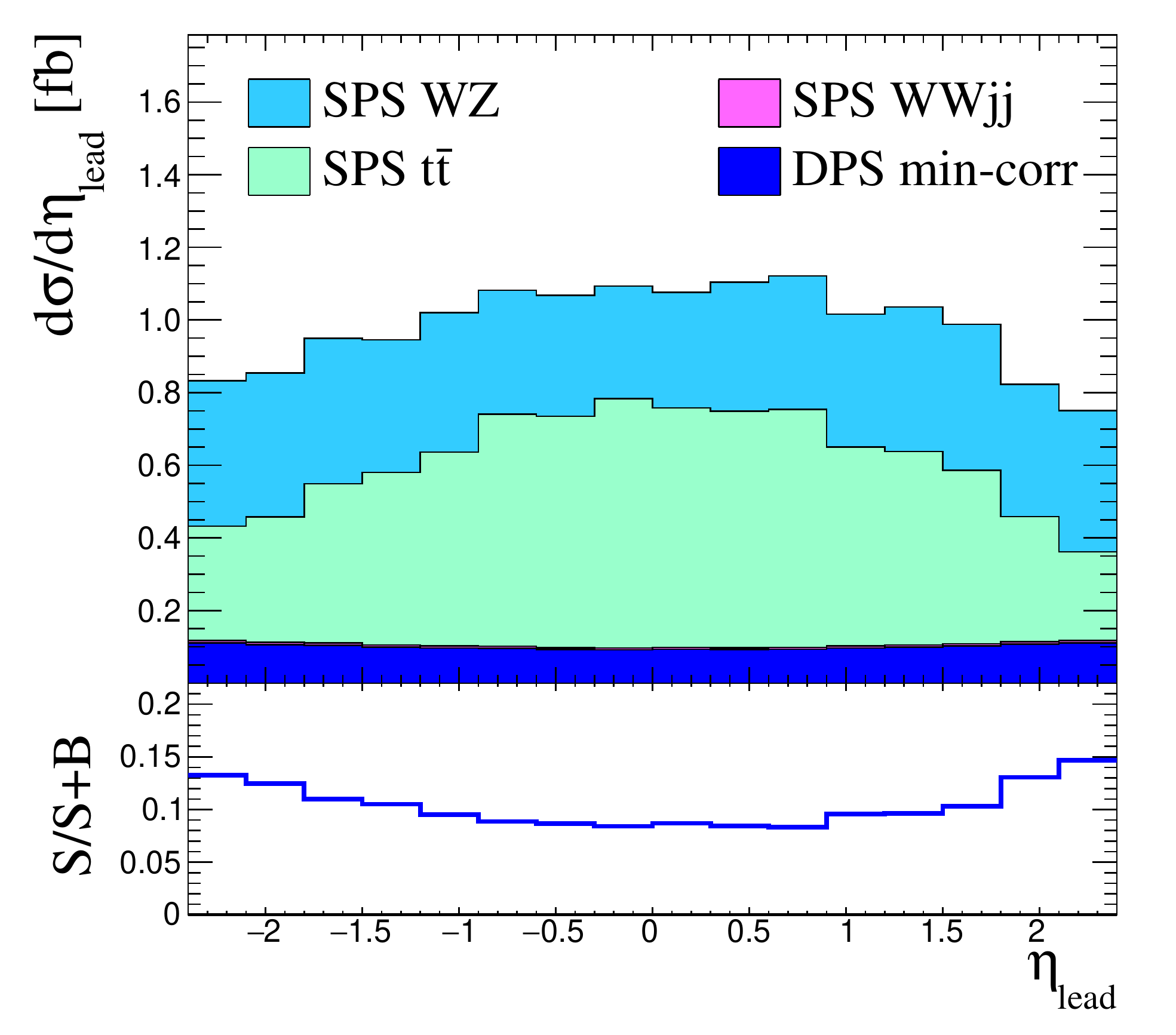}} 
\subfigure[]{\includegraphics[width=0.42\textwidth]{%
    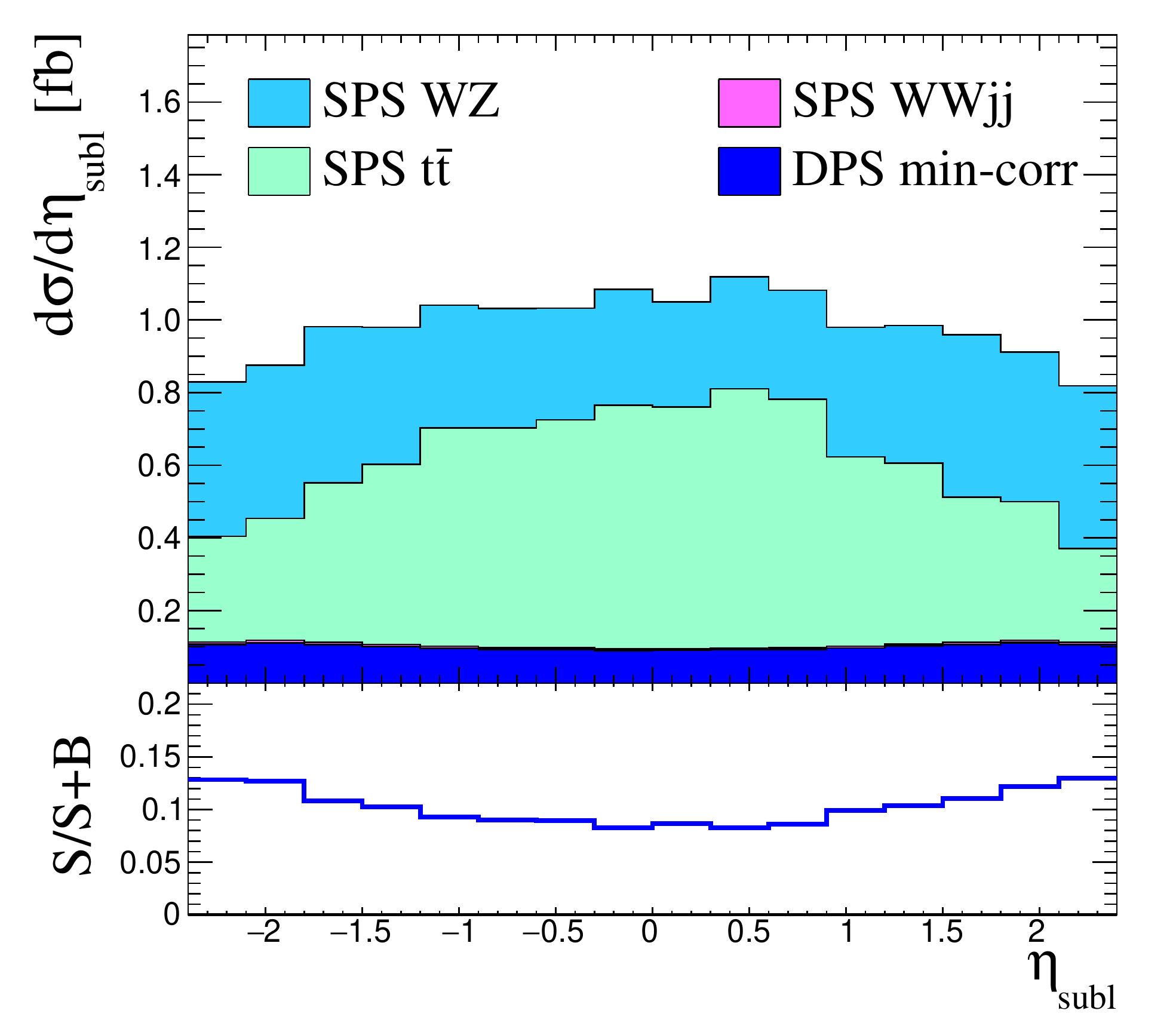}} 
\caption{\label{fig:4_muon_eta} The distributions of the rapidity of the leading (a) and sub-leading (b) muon. The signal contribution to the total yield is drawn at bottom 
to optically capture its shape. }
\end{center}
\end{figure}

In this subsection, we turn to the analysis of the processes whose signature contains a pair of positively charged muons, which constitute the relevant background to SSW production, 
and we identify the phase-space region suitable for correlation  measurements.

The major contributions to the background of same-sign muon-pair
production come from SPS processes such as diboson production and heavy flavor production, the latter represented here by the (dominant) $t\bar{t}$ process.
For other detector-related backgrounds, such as single Drell-Yan process, where one of the muon charges is mis-measured, we do not provide any quantitative
prediction and assume them negligible.

The SPS production of a pair of gauge bosons is the most direct background process. We distinguish three
types of processes: $ZZ$, $WZ$ and same-sign $WW$, where $Z$ stands for both $Z$ boson and virtual photon. The same-sign $WW$ in SPS production is strongly suppressed
by the presence of two additional strong vertices at the lowest order diagram and we denote it as $WWjj$ process. 
In the $t\bar{t}$ process, one lepton is generated in the first top decay, and another
lepton, with the same sign, typically arises from a bottom quark emitted by the
other top quark. Since we aim to remove these types of events as much
as possible, there is no real need to go through all possible flavors, as
the top quark has the largest chance to produce a hard muon. Diboson
samples were obtained via a combination of MadGraph5\_aMC@NLO at LO \cite{Alwall:2014hca} and Pythia 8.223, whereas the $t\bar{t}$ data sample was obtained using Herwig 7.1.2.

The Baseline Selection (\ref{selection0}) is a theoretical landmark needed to set the stage for a more realistic analysis of same-sign di-muon events. 
Actually, LHC-based experiments plan to expand their inner tracker acceptance far behind the current value of 2.5 in rapidity \cite{Collaboration:2272264,Collaboration:2285585}. This acceptance enlargement  would make our Baseline Selection potentially more realistic for future experimental measurements and might provide an excellent opportunity to measure correlations, as shown in Section~\ref{Sec:parton}. 
However, in the following we restrict our event selection to mimic the acceptances of the ATLAS \cite{Aad:2008zzm} and CMS \cite{Chatrchyan:2008aa} detectors (current state), 
which have the potential to perform the suggested measurements at present or near future. The aim is to obtain data as signal-pure as possible, unlike the latest W-pair DPS 
measurement by CMS \cite{Sirunyan:2019zox}, whose aim was to measure the total DPS cross section.

The first step towards the identification of an optimal phase-space region for the signal, in the current experimental set-up, is to limit the absolute values of the rapidity of muons to a maximum of 2.4, which is the acceptance of the trigger 
chambers of the muon spectrometers. Further, one needs to increase the minimal threshold of muon transverse momenta up to 15 GeV, and secure
the basic spatial separation of muons. We also apply an upper limit on the muon transverse momenta in order to additionally isolate the signal. 
We follow the CMS strategy and veto events with a third muon with transverse momentum
larger than 5 GeV. The last set of cuts suggests to use also hadron calorimeters to restrict the energy of 
Underlying Event jets from above and missing energy from below. Jets are defined using the default anti-$k_\st$ algorithm of FastJet \cite{Cacciari:2011ma} with a pseudo-radius $R = 0.4$.

Let us note here that the experimentally measured muons cannot, of course, be identified as ``first" and ``second" muon, as in Section~\ref{Sec:parton}, 
since the information on their origin is unaccessible. 
From this moment, we will identify the measured muon pair as composed by a leading ($\mu_\text{lead}$) and a sub-leading ($\mu_\text{subl}$) muon, 
i.e. the hardest and second hardest muons measured respectively, whenever meaningful. Otherwise, we keep indices 1 and 2 for simplicity.

To summarize, our kinematical cuts labeled as Final Selection are:
\begin{align}\label{eq:sel}
	|\eta_i| < 2.4 \; , \quad 25~\text{GeV} < k_{\st}^\text{lead} < 50~\text{GeV} \,, \quad 15~\text{GeV} &< 		k_{\st}^\text{subl} < 40~\text{GeV}\,, \quad
	k_{\st}^{\mu_3} < 5~\text{GeV} \,, \nn\\
	 \quad \slashed{E}_T > 20~\text{GeV} \,, \quad dR(\mu_1, \mu_2) > 0.1 \,, \quad k_\st^\text{jet1} &< 50~\text{GeV} \,, \quad k_\st^\text{jet2} < 		25~\text{GeV}\,,
\end{align}
where $\eta_i$ is the rapidity of the muon $i$, $k_{\st}^\text{lead}$ ($k_{\st}^\text{subl}$) the transverse momentum of the leading (sub-leading) muon, $k_{\st}^{\mu_3}$ the transverse momentum of a third muon, $\slashed{E}_\st$ the missing transverse energy, and $k_\st^\text{jet1}$ and $k_\st^\text{jet2}$ are the transverse momenta of the two hardest jets. 
$dR = \sqrt{ (\phi_1 - \phi_2)^2 + (\eta_1 - \eta_2)^2 }$, where $\phi_i$ is the azimuthal angle of 
$\mu_i$, measures the distance between the two muons. On top of this, we apply b-tagging veto with efficiencies 
$75\%$ for $k_\st^\text{jet} \in \{25-30 \}$ GeV, $80\%$ for $k_\st^\text{jet} \in \{30-40 \}$ GeV, and $85\%$ 
for $k_\st^\text{jet} \in \{40-50\}$~GeV \cite{Aaboud:2018xwy,Sirunyan:2017ezt}. 
A b-tagged jet is a shortcut for a jet containing hadrons deriving from the fragmentation of b-quarks (see, for instance, \cite{CMS-PAS-BTV-15-001,ATLAS:2012aoa} for
the extra information on the adopted procedure and efficiency).

With these cuts, the cross sections are given in Table~\ref{tab:xsec}. They report a good suppression of the $WWjj$ and $ZZ$ backgrounds. 
Both $WZ$ and $ZZ$ processes can be substantially suppressed by vetoing events containing a third muon. While the $ZZ$ contribution practically vanishes, 
we notice that the $WZ$ background is still dominant with respect to the signal, and the situation is similar for the $t\bar{t}$. 
Let us note that the optimal phase-space region for the measurement in the future era of LHC has to be found through more sophisticated methods of multivariate 
analysis performed by the experimental collaboration.  For the purpose of this paper, we assume that such a dedicated analysis will significantly 
improve the background suppression. A naive application of a forest of decision trees is capable of performing the suppression of the $WZ$ and $t\bar t$ background 
while leaving the signal almost unchanged, see Section \ref{Sec:Corr}. 

We now show a selection of distributions of kinematical variables at the level of Final Selection. 
These variables (among others) have the potential to discriminate signal from background in the decision tree diagnostics.
All the plots contain in their bottom part the distribution of the ratio of signal over signal plus background (S/S+B), to show where the signal process is more significant.
For simplicity, we show only min-corr signal scenario in the figures, since here we focus on the main differences between signal and background, 
which are larger than the differences among the correlation models. 

\begin{figure}
\begin{center}
\subfigure[]{\includegraphics[width=0.42\textwidth]{%
    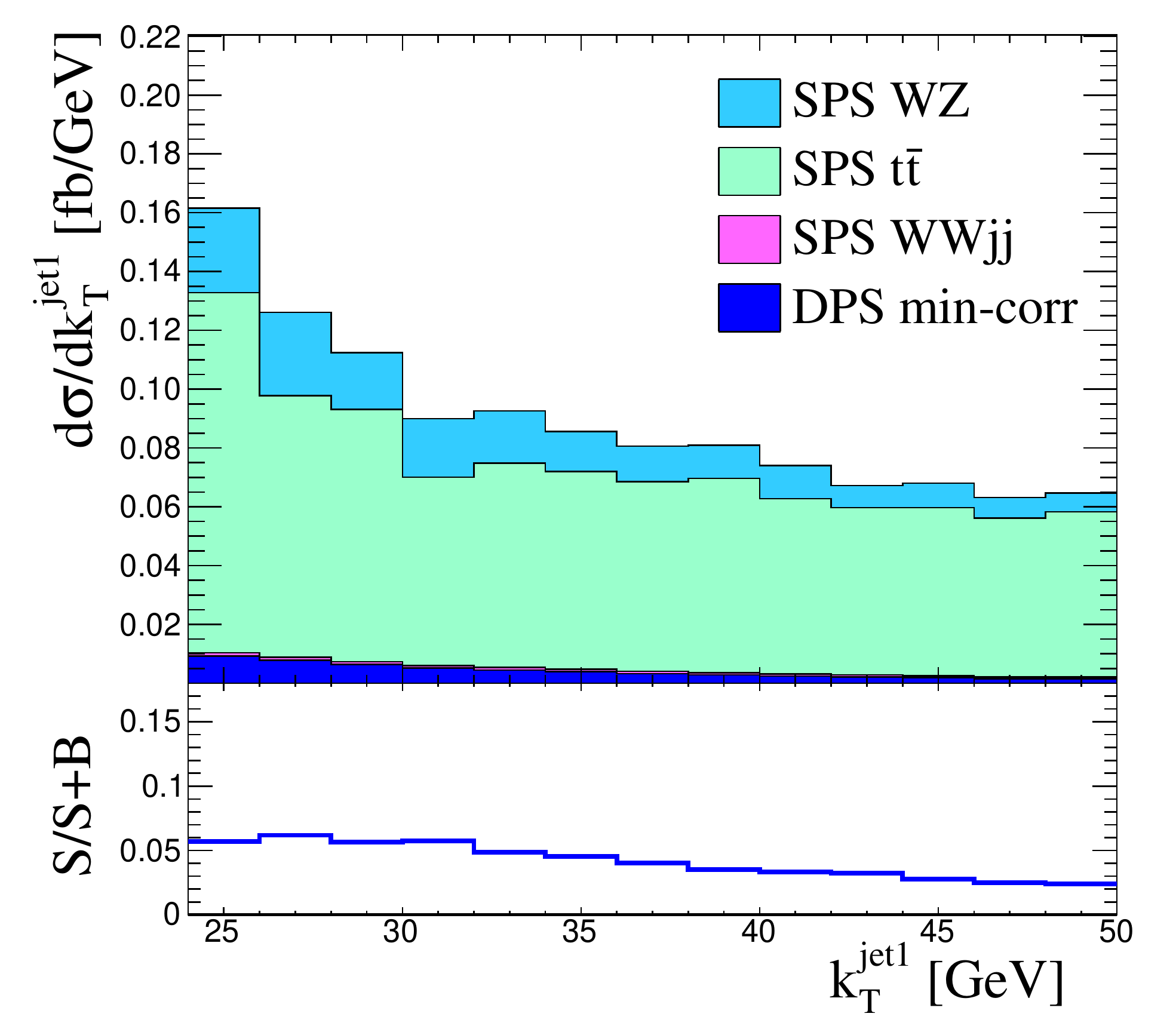}} 
\subfigure[]{\includegraphics[width=0.42\textwidth]{%
    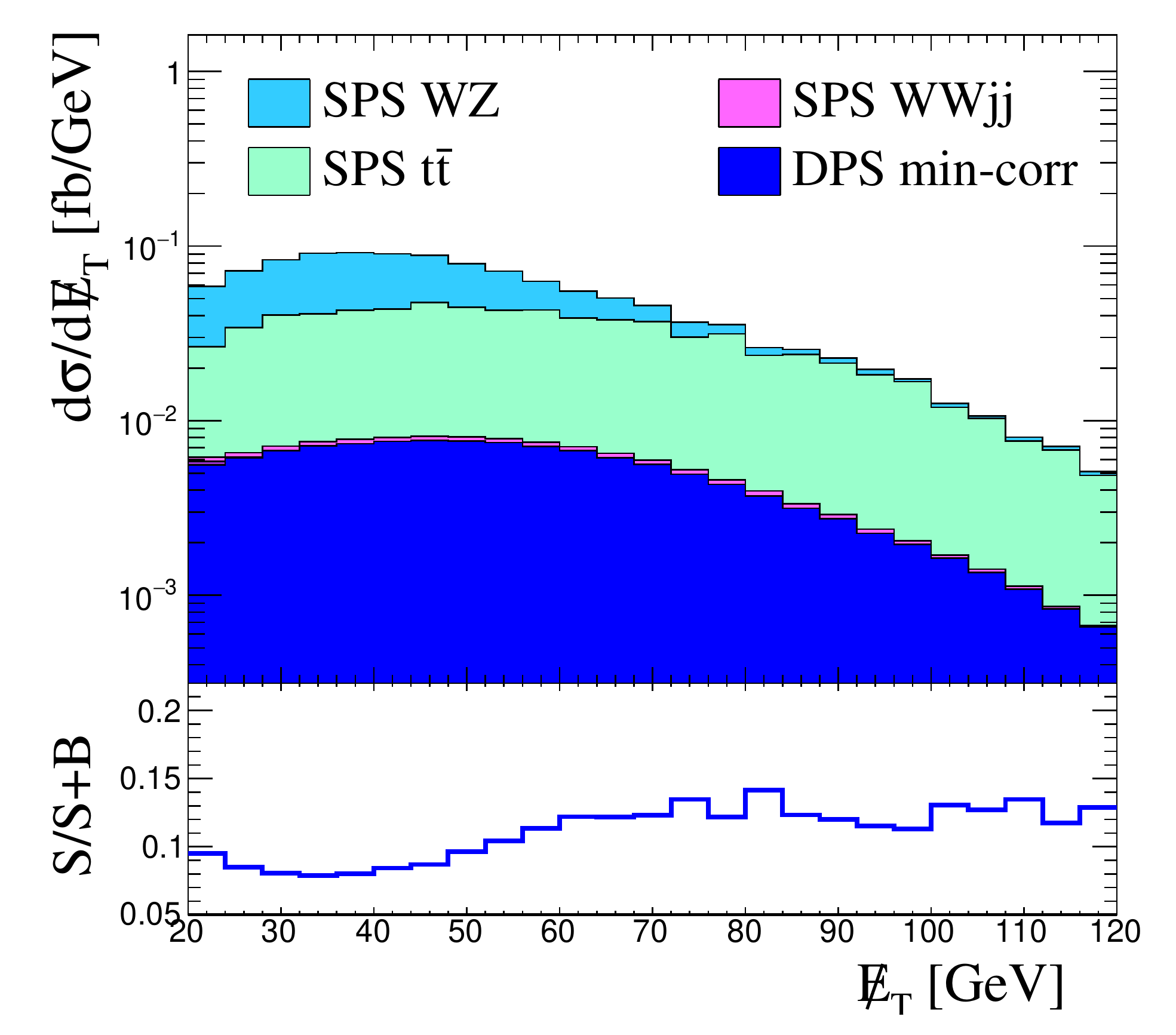}} 
\caption{\label{fig:4_jet_miss} An example of usage of Underlying Event for signal-background discrimination: (a) the distribution of transverse momentum of the leading jet and (b) the distribution of the 
missing transverse energy in the event.}
\end{center}
\end{figure}

\begin{figure}[t]
\begin{center}
\subfigure[]{\includegraphics[width=0.42\textwidth]{%
    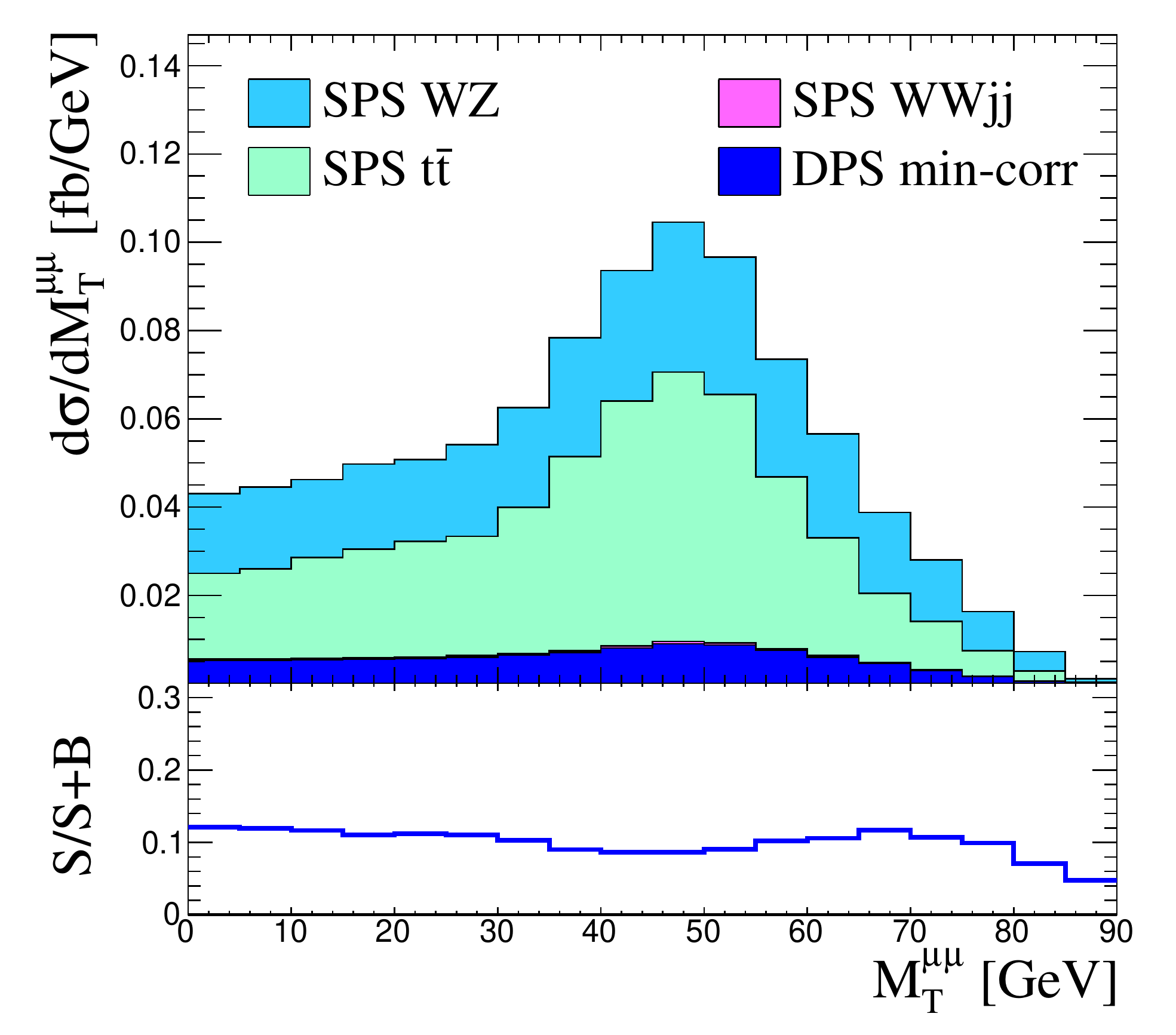}} 
\subfigure[]{\includegraphics[width=0.42\textwidth]{%
    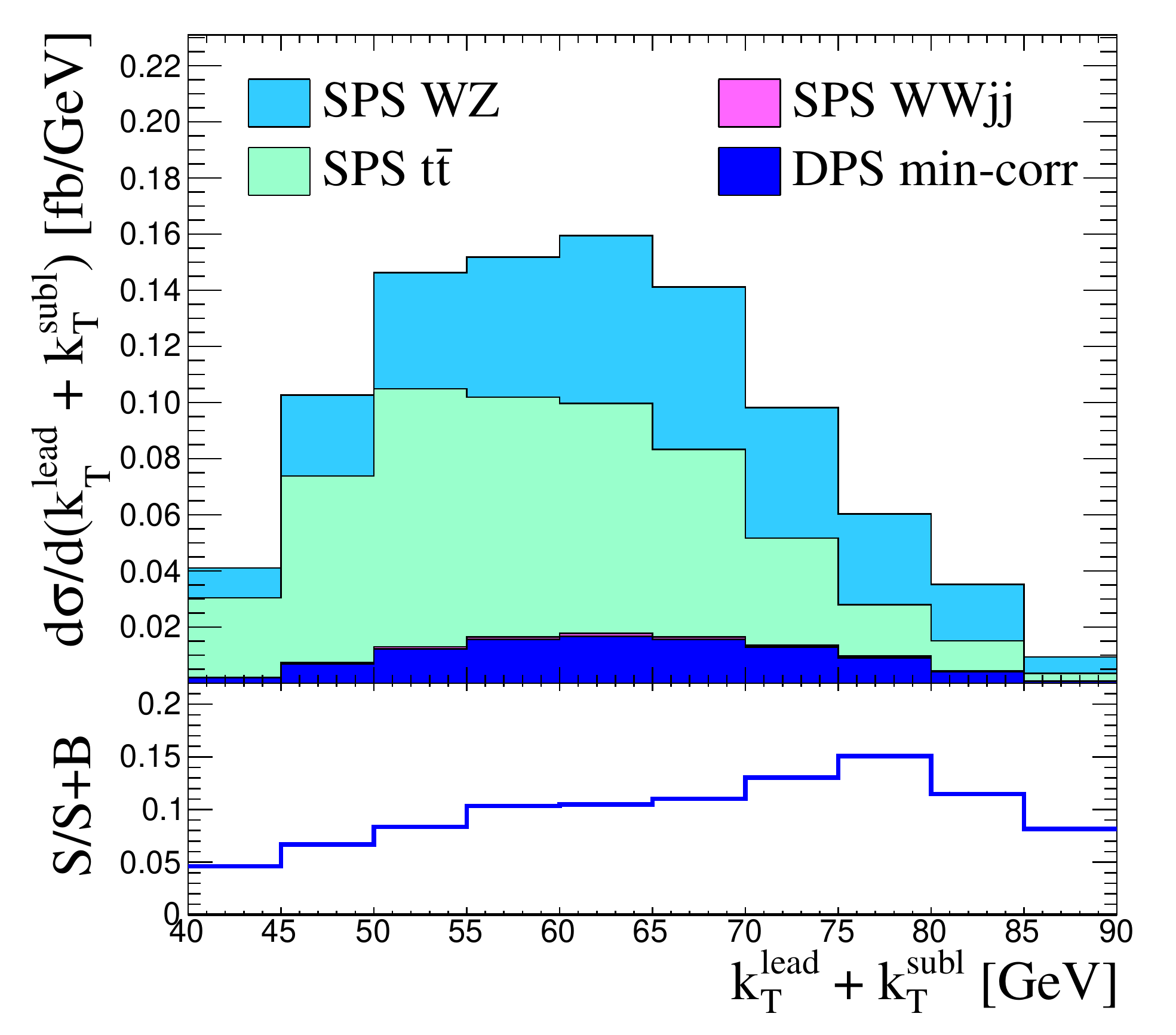}} 
\caption{\label{fig:4_mupair_mtpt} An example of di-muon characteristics for signal-background discrimination: (a) the distribution of transverse mass of the muon pair and (b) the distribution of scalar sum of transverse momenta of the two muons.}
\end{center}
\end{figure}

\begin{figure}[t]
\begin{center}
\subfigure[]{\includegraphics[width=0.42\textwidth]{%
    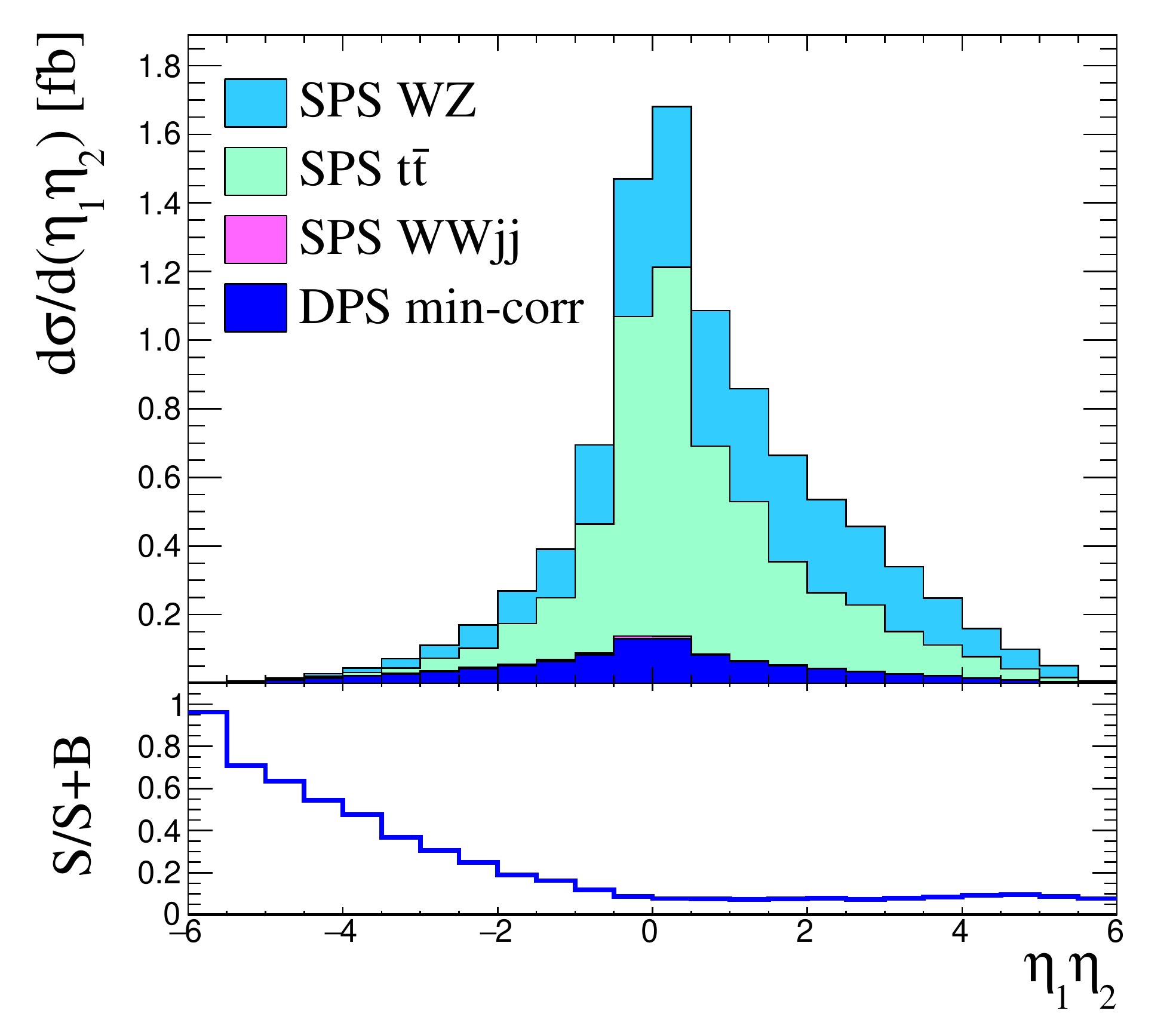}} 
\subfigure[]{\includegraphics[width=0.42\textwidth]{%
    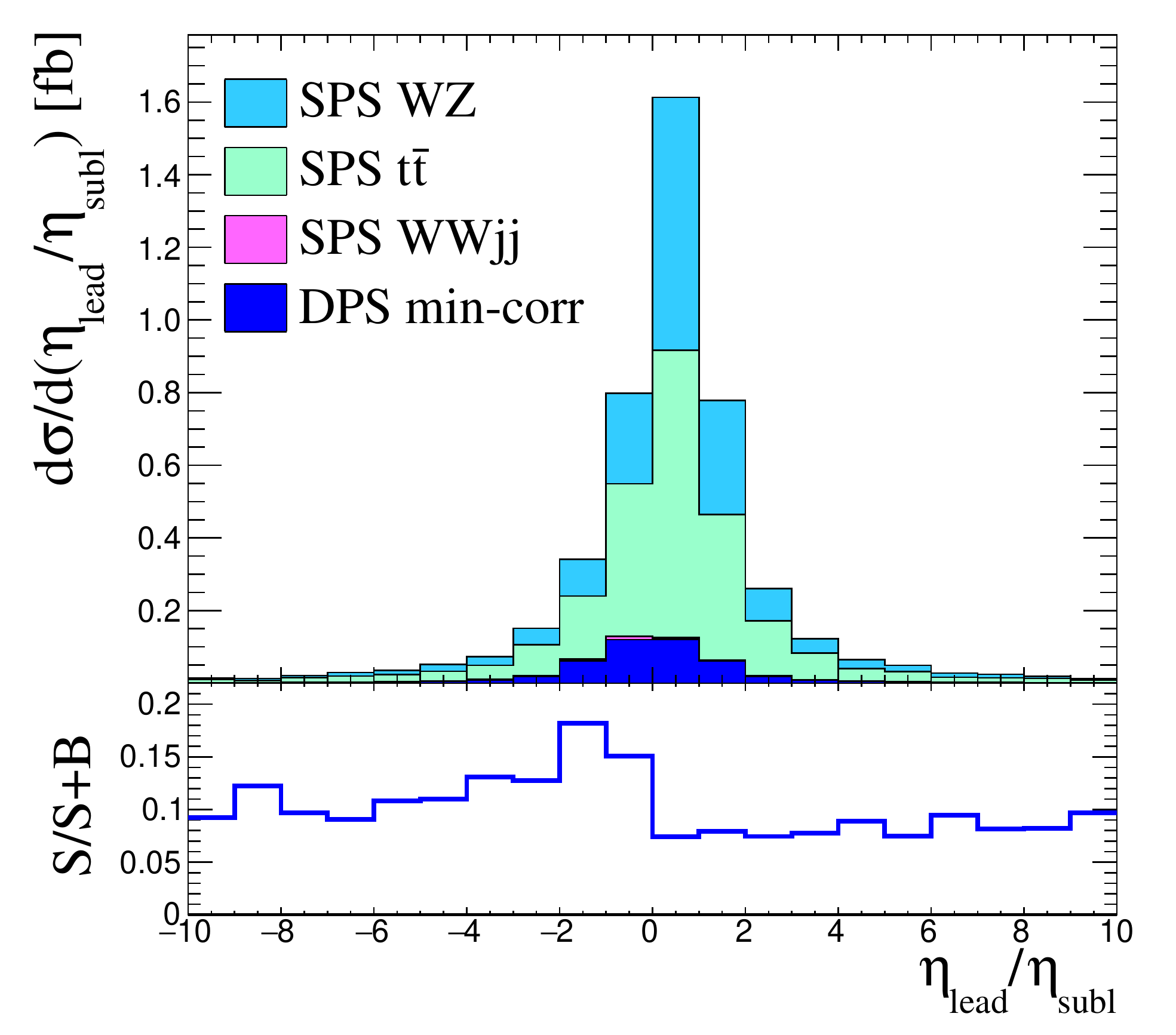}} 
\caption{\label{fig:4_mupair_etaprod} The distributions based on muon rapidity has the highest signal-background discriminant power. Here, (a) shows the distribution of the product of the muon rapidities and (b)
shows the ordered ratio of them, i.e. leading muon $\eta$ over the sub-leading muon $\eta$. }
\end{center}
\end{figure}

The various signal distributions are compared in Section \ref{Sec:Corr}. 
The distributions of transverse momenta of the leading and sub-leading muons can be seen in Fig.~\ref{fig:4_muon_kt} 
and their rapidity in Fig.~\ref{fig:4_muon_eta}. The two peaks of the signal in  Fig.~\ref{fig:4_muon_kt}  greatly help suppress the background contributions. 
Especially the $t\bar t$ process is suppressed to a large extent by the lower threshold on transverse momentum of the sub-leading muon.
Unlike the transverse momenta, the rapidity of signal muons slightly rises at the edges of the acceptance, see Fig.~\ref{fig:4_muon_eta},
and, therefore, we loose lots of signal events due to the detector acceptance limits.
The behavior of Underlying Event for the studied processes is indicated in Fig. \ref{fig:4_jet_miss},
where the distributions of the leading jet transverse momentum and missing transverse energy is depicted. 
For the leading jet, we can see a non-negligible contribution from the signal, though sharply falling down. 
At the first sight, the cut on 50 GeV is a good trade-off for signal purity.
However, we do not want to decrease the jet cuts too much (even if experimentally possible)
to keep the sizable amount of signal events containing also low-$p_\st$ Underlying Event jets. 
In particular, we have found that the cuts on transverse momenta of Underlying Event jets effectively remove the $WWjj$ background.
Regarding the missing transverse energy, we point out that it falls down quicker for the $WZ$ process than for the signal one 
and becomes highly suppressed above 70 GeV. This might be used if the higher signal purity is desired.
Fig.~\ref{fig:4_mupair_mtpt} shows the distributions of the transverse mass of the muon pair and the scalar sum of the muon transverse momenta, which
have only limited discrimination power individually. However, further multivariate analysis 
could combine many distributions with similar characteristics to substantially separate the signal.
Fig.~\ref{fig:4_mupair_etaprod} depicts two ways how to combine the muon rapidities into observables sensitive to correlations: 
their product ($\eta_1\eta_2$) and their ordered ratio ($\eta_{\text{lead}}/\eta_{\text{subl}}$).
One can observe more peaked background distributions with significant asymmetries, a pattern that can additionally aid the separation.

\begin{table}\centering
\begin{tabular}{ c|c } 
 & $\sigma$ [fb] \\
 \hline
  DPS $W^+W^+$ & 0.48 - 0.51\\ 
 $W^+ W^+ jj$ & 0.03 \\ 
 $W^+Z$ & 1.77  \\ 
 $ZZ$ & 0.00 \\
 $t\bar{t}$ & 2.46
\end{tabular}
\caption{\label{tab:xsec}Signal and background cross sections in fb for the production of two positively 
charged muons for the Final Selection \eqref{eq:sel}. The signal cross section depends on the correlation 
scenario and varies $\pm3\%$ around 0.495 fb.}
\end{table} 

\begin{table}\centering
\begin{tabular}{l|l|c|c|c|c|c|c } 
Variable  & Selection      & H7	  & H7 mix & min-corr & long-corr & pos-pol & mix-pol  \\
 \hline
\multirow{3}{*}{$A$} & Baseline PL  & 0.012 & 0.000  & 0.001    & 0.011     & -0.054  & 0.121 \\ 
		     & Baseline HL  & 0.011 & 0.000  & 0.002    & 0.010     & -0.051  & 0.115 \\ 
		     & Final HL     & 0.009 & -0.001 & 0.002    & 0.004     & -0.036  & 0.073 \\ 
\hline
\multirow{3}{*}{$S_{lin} (\Sigma_{\eta})$} & Baseline PL  & -0.088 & -0.083 & -0.081 & -0.086 & -0.069 & -0.110 \\ 
		                           & Baseline HL  & -0.089 & -0.084 & -0.082 & -0.087 & -0.071 & -0.109 \\ 
		                           & Final HL     & -0.045 & -0.044 & -0.040 & -0.043 & -0.038 & -0.045 \\ 
\hline
\multirow{3}{*}{$S_{lin} (\Delta_{\eta})$} & Baseline PL  & -0.084 & -0.084 & -0.081 & -0.082 & -0.091 & -0.060 \\ 
		                           & Baseline HL  & -0.085 & -0.084 & -0.082 & -0.083 & -0.091 & -0.063 \\ 
		                           & Final HL     & -0.045 & -0.044 & -0.040 & -0.042 & -0.040 & -0.041 \\ 
\hline
\multirow{3}{*}{$S_{lin} (\Sigma_{\eta}/\Delta_{\eta})$} & Baseline PL  & -0.016 & 0.001 & -0.001 & -0.014 & 0.075 & -0.170 \\ 
		                                     & Baseline HL  & -0.015 & 0.001 & -0.001 & -0.014 & 0.072 & -0.165 \\ 
		                                     & Final HL     & -0.014 & 0.001 & -0.003 & -0.009 & 0.069 & -0.140 \\ 
\hline
 \end{tabular}
\caption{\label{tab:allnumbers} Results for the rapidity asymmetry $A$ and slopes of linear fit to three correlation sensitive observables. 
All three event selections are shown for all four observables to compare its evolution through the considered steps: transition from parton level (PL) to hadron level (HL)
and transition to restrictive cuts labeled as Final Selection.}
\end{table}

We include all the results for the Final Selection in Table \ref{tab:allnumbers}, which can finally be thoroughly discussed. 
The  first column contains the four variables especially suitable for quantifying the effects of 
parton correlations, as described in Section \ref{Sec:parton}.
For each variable, we report the results at the parton level (selection Baseline PL) and hadron level (selections Baseline HL and Final HL). 

We can see that moving from the looser to the more restrictive cuts reduces significantly the differences among the models, but there is still 
a good chance to distinguish them. For instance, the original maximal asymmetry $A$ obtained for mix-pol model
is 0.12 (at PL as well at HL), while for the Final Selection it is only 0.07, i.e. it is reduced to 63$\%$ of the original value.
The asymmetries for other scenarios are reduced to 40$\%$ (long-corr model) and to 71$\%$ (pos-pol model). One should note
that the final cuts do not produce any artificial asymmetry and that the models with minimal correlations still produce negligibly small 
asymmetry $A$, namely 0.002 for min-corr model, caused primarily by statistical fluctuations. The $S_{lin} (\Sigma_{\eta}/\Delta_{\eta})$ is actually also very promising, as it remains
almost zero for the min-corr model and changes its value only for the other models to 65-95$\%$ with respect to the Baseline HL selection. 
The other two slopes, $S_{lin} (\Sigma_{\eta})$ and $S_{lin} (\Delta_{\eta})$ have limited discriminating power, since they directly depend on the actual uncertainties of the experimental data. Their absolute values
drop to almost a half due to the total cross section reduction. If we calculate the relative differences
between the most different models (min-corr and mix-pol), for the $S_{lin} (\Sigma_{\eta})$ we get 0.33 for Baseline HL selection and 0.13 for the Final HL (i.e. the drop 
is even larger than by half). For the $S_{lin} (\Delta_{\eta})$ slope the analogous relative differences decrease even more, from 0.23 to 0.03.

% !TEX root = WWLongPaper.tex
%%%%%%%%%%%%%%%%%%%%%%%%%%%%%%%%%%%%%%%%
\section{Measuring correlations in DPS}
\label{Sec:Corr}
%%%%%%%%%%%%%%%%%%%%%%%%%%%%%%%%%%%%%%%%

Now we examine in details how to measure correlations between the two partons inside a proton through the DPS cross section. 

When not explicitly stated otherwise, the results in this section are for our Final Selection \eqref{eq:sel} and the values for the total cross sections are in Table \ref{tab:xsec}.
The two remaining relevant backgrounds are the $WZ$ and $t\bar{t}$, which we now discuss separately. There are specific techniques to suppress both of these 
backgrounds which require dedicated work in connection with performing the measurements. 
A detailed examination of all these possibilities lies outside the scope of this work and we limit ourselves to explain the path towards this objective. 

For $t\bar{t}$, demanding tight isolation of the produced muons is a very strong discriminant to separate prompt muons from muons produced by meson decays. 
Using vertex localization to further discriminate between these two cases can additionally aid the separation, and improvements on b-tagging can also help \cite{Sirunyan:2018fpa,Aad:2016jkr,Chatrchyan:2012xi}.  
Based on our investigations we assume that this type of discrimination, in combination with data driven subtractions, can reduce the top background to 1\% 
of the cross section in Table \ref{tab:xsec}, with only a minor impact on the signal. For example, we could reduce the $t\bar{t}$ background by more than 95\%, keeping more 
than 90\% of the signal through crude muon isolation requirements (i.e.\ limiting the scalar transverse momentum sum of particles in a cone around the muon). 

As already announced, $WZ$ background can be effectively suppressed through methods of 
multivariate analysis. For instance, with a simple application of a forest of decision trees, we could enhance the signal to background ratio to about 1, 
with a signal cross section around 0.3 fb. We assume that a dedicated analysis could achieve a $WZ$ background suppression to a level of one third of the signal. 
The contribution of the remaining $WZ$ background to the asymmetry can be subtracted by theoretical calculations aided by data driven methods.
The $WZ$ production is a relatively clean process theoretically and the calculations for the total cross section have already
been made with high precision \cite{Grazzini:2017ckn,Grazzini:2016swo}. 

\begin{figure}[t]
\begin{center}
\subfigure[]{\includegraphics[width=0.42\textwidth]{%
    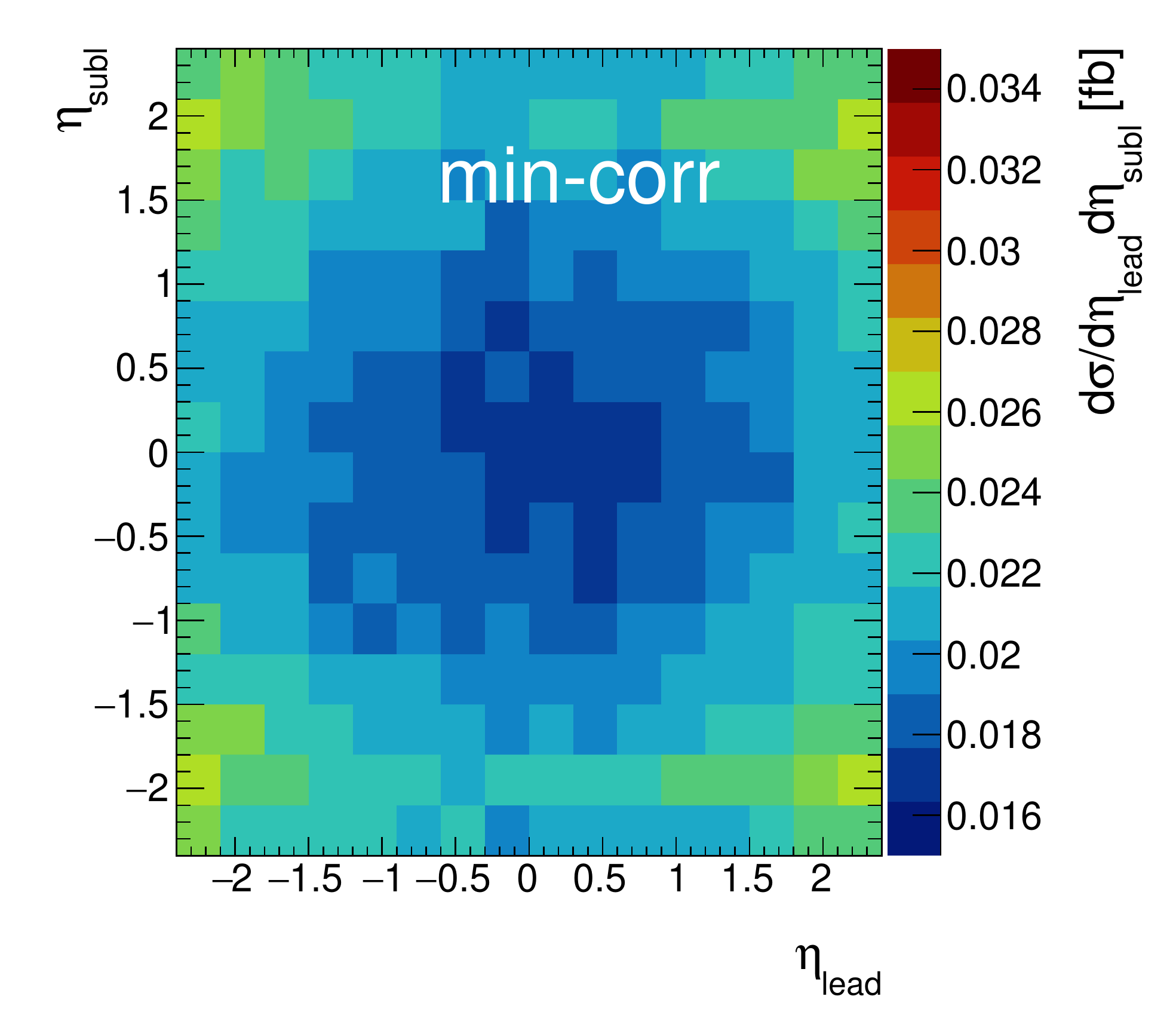}}
\subfigure[]{\includegraphics[width=0.42\textwidth]{%
    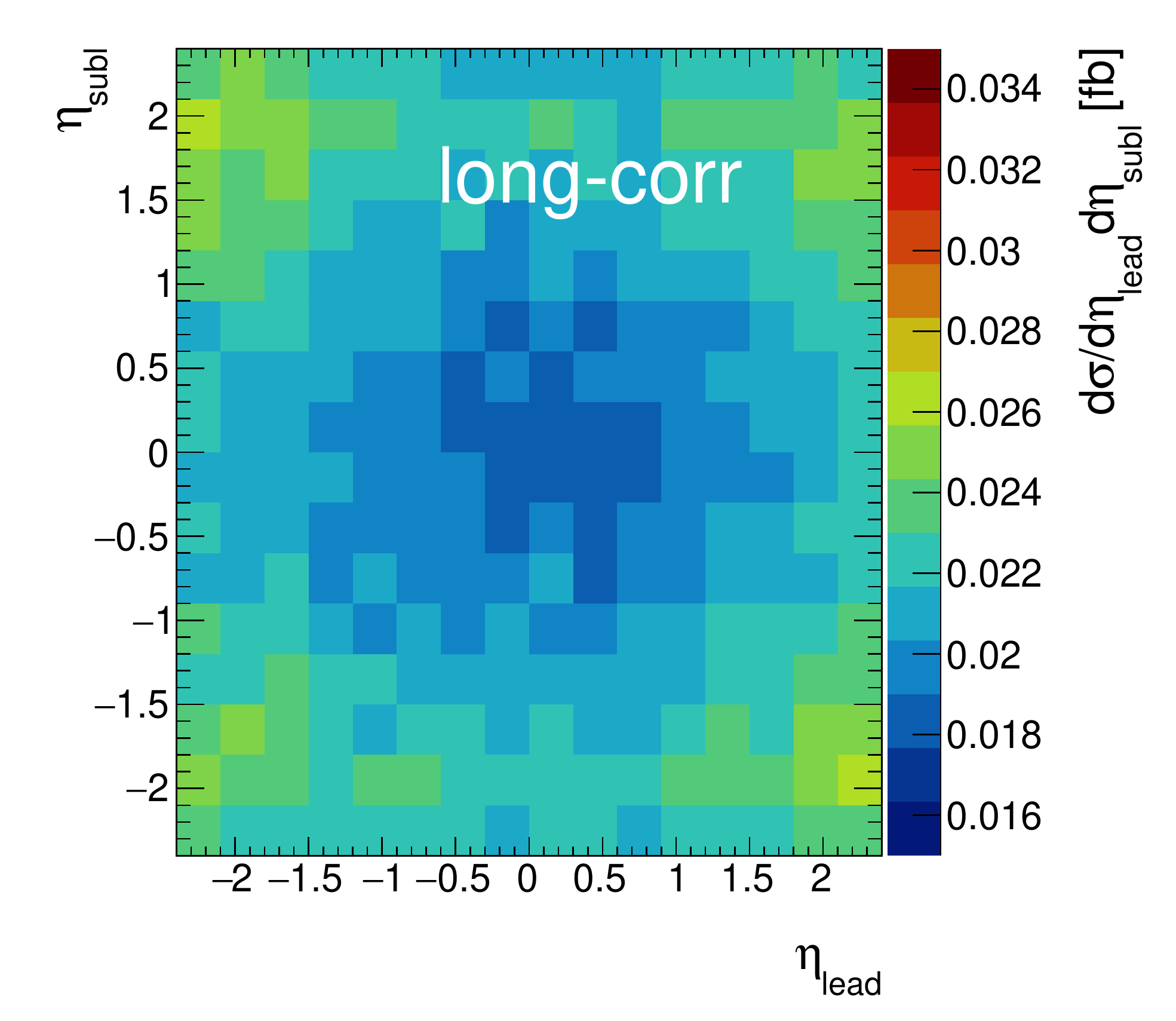}}
\subfigure[]{\includegraphics[width=0.42\textwidth]{%
    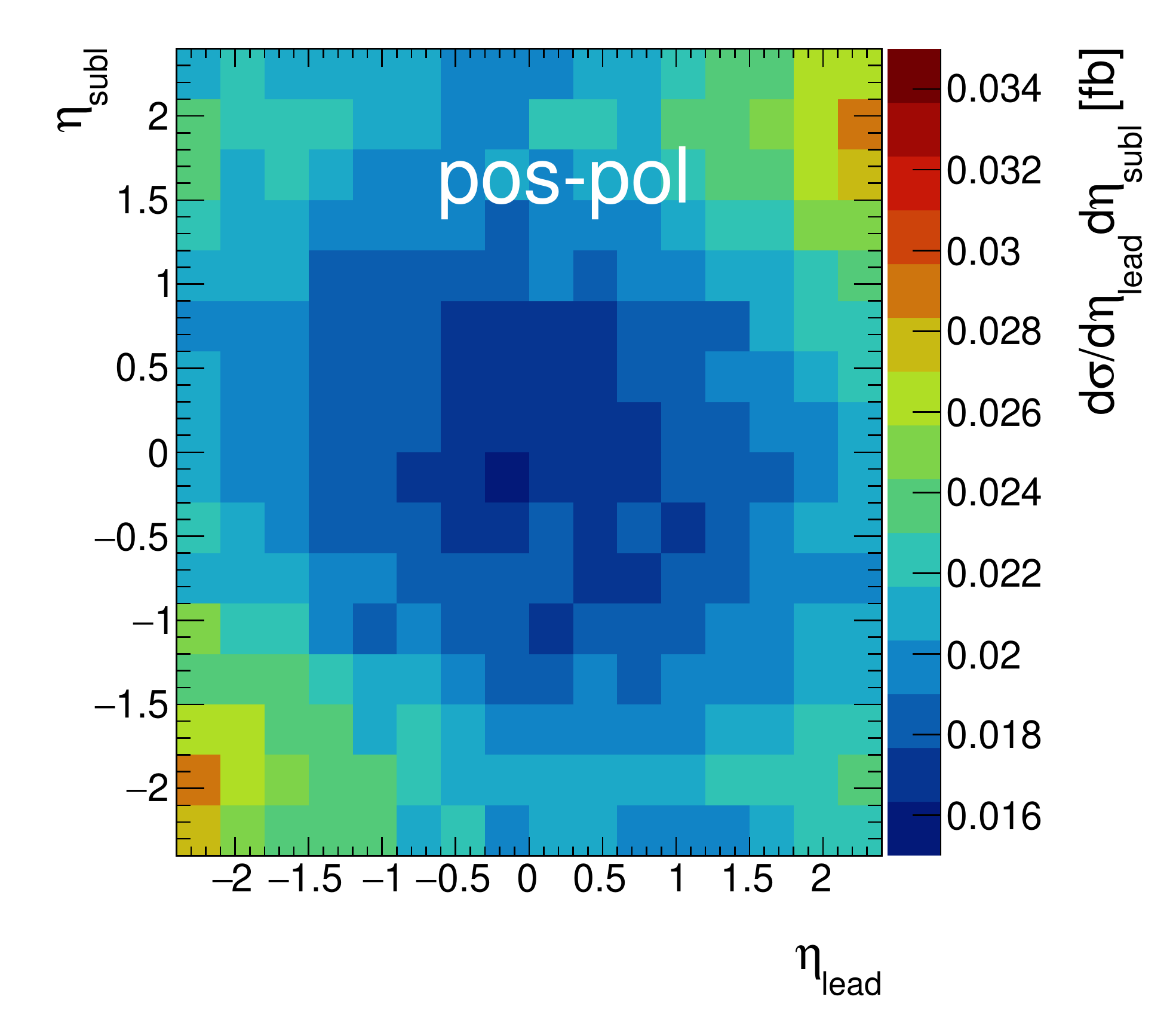}}
\subfigure[]{\includegraphics[width=0.42\textwidth]{%
    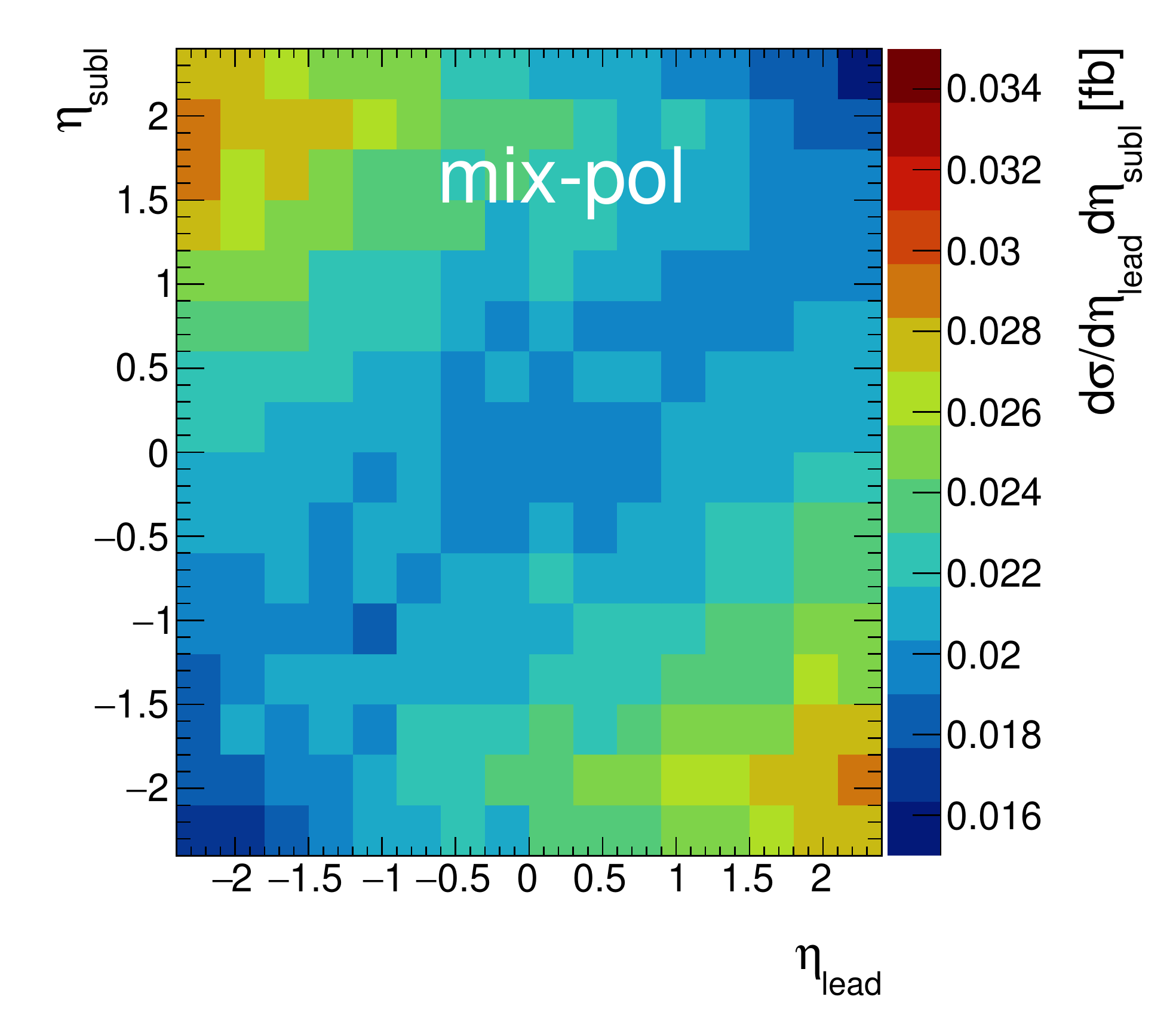}}

\caption{\label{fig:4_etaplane} The 2D rapidity plane similar to Fig. \ref{fig:0_etaeta_2}, here with final-state muons (at HL) and within the phase space \eqref{eq:sel}. 
    Top left: minimally correlated scenario. Top right: longitudinal correlation. Bottom left: positive polarization. Bottom right: mix polarization.
}
\end{center}
\end{figure}

\begin{figure}[t]
\begin{center}
\subfigure[]{\includegraphics[width=0.42\textwidth]{%
    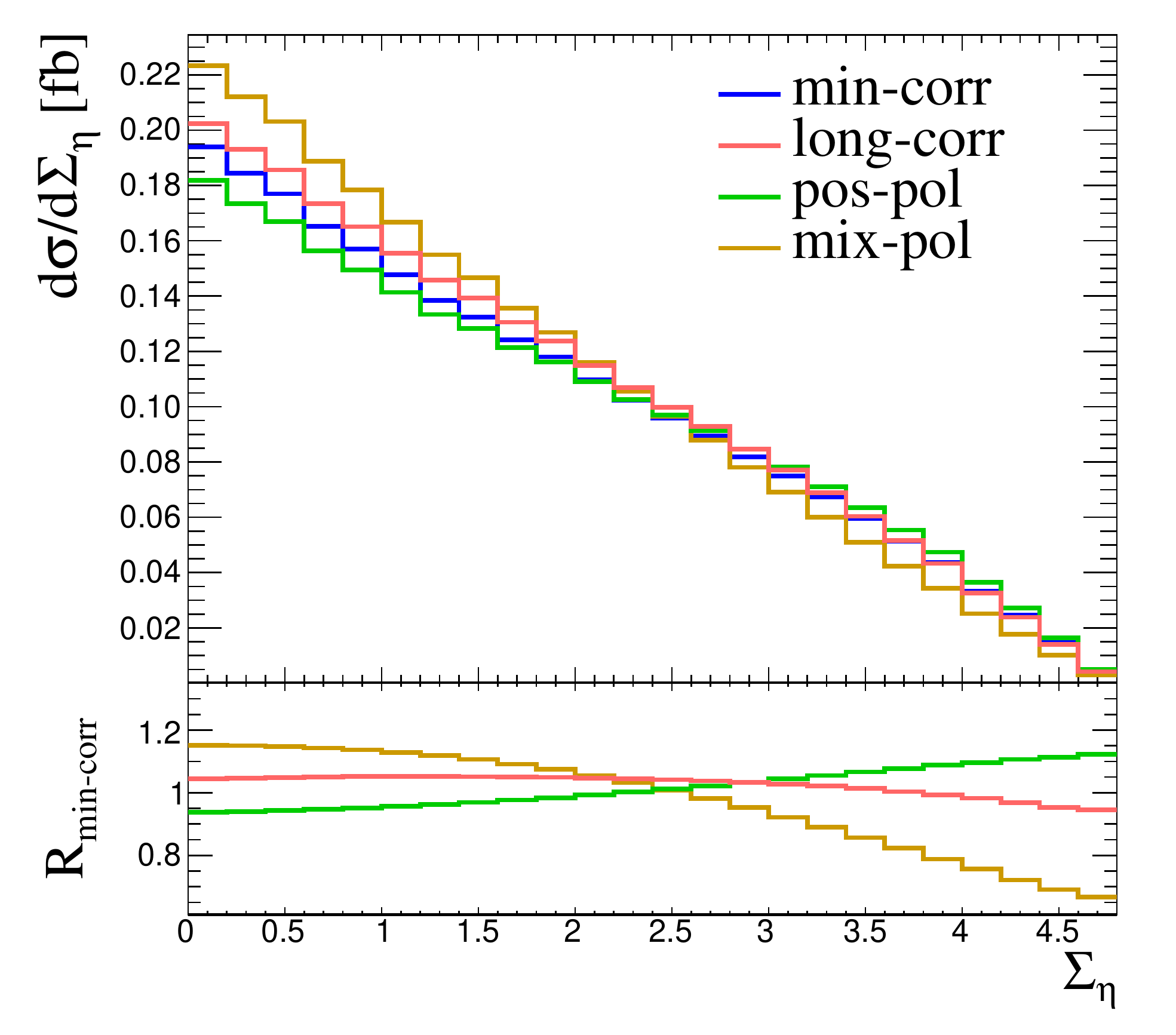}} 
\subfigure[]{\includegraphics[width=0.42\textwidth]{%
    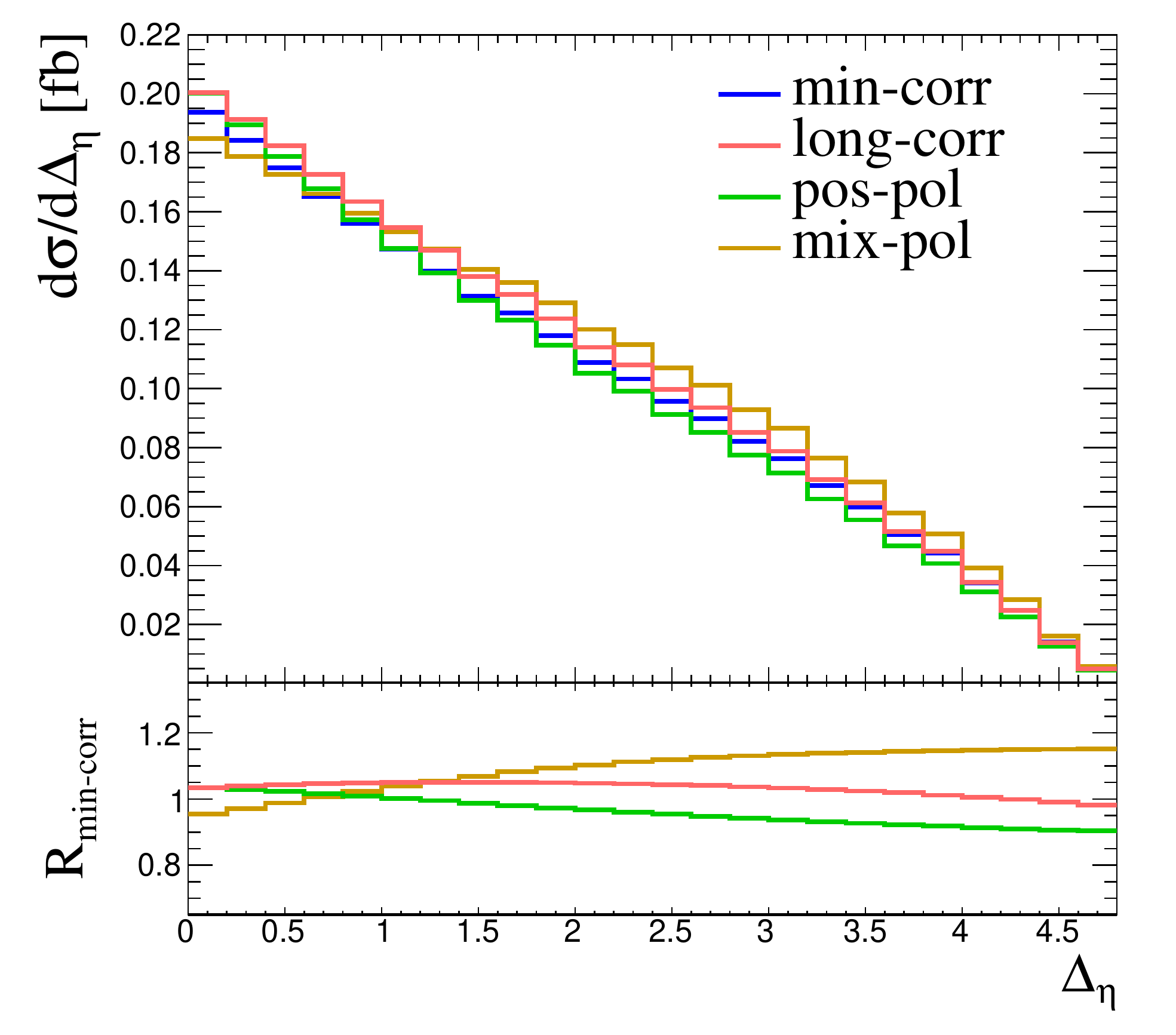}} 
    \subfigure[]{\includegraphics[width=0.42\textwidth]{%
    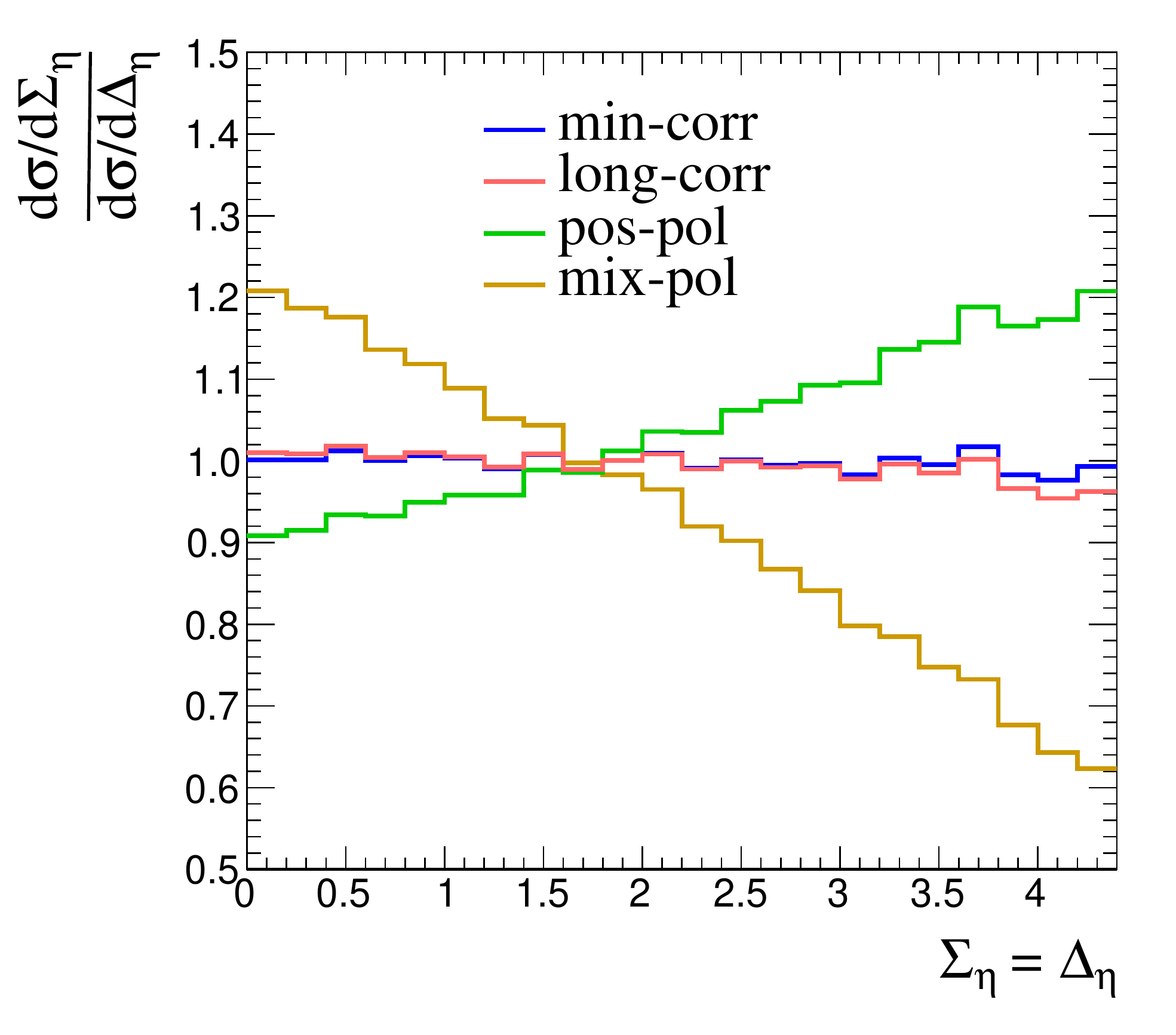}}      
    \subfigure[]{\includegraphics[width=0.42\textwidth]{%
    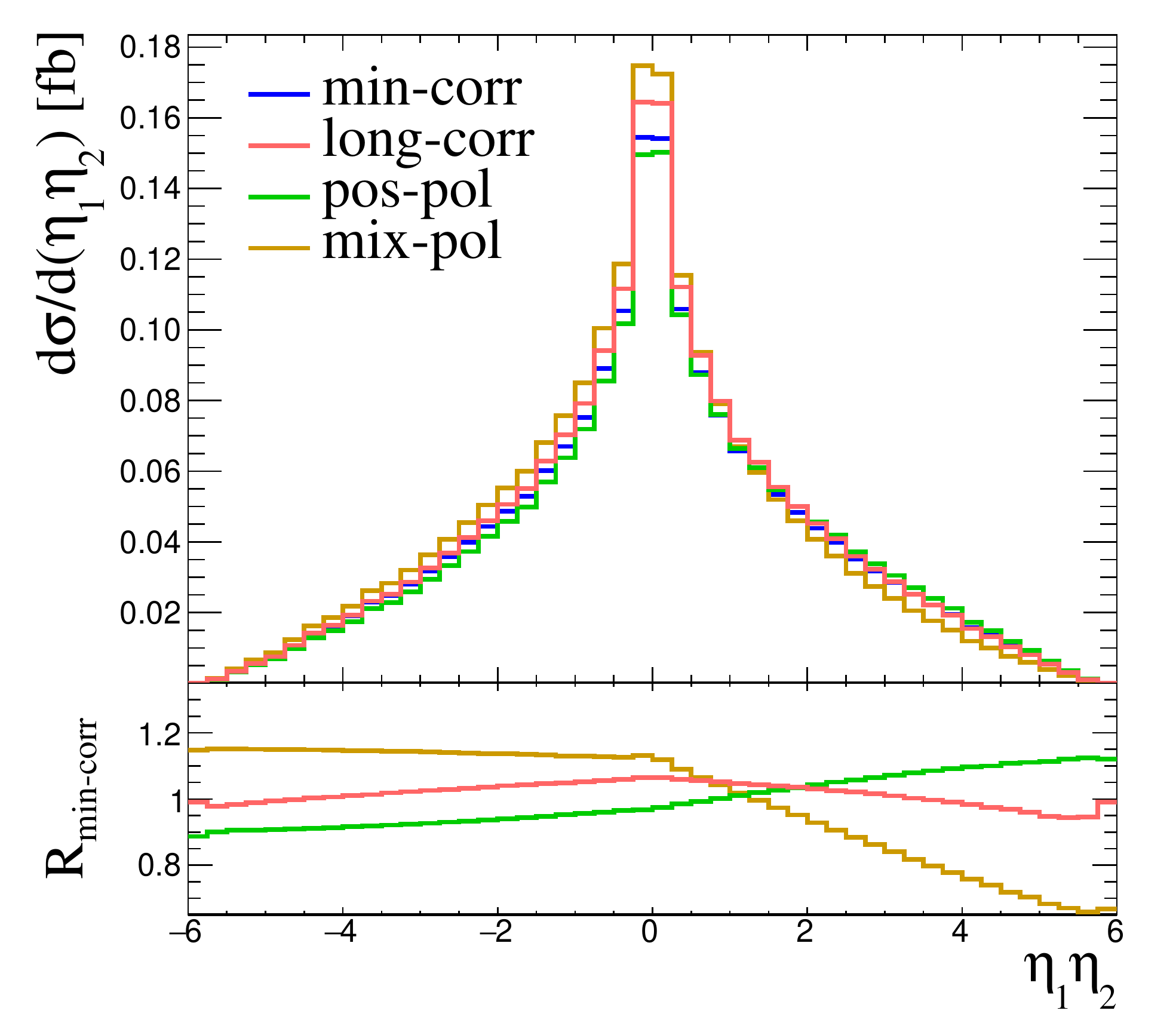}
    \label{fig:4_EtaProduct}}
\caption{\label{fig:SumDiff} Upper panels: the distributions in the muon rapidity sum $\Sigma_\eta$ (a) and difference $\Delta_\eta$ (b) for the different models with the ratio to the min-pol result.
Lower panels: the distributions in the correlation sensitive cross section ratio (c) and rapidity product (d). In absence of correlations, the cross section ratio is constant while the rapidity product is symmetric around the origin.
Distributions are analogous to those in Fig. \ref{fig:0_mupair_etaDifSum}, here with HL muons within the phase space \eqref{eq:sel}.}
\end{center}
\end{figure}

\begin{figure}[t]
\begin{center}
\subfigure[]{\includegraphics[width=0.35\textwidth]{%
    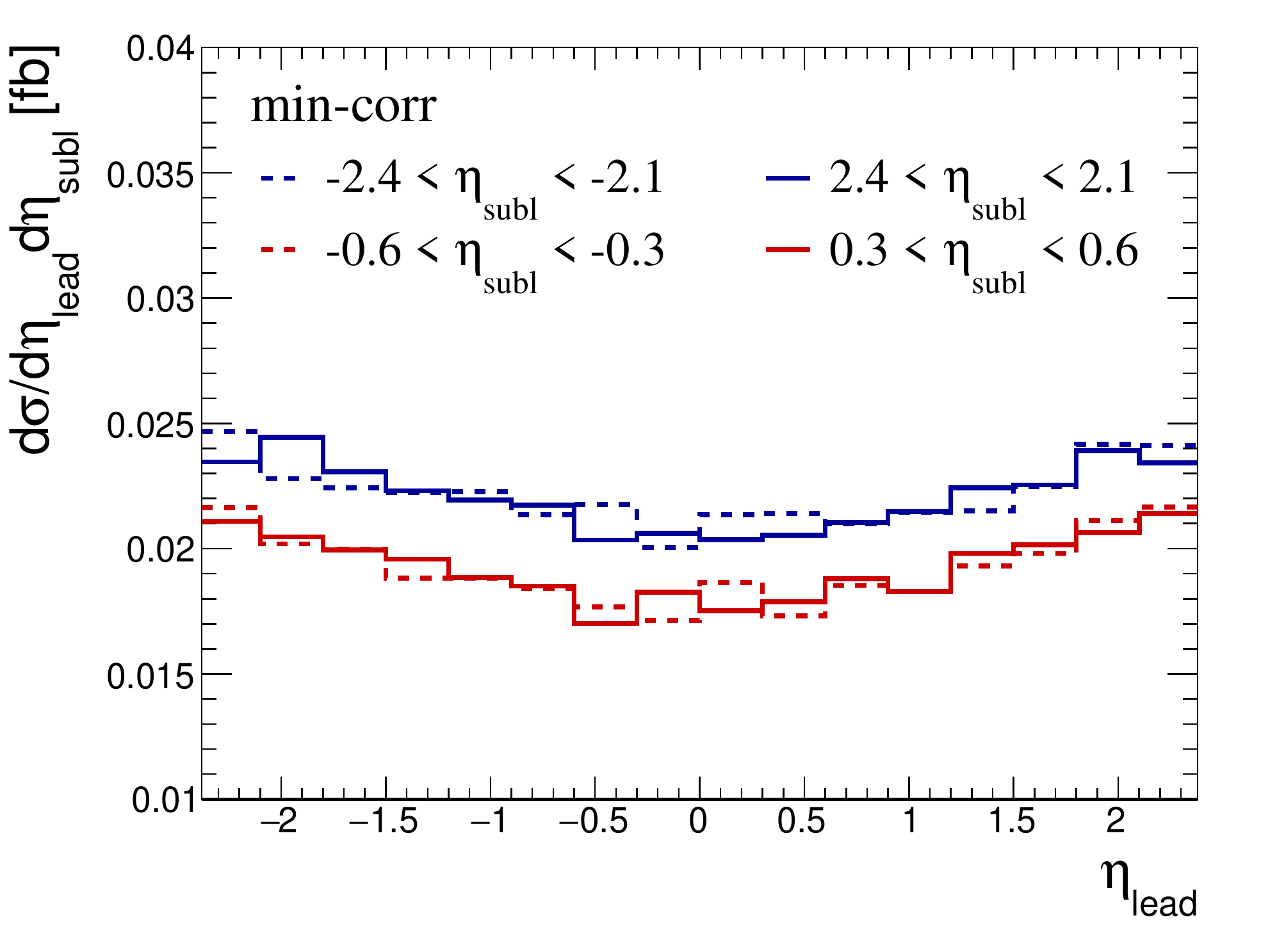}}
\subfigure[]{\includegraphics[width=0.35\textwidth]{%
    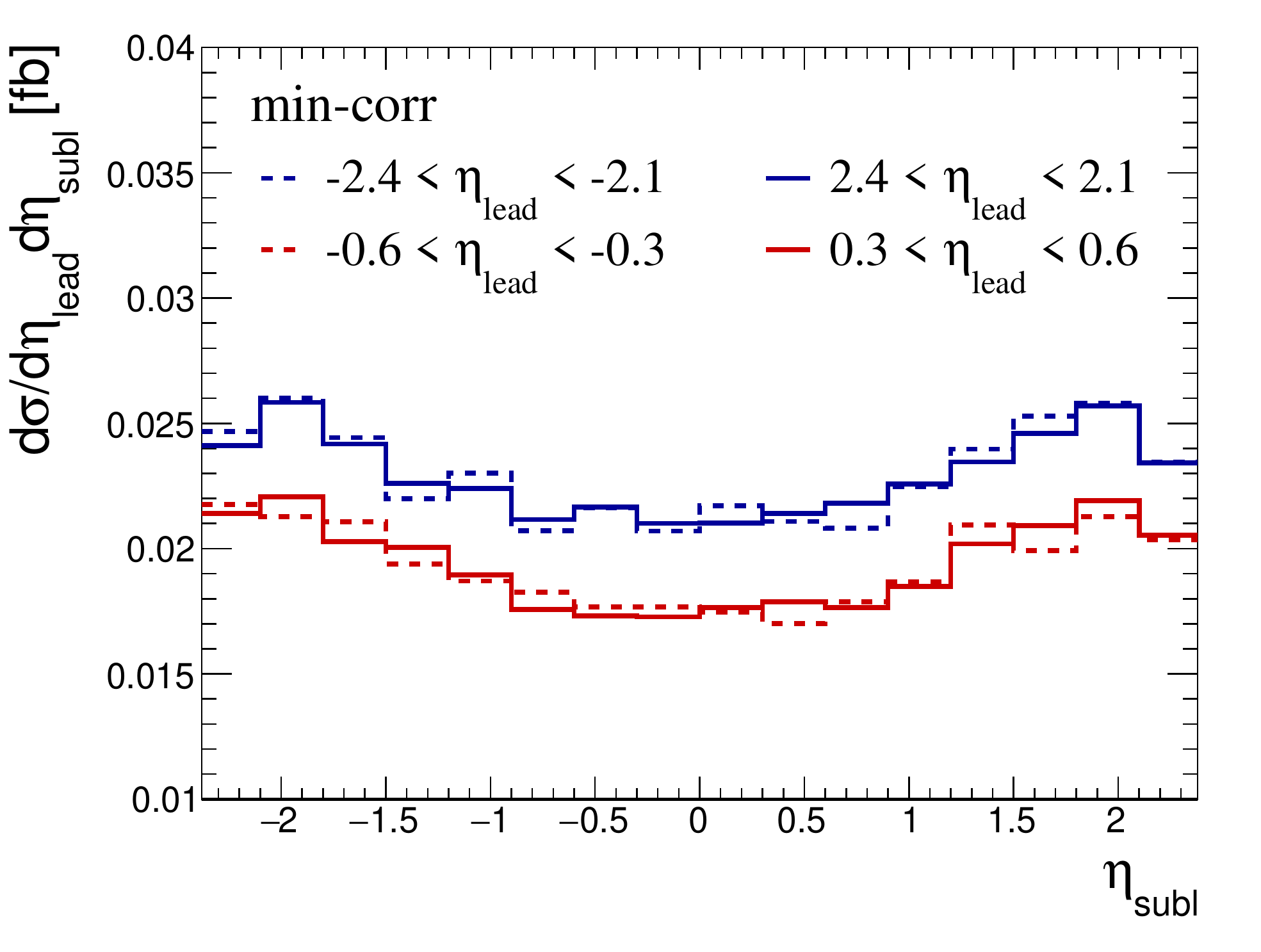}}    
\subfigure[]{\includegraphics[width=0.35\textwidth]{%
    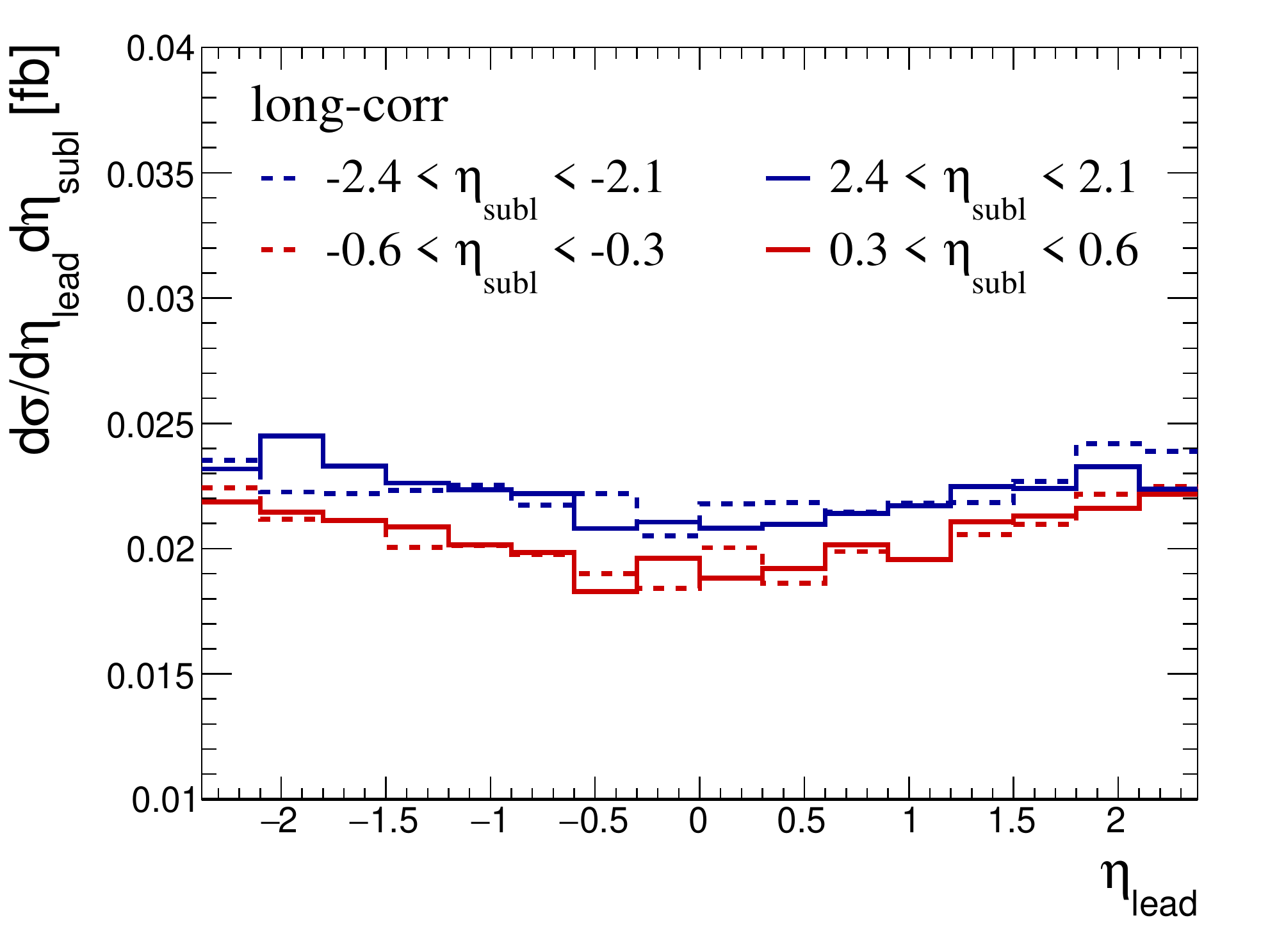}}
\subfigure[]{\includegraphics[width=0.35\textwidth]{%
    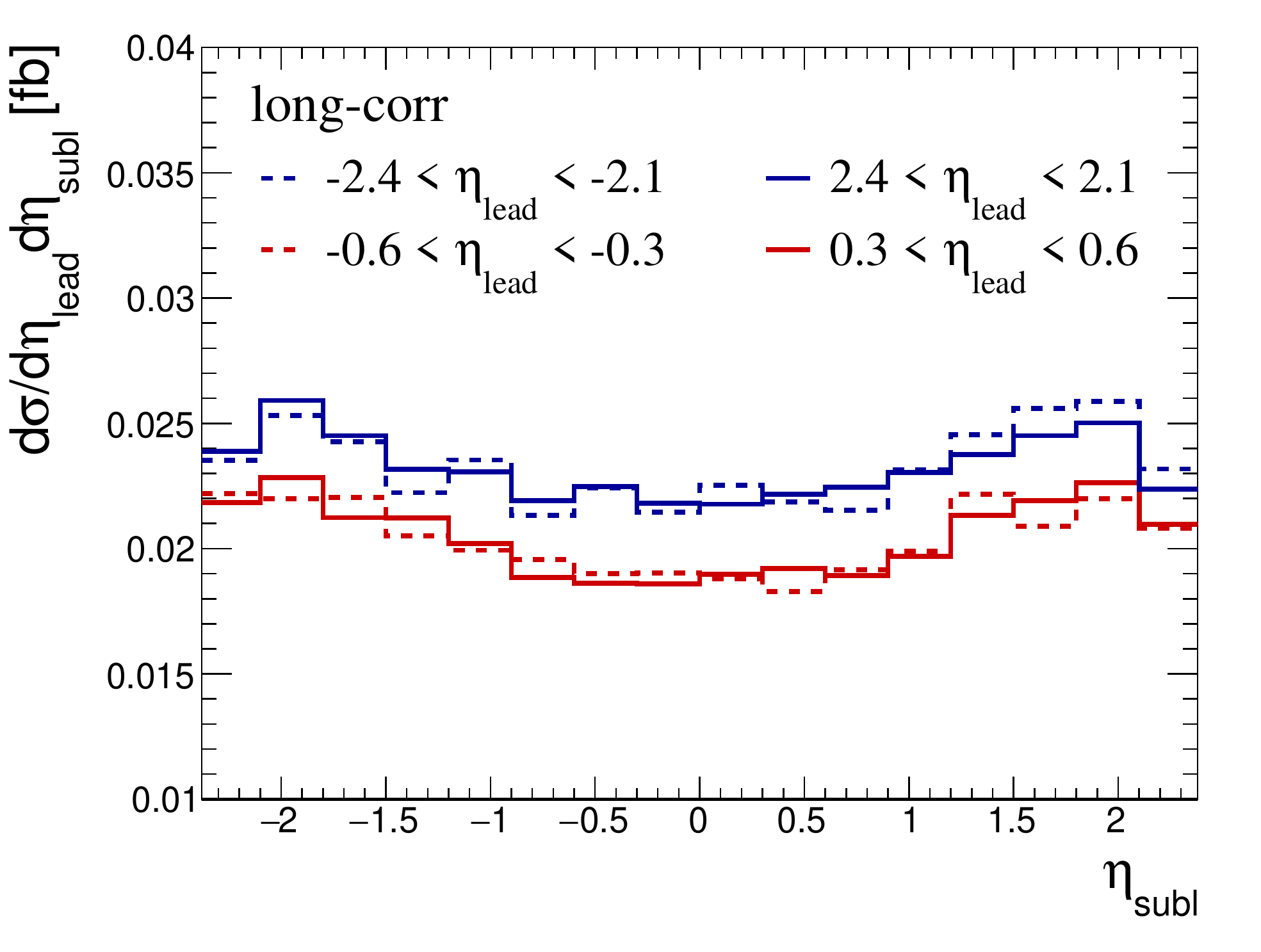}}    
\subfigure[]{\includegraphics[width=0.35\textwidth]{%
    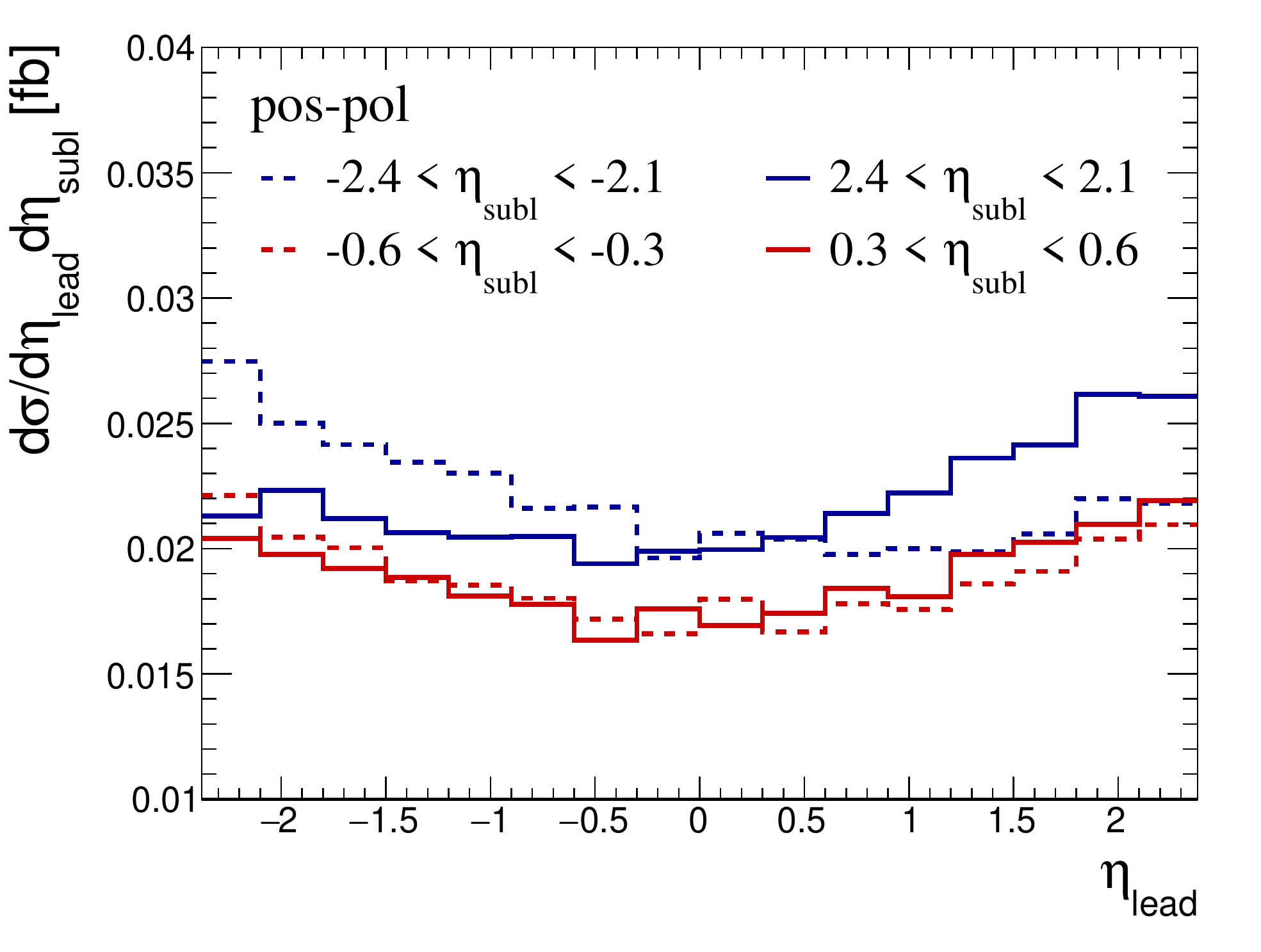}}
\subfigure[]{\includegraphics[width=0.35\textwidth]{%
    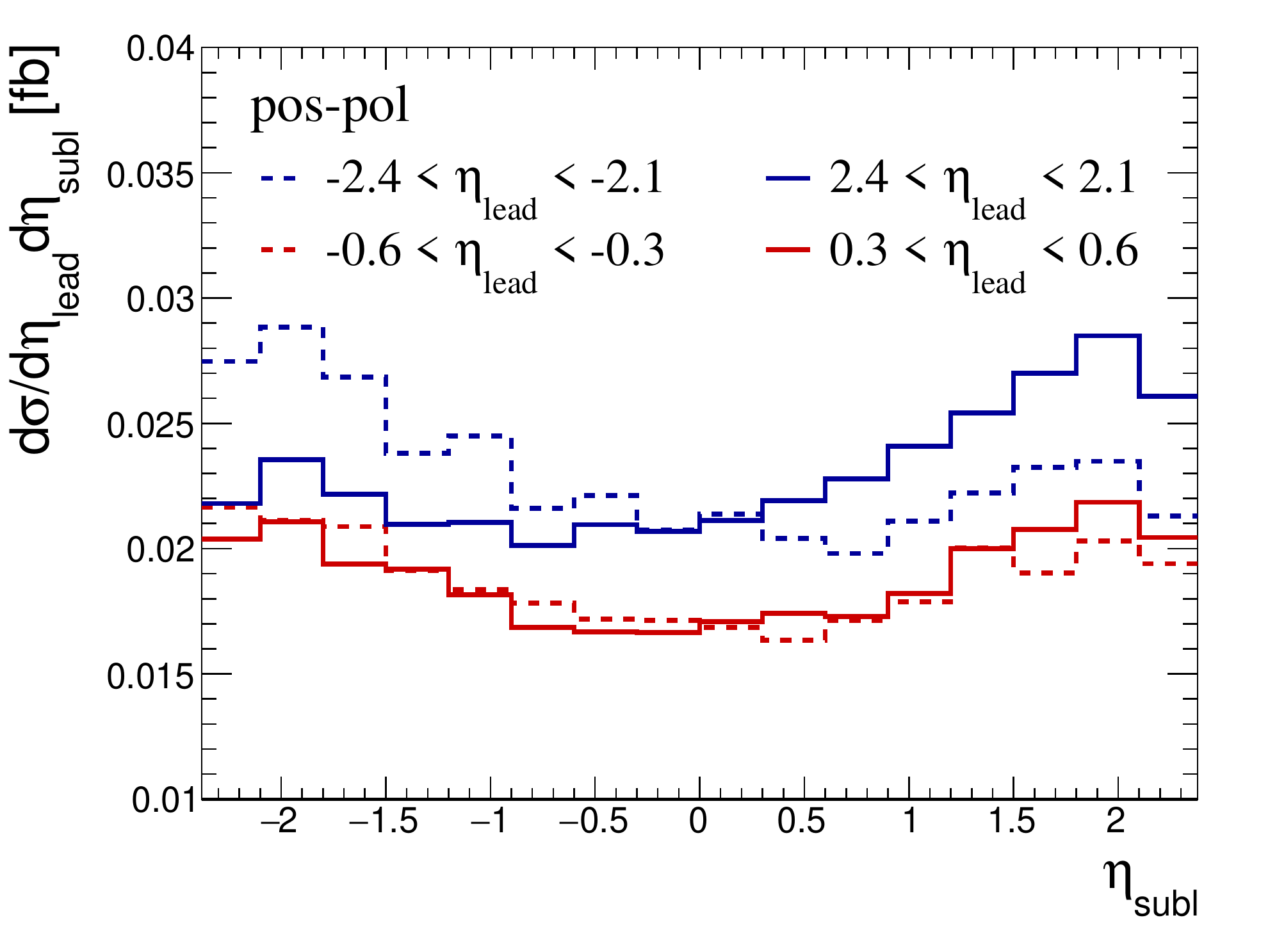}}    
\subfigure[]{\includegraphics[width=0.35\textwidth]{%
    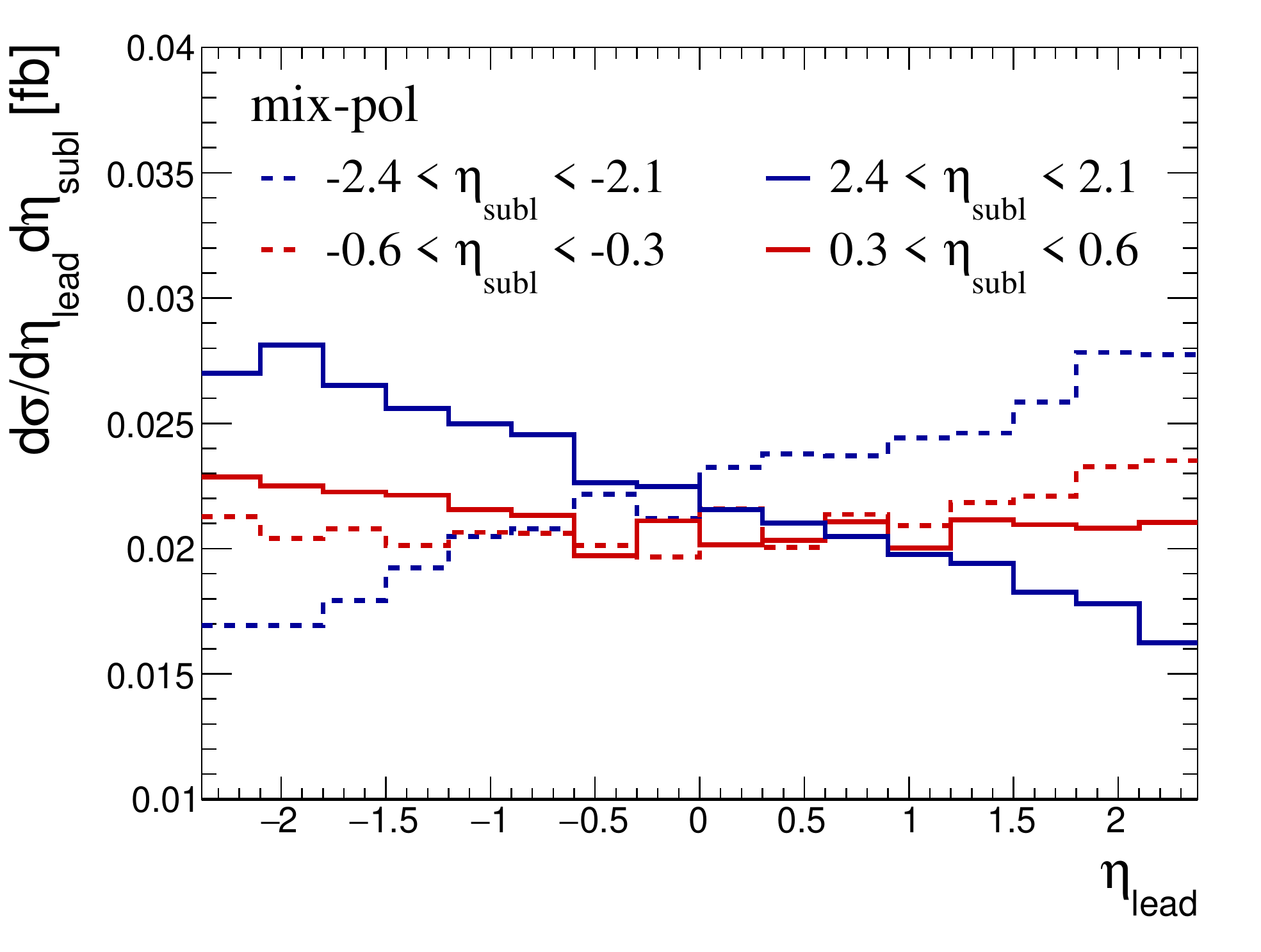}}
\subfigure[]{\includegraphics[width=0.35\textwidth]{%
    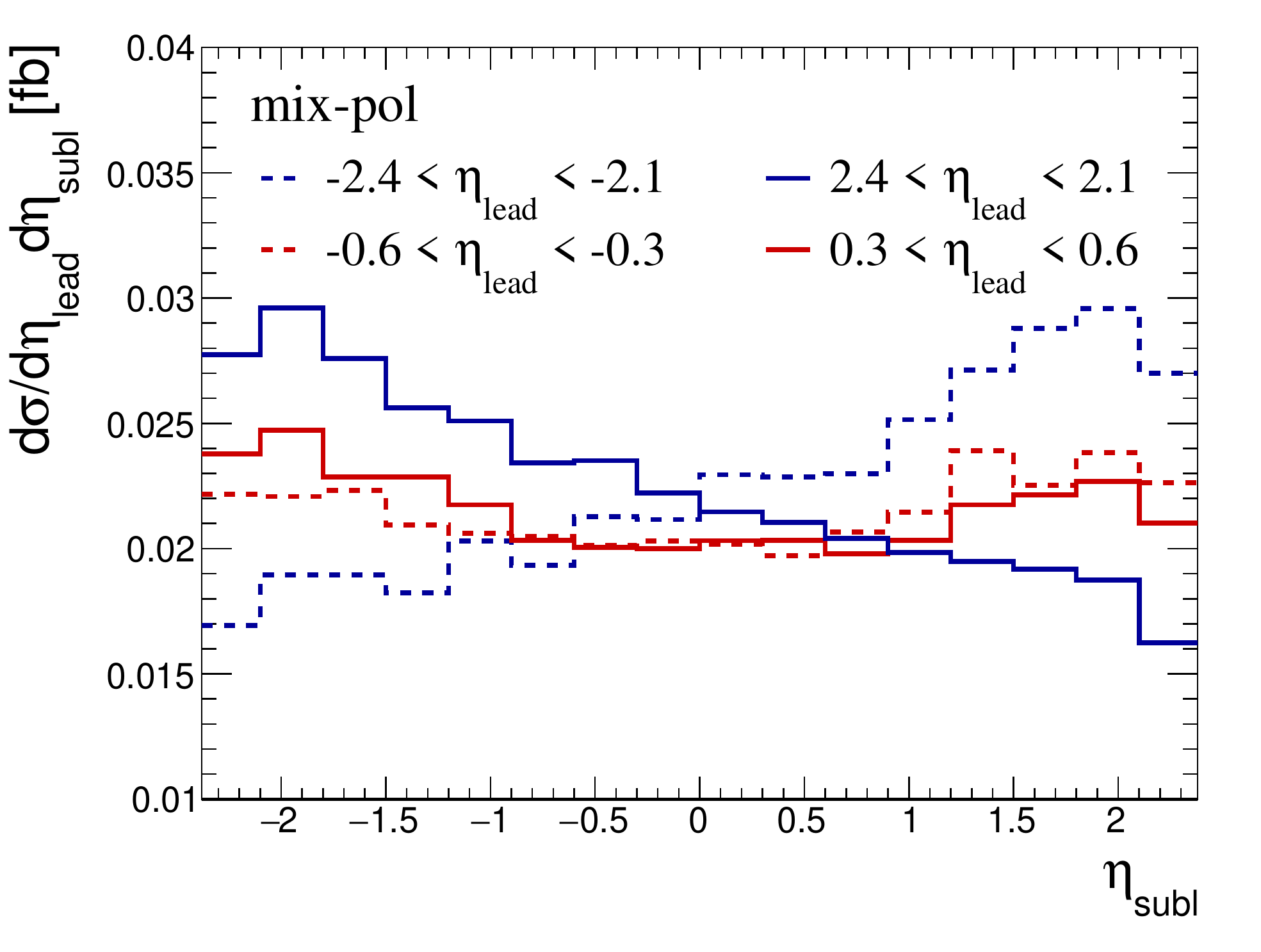}}    
\caption{\label{fig:slices} Double-differential distributions in rapidity of one of the muons (left: leading muon, right: subleading muon), for different ranges of the second muon rapidity separately 
for each correlation scenario, as indicated in the figures. Plots are similar to those in Fig. \ref{fig:0_etaslice_2}, here with HL muons within the phase space \eqref{eq:sel}.}

\end{center}
\end{figure}

\begin{figure}[t]
\begin{center}
\subfigure[]{\includegraphics[width=0.42\textwidth]{%
    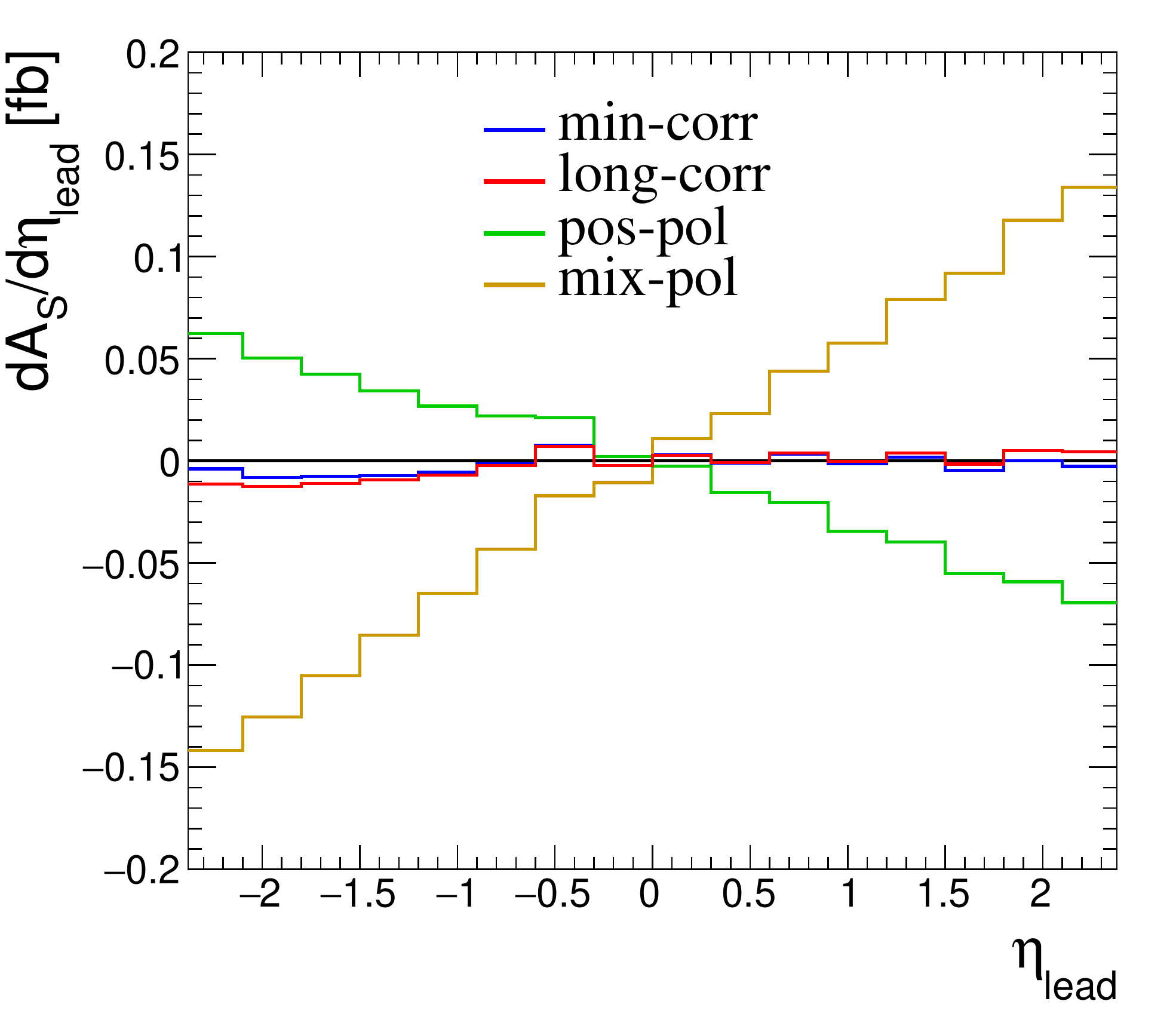}}
\subfigure[]{\includegraphics[width=0.42\textwidth]{%
    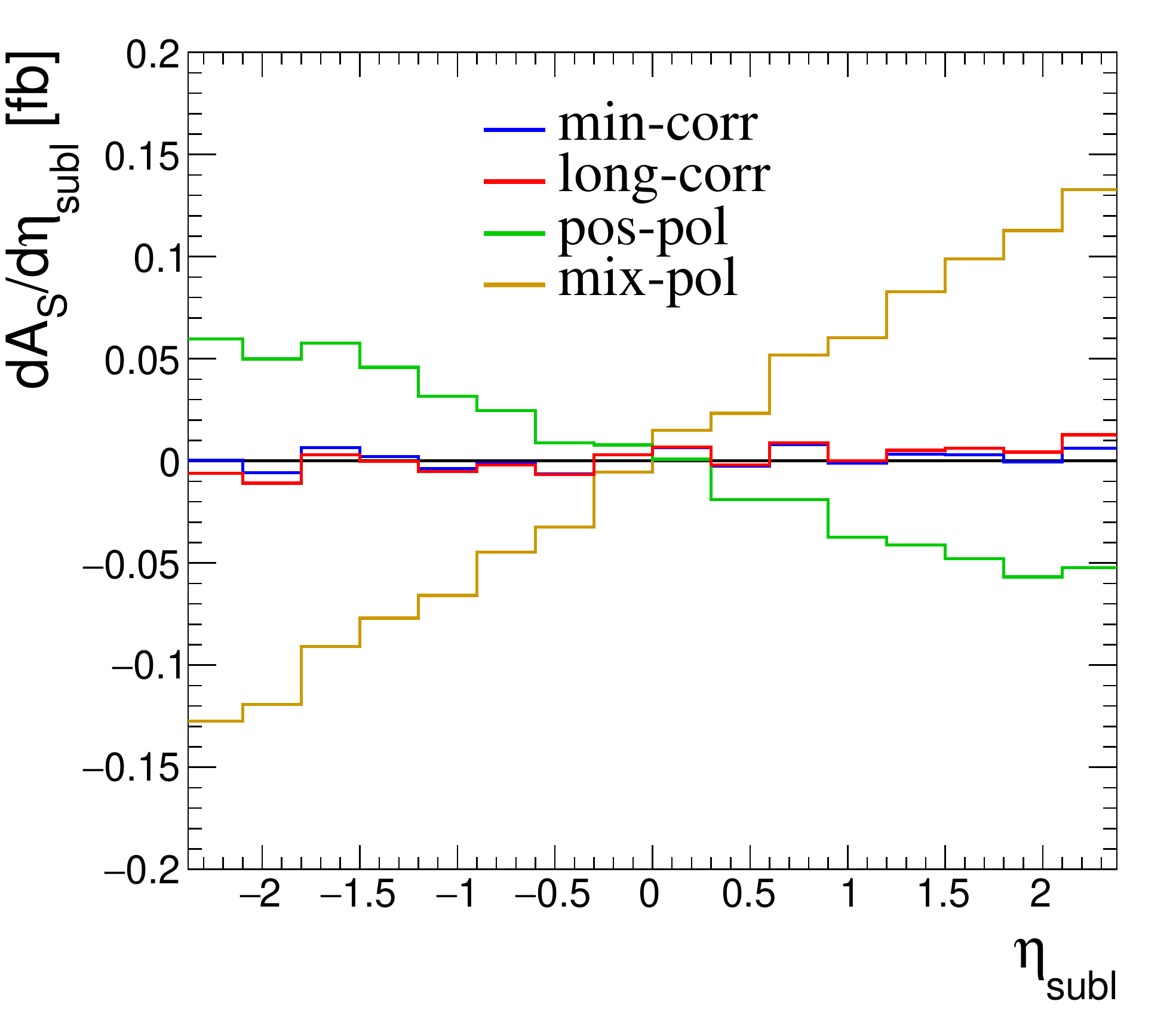}}
\caption{\label{fig:slice_asym} The asymmetry constructed from the rapidity slicing as defined in \eqref{eta-asymm}, similar analogous to Fig. \ref{AsymmSlice}, 
as a function of leading muon rapidity (left) or subleading muon rapidity (right). Here with HL muons within the phase space \eqref{eq:sel}.}

\end{center}
\end{figure}

With this (close to) pure DPS sample, detailed studies of the different correlations can be performed. In the following, we show the variables described in Section \ref{Sec:parton} 
  at the hadron level (HL) using the Final Selection, i.e. in the case of realistic measurements. Figures \ref{fig:4_etaplane}-\ref{fig:slice_asym} discussed in this section correspond to Figures \ref{fig:0_etaeta_2}-\ref{AsymmSlice} 
of Section \ref{Sec:parton} related to the parton level analysis.
All the qualitative pictures discussed in Section \ref{Sec:parton} are still valid. Therefore, we refer to the previous text for a thorough explanation of these variables. 

The  variables considered for the detection of parton correlations are based on combinations of the two muon rapidities. The full pictures of the double differential 
cross sections in the two muon rapidities ($\eta_{\text{lead}}$ and $\eta_{\text{subl}}$), for the different scenarios at HL, are shown in Fig.~\ref{fig:4_etaplane}. 
The corresponding plots are shown at PL in Fig.~\ref{2DPlots} but with a change of variables to $\eta_1$ and $\eta_2$. 
No correlations implies symmetries around zero rapidity for both of the two rapidities, and this is also the case for the min-corr scenario of Fig.~\ref{fig:4_etaplane} (a). 
The cross section reaches its maximum value when both of the two rapidities are around $\pm 2$. Once correlations are introduced in long-corr, we can see how the symmetry 
between positive and negative rapidities is broken and more muons are produced with rapidities of opposite sign rather than with same sign. This is true also for mix-pol 
scenario and to a much larger extent. In this case, the broken symmetry is clearly visible in Fig.~\ref{fig:4_etaplane} (d), where there is almost twice as 
many muons produced in the peak region of opposite- compared to same-sign rapidity. In the pos-pol scenario in Fig.~\ref{fig:4_etaplane} (c), there is instead an 
abundance of muons created with same-sign rapidities.

It is important to be careful with the choice of phase-space cuts for the event selection when reading off correlation effects from the rapidity distributions. For instance, a cut on the invariant mass of the final-state muons creates an artificial imbalance in the distributions. Such a selection was used when discussing correlations in~\cite{Ceccopieri:2017oqe}, where the kinematical cuts used by the CMS collaboration~\cite{Sirunyan:2017hlu} were adopted.

Our golden variable to measure correlations is the asymmetry $A$ and, in its less inclusive version, the distribution of the rapidity product, 
previously explored in studies of SSW production in DPS, see e.g. \cite{Gaunt:2010pi,Ceccopieri:2017oqe}. 
The cross section differential in the rapidity product is shown in Fig. \ref{fig:4_EtaProduct}, to be compared with Fig. \ref{EtaProduct}. 
The  shape dependence of this variable on the correlation scenario remains.
From the integration over the $\eta_1\eta_2$ range in Fig. \ref{fig:4_EtaProduct}, separately for positive and negative $\eta_1 \eta_2$, 
we have constructed the asymmetry $A$ of eq. \eqref{eta-asymm}. As extensively stressed earlier, these two cross sections have to be exactly 
equal in absence of correlations. A non-zero measurement of this variable would be a direct indication of correlations. 
%%%%%%%%%%%%%%
%\iffalse
%The results for the difference scenarios are given in Table~\ref{tab:as-nocut}.
%
%%
%\begin{table}[H]\centering
%\begin{tabular}{ c|c|c|c|c|c|c }
%&  \multicolumn{3}{c|}{\bf min-corr} & \multicolumn{3}{c}{\bf long-corr}  \\
%Selection &  $\sigma$ [fb] & $A$ & $S_{lin} (\Sigma_{\eta}/\Delta_{\eta})$ & $\sigma$ [fb] & $A$ & $S_{lin} (\Sigma_{\eta}/\Delta_{\eta})$ \\
% \hline
%Baseline FS & 1.66 & 0.00 & -0.00 & 1.67 & 0.01 & 0.01 \\
%Final FS                  & 0.48 & 0.00 & 0.00 & 0.50 & 0.00 & 0.01 \\
%\hline
%\hline
%&   \multicolumn{3}{c|}{\bf pos-pol} &  \multicolumn{3}{c}{\bf mix-pol} \\
%&  $\sigma$ [fb] & $A$ & $S_{lin} (\Sigma_{\eta}/\Delta_{\eta})$ & $\sigma$ [fb] & $A$ & $S_{lin} (\Sigma_{\eta}/\Delta_{\eta})$ \\
%\hline
%Baseline FS & 1.66 & -0.05 & 0.08 & 1.67 & 0.12 & -0.17 \\
%Final FS           & 0.47 & -0.04 & 0.07 & 0.51 & 0.07 & -0.14  \\
%\end{tabular}
%\caption{\label{tab:as-nocut} Asymmetry between same vs opposite hemisphere muon production for the different scenarios.}
%\end{table}
%%
%\fis
%%%%%%%%%%%%%%%
The results for this asymmetry in the mix-pol scenario using the Final Selection have already been presented in \cite{Cotogno:2018mfv}. 
Both the polarized scenarios produce clear asymmetries, although of opposite sign, see again Table \ref{tab:allnumbers}. The asymmetry for min-corr is zero and for long-corr close to zero. 
The inclusive nature of this variable, in combination with the quite large asymmetries generated by the models of polarization, makes it a promising 
candidate for first measurements of spin correlations between two partons inside a proton. It is important, however,  to keep in mind that even a precise 
measurement of a zero asymmetry would be interesting, as it would put severe limits on the correlations and on models for DPDs. 

\begin{table}[t]\centering
\begin{tabular}{ c|c|c|c|c|c } 
 $|\eta_i|$& $ > 0.0$  &  $ > 0.3$ & $ > 0.6$ & $ > 0.9$ & $ > 1.2$ \\
 \hline
$A_{\text{min-corr}}$            & 0.00 & 0.00 & 0.00 & 0.00 & 0.00 \\
$\sigma_{\text{min-corr}}$ [fb]  & 0.48 & 0.37 & 0.28 & 0.20 & 0.13 \\
\hline
$A_{\text{long-corr}}$           & 0.00 & 0.01 & 0.01 & 0.01 & 0.01 \\
$\sigma_{\text{long-corr}}$ [fb] & 0.50 & 0.39 & 0.29 & 0.21 & 0.14 \\
\hline
$A_{\text{pos-pol}}$             & -0.04 & -0.05 & -0.05 & -0.06 & -0.07 \\
$\sigma_{\text{pos-pol}}$ [fb]   & 0.47  & 0.37  & 0.28  & 0.20  & 0.13 \\
\hline
$A_{\text{mix-pol}}$             & 0.07 & 0.09 & 0.11 & 0.14 & 0.16 \\
$\sigma_{\text{mix-pol}}$ [fb]   & 0.51 & 0.39 & 0.29 & 0.20 & 0.13 \\
\end{tabular}
\caption{\label{tab:as} Asymmetries and DPS cross sections for different cuts on the rapiditiy of the individual muons. The larger part of the central detector is removed, 
the higher values of asymmetry can be reached.}
\end{table} 

By imposing additional cuts on the rapidities of the two muons, it is possible to increase the asymmetry further, at the price of reducing the size of the cross section. 
In particular, cutting out the central rapidity regions increases $A$. This is demonstrated for the different scenarios in Table \ref{tab:as}. 
The optimal trade-off between increasing the asymmetry and decreasing the cross section should be investigated in detail when performing the measurement. 
The asymmetry could be further enhanced, for instance by including smaller transverse muon momenta. 
In addition, we have found that even the simple decision tree analysis, which was performed to explore the potential power of a multivariate analysis to suppress the $WZ$ background, naturally causes a small enhancement of the asymmetry as a by-product.  A full fledged statistical analysis of this kind can simultaneously enhance the signal to background ratio and optimize the asymmetry.
In Section \ref{Sec:significance} we discuss in more detail the actual feasibility of this measurement at the LHC.

Here, we would like to note that a large asymmetry $A$ was found in \cite{Cabouat:2019gtm}, due to longitudinal correlations and splitting contributions to DPS.  However, the comparison must be done with care. In particular, the phase-space cuts on rapidity have a very large impact on the asymmetry. Further, cuts to suppress the single parton scattering contribution to the SSW production will reduce also the DPS splitting case. In any case, the result remains interesting in its own right, and the effects could combine to increase the asymmetry further.

The cross section distribution in the sum $\Sigma_\eta$ and the difference $\Delta_\eta$ of the muon rapidities for our four scenarios are given in the two upper plots of Fig. \ref{fig:SumDiff},  
to be compared to Fig. \ref{SumDiff}. From these two variables, we have constructed the bin-by-bin ratio $(d\sigma/d\Sigma_\eta)/(d\sigma/d\Delta_\eta)$, which we show in Fig.~\ref{fig:SumDiff} (c).  
The shape differences described earlier in the text
of Section \ref{Sec:parton} are all still clearly visible for the polarized distributions, thus correlations are not washed out. The distributions in the sum (difference) of rapidity, as well as their slopes, are sensitive to the details of the unpolarized DPS cross section. This makes them less suitable for measuring correlations. However, the ratio,  $S_{\text{lin}}(\Sigma_\eta/\Delta_\eta)$, related to the slope of the curves in Fig. \ref{fig:SumDiff}(c), remains a very promising variable to look at to constrain the size of parton correlation in DPS.

%
% IF YOU UNCOMMENT THIS, COMPILING CRASHES .... S
%\afterpage{\clearpage}

The last remaining variable we discuss, in which the effects of correlations are visible,
is the rapidity dependence on one of the muons in different rapidity slices of the other muon. 
This is shown in Fig. \ref{fig:slices}, to be compared to Fig. \ref{fig:0_etaslice_2}. One needs to bare in mind that, when dealing with distributions involving the rapidity of one single muon, 
we need to move from the identification of $\mu_1$ and $\mu_2$ (muon coming from the first and the second hard scattering) to the labeling $\mu_{\text{lead}}$ and $\mu_\text{subl}$ 
(the two hardest muons reaching the detector).  Therefore, there is the need for showing the distributions of both $\eta_{\text{lead}}$ and $\eta_{\text{subl}}$ in Fig. \ref{fig:slices}, 
since they are not equal. The uncorrelated scenario still has the same shape regardless of the rapidity slice, and this behavior is observed through the min-corr scenario in Fig. \ref{fig:slices} (a). 
The long-corr scenario shows some minor dependence, but is similar to min-corr. Once again, the two polarized scenarios lead to large correlation effects in opposite directions, 
as shown in Fig. \ref{fig:slices} (c) for the pos-pol model and (d) for mix-pol. 
From this type of rapidity slicing we constructed the differential asymmetry in \eqref{eta-asymm}. The results for $d A_S/d \eta_B$ is shown in Fig. \ref{fig:slice_asym}, 
to be compared with Fig. \ref{AsymmSlice}. Also for this variable, the min-corr and long-corr show results similar to the uncorrelated (which equals zero), while the two polarized models produce sizable differences. 

 \subsection{Significance of correlation measurements}
 \label{Sec:significance}
 
 We now try to quantify the sensitivity of ATLAS and/or CMS to the measurement of the asymmetry $A$, defined in \eqref{asymmetry}. 
To this end, we assess how large significance can be reached with respect to the measurement of exact zero. 
 
Based on our discussion above, we assume that the 0.29 fb cross section with $A=0.11$ (result from Table \ref{tab:as}) can be reached with the (improved) signal to background ratio
$S/B = 3$ ($WZ$ background), and that the  contribution of the remaining $WZ$ background to the asymmetry can be subtracted by a precise theoretical 
calculation, on which we assume a further 10\% uncertainty. 

Let us assume a Poissonian uncertainty on the number of DPS events with the two muons in the same/opposite hemisphere. Corresponding Gaussian distributions of the signal cross sections 
in the two hemispheres then represent statistical fluctuations, and we use them to test how many standard deviations the measured asymmetry differs from zero. 
In order to estimate the effect of our assumptions on the size of the background, we add a scaling parameter $b$ to the uncertainty calculation for the number of events per hemisphere (after the background subtraction)
\begin{equation}
\Delta N = \sqrt{ [N_{WW}+b(N_{WZ} +N_{\text{top}})] + (b \Delta N^{\text{theo.}}_{WZ})^2},
\end{equation} 
where $N_X$ is the number of events from process $X$, and $\Delta N^\text{theo.}_{WZ}$ is the theoretical uncertainty on the subtraction. 
Fig. \ref{fig:sigmas} shows the significance of a measurement as a function of the integrated luminosity.
Here, the central predictions (b = 1) are drawn as solid lines, while the colored uncertainty bands are created by 
variating the parameter $b$ between 1/2 and 2, to indicate the sensitivity of our predictions on the totality of the background assumptions. 

\begin{figure}[tb]
\begin{center}
\subfigure[]{
   \includegraphics[width=0.45\textwidth]{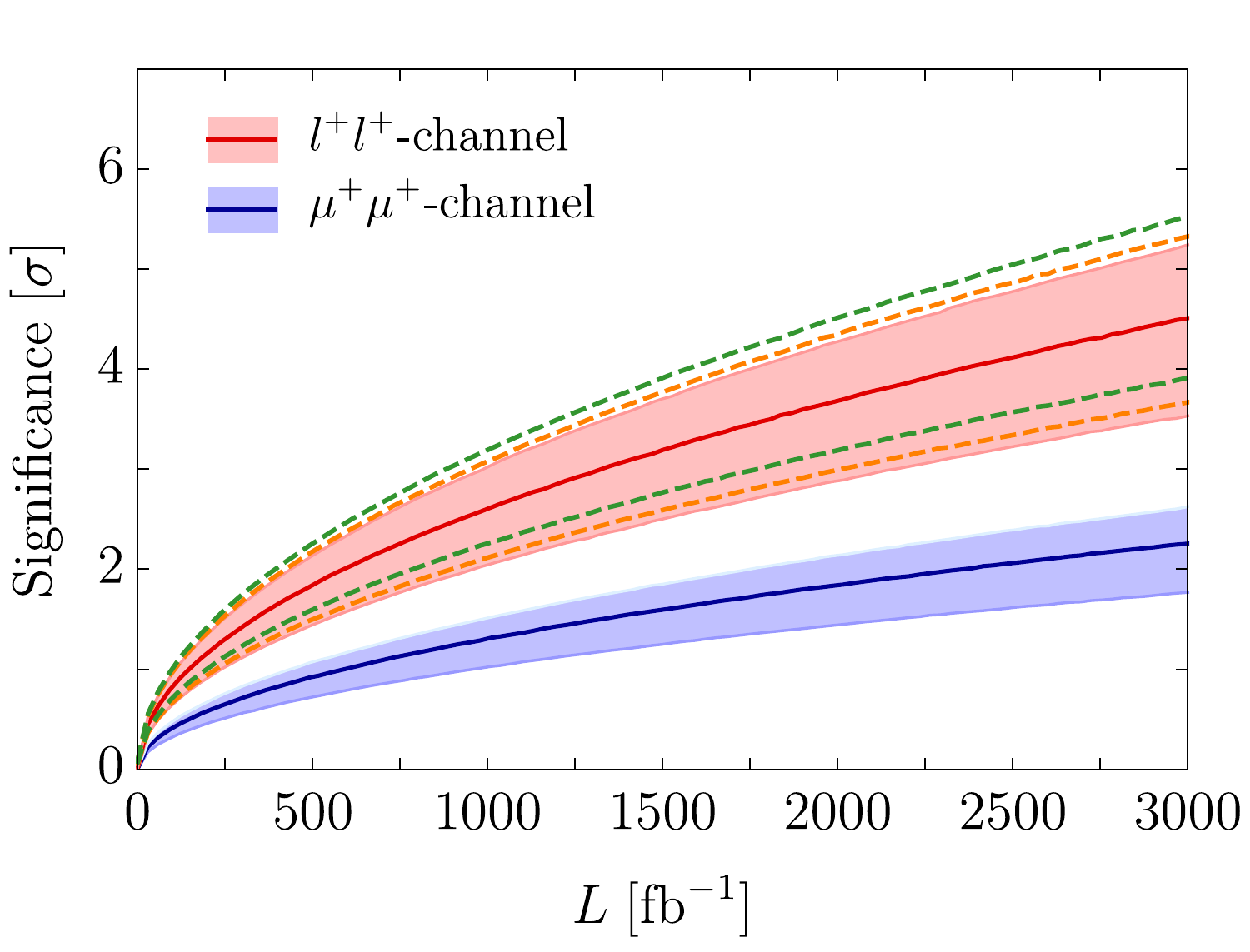}    \label{fig:sigmas}
 }   
\subfigure[]{   \includegraphics[width=0.45\textwidth]{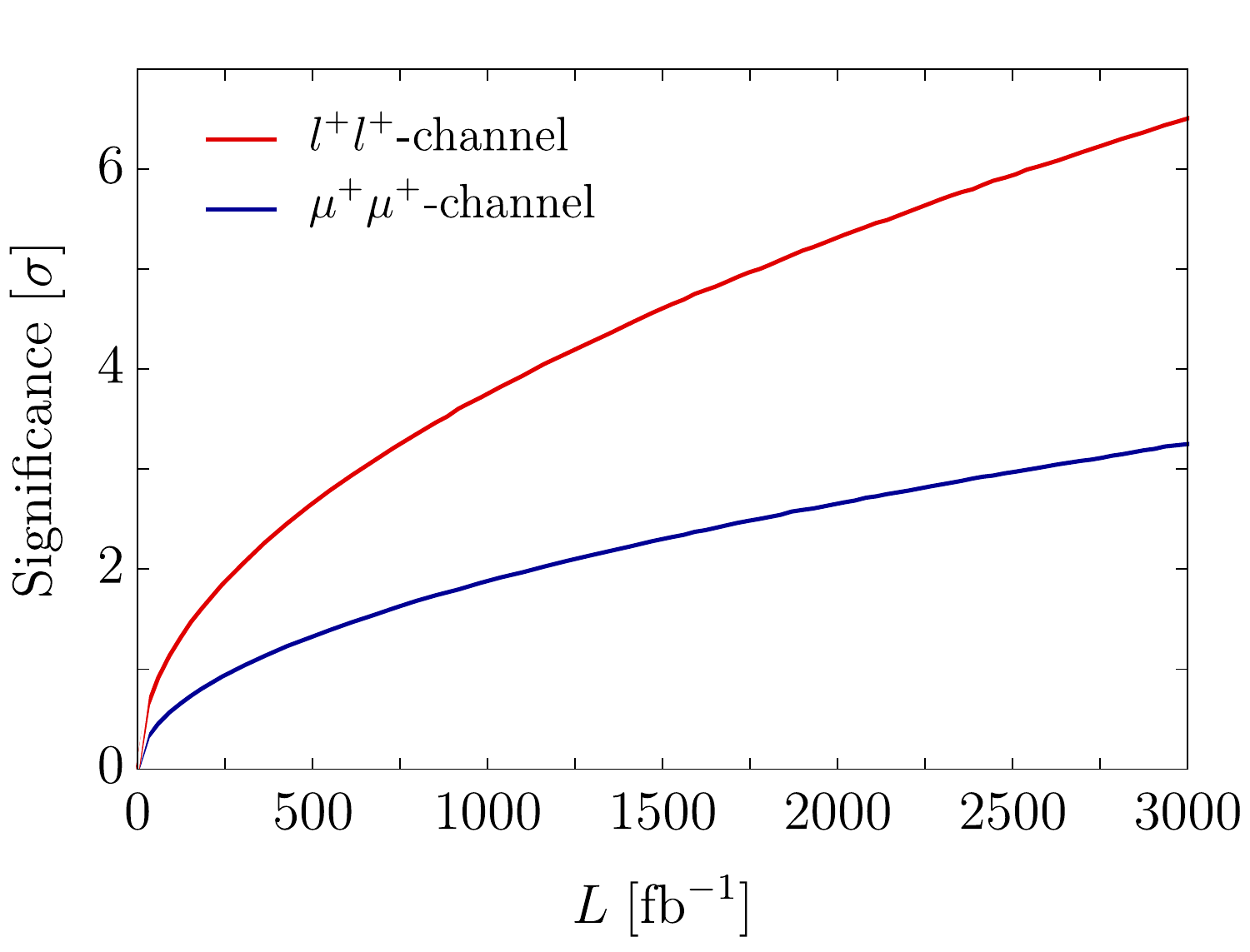}    \label{fig:sigmasModel}
}
   \end{center}
   \caption{(a) Estimate of the significance of the asymmetry of 0.11 from zero for a signal cross section of 0.29 fb, as the distance in standard deviations.
   Blue line/band corresponds to $\mu^+\mu^+$ only while the red line/band includes all positively charged combinations of two light leptons ($e^+$ and $\mu^+$). 
   Dashed curves show the sensitivity of the central red curve to changes of the asymmetry by 20\% (orange dashed curves) 
   and the magnitude of the DPS cross section by a factor of $3/2$ or $3/4$ (green dashed curves). (b) Estimate of the significance of the asymmetry of 0.11 from the pos-pol value of -0.05. Blue line corresponds to $\mu^+\mu^+$ only while the red line includes all positively charged combinations of two light leptons ($e^+$ and $\mu^+$). }
   \label{fig:sigmas}
  \end{figure}
Through an experimental measurement in the $\mu^+\mu^+$ channel alone, a more than 2-sigma indication can be obtained with the full integrated luminosity of 3000 fb$^{-1}$ 
of the high-luminosity LHC (HL-LHC) \cite{Apollinari:2017cqg}. A detailed investigation of the background processes 
corresponding to one or both muons replaced by electrons is beyond the scope of this article. Under the assumption that a similar sensitivity can be 
achieved, the DPS signals are equal and, therefore, effectively enhanced by a factor of 4 including all combinations of positively charged muons and electrons 
($\mu^+\mu^+$, $e^+\mu^+$, and $e^+e^+$). The integrated luminosity necessary to measure the asymmetry then significantly decreases, as shown in Fig.~\ref{fig:sigmas}.
In particular, 400 fb$^{-1}$ would give a 2-sigma hint, 1500 fb$^{-1}$  a more than 3-sigma observation and the full integrated luminosity would lead to a 
measurement approaching 5-sigma.  Fig.~\ref{fig:sigmas} additionally  demonstrates how the significance of a measurement would be affected by a change to 
the absolute magnitude of the DPS cross section (changed by a factor of 3/4 or 3/2) and to a change of the asymmetry itself (changed by a factor of 0.8 or 1.2). 
In addition, a combined measurement by CMS and ATLAS as well as including negatively charged leptons would further increase the sensitivity of the measurement. 
This implies that  first indications of spin correlations can be possibly seen even before the start of the high-luminosity LHC.

It is further possible to ask when experimental measurements could start to discriminate between the range of models for DPSs. 
In order to give an indication of this, we can compare the hypothetically measured asymmetry in the mix-pol scenario to the value obtained in 
the pos-pol scenario. This means repeating the exercise above but counting the number of standard deviations away from the pos-pol value $-0.05$ (instead of zero). 
The results, central values only, are shown in Fig.~\ref{fig:sigmasModel}. A 3-sigma discrimination is possible with the $\mu^+\mu^+$ channel alone. 
The combined lepton flavor measurement could reach 3-sigma with about 600 fb$^{-1}$ and 5-sigma with around 2000 fb$^{-1}$. 	   

\afterpage{\clearpage}

% !TEX root = WWLongPaper.tex
%%%%%%%%%%%%%%%%%%%%%%%%%%%%%%%%%%%%%%%%
\newpage
\section{Impact of correlations on extracted DPS yield}
\label{template}
%%%%%%%%%%%%%%%%%%%%%%%%%%%%%%%%%%%%%%%%

The extractions of the DPS cross section in experiments \cite{
Akesson:1986iv,Alitti:1991rd,Abe:1993rv,Abe:1997xk,Abe:1997bp,Abazov:2009gc,Aaij:2012dz,Aad:2013bjm,
Chatrchyan:2013xxa,Abazov:2014fha,Aad:2014kba,Aaboud:2016dea,
Aaij:2015wpa,Abazov:2015nnn,Aaij:2016bqq,Sirunyan:2017hlu,Abazov:2014qba,
Abazov:2015fbl,Aaboud:2016fzt,Khachatryan:2016ydm,Aaboud:2018tiq,Sirunyan:2019zox} 
are usually based on fitting signal and background templates to data. 
The templates for the DPS signal are typically obtained from Monte Carlo generators or from data-driven methods assuming no-correlations, 
i.e. combining the measured differential cross sections for the individual subprocesses.
Here, we explore the impact of parton correlations on attempts to extract the cross section for double parton scattering. 
In other words, we try to quantify the naivety of the no-correlation assumption.

The template fits (or multivariate analyses), used for DPS cross section measurement by collaborations, deal with many observables. 
However, to illustrate the potential problems, we restrict ourselves to a single distribution, namely the product of muon rapidities. 
This was one of the variables in the multivariate analysis of \cite{Sirunyan:2019zox}. With the event selection \eqref{eq:sel}, we generate data based on the sum of DPS $WW$ signal 
(for both min-corr and mix-pol correlation models) as well as $t\bar{t}$ and $WZ$ backgrounds. We then make a simple template fit, to extract the size of the DPS signal in the form of
\begin{equation}
\sigma^\text{tot} = a \sigma_\text{DPS} + b \sigma_{WZ} + c \sigma_{t\bar{t}},
\end{equation}
where the parameter $a$ provides the (relative) size of the DPS contribution to the cross section, and the parameters $b$ and $c$ scale the size of the respective backgrounds.
We first generate pseudo-data based on the correlated (mix-pol) DPS model (Scenario 1) and then make the extraction based on the uncorrelated (min-corr) model. 
The result for the parameters of the fit is: $a = 1.23$,  $b = 0.97$ and $c = 0.99$. This translates to a DPS fraction in the data sample $f_{\text{DPS}}$ = 12\%, where $\sigma^\text{tot}$ = 4.74 fb. 
If we instead generate data based on the min-corr DPS model (Scenario 2) and make the cross section extraction based on the mix-pol DPS model, 
we obtain $a = 0.86$, $b=1.03$, $c=0.99$. This corresponds to DPS fraction $f_{\text{DPS}}$ = 9\%, where $\sigma^\text{tot}$ = 4.71 fb. 
We thus see a difference of 0.15 fb in the size of the extracted DPS cross section, i.e. a difference of 30\%. 
The corresponding values for the fiducial cross sections in the two scenarios, as well as the corresponding values for an extracted $\sigma_{\text{eff}}$ are shown in Table \ref{tab:extract}. 

\begin{table}
\centering
\begin{tabular}{ c|c|c } 
 & DPS $W^+W^+$ [fb] & $\sigma_{\text{eff}}$ [mb]\\
 \hline
  Scenario 1 & 0.59 & 12.2\\ 
  Scenario 2 & 0.44 & 16.4\\ 
\end{tabular}
\caption{\label{tab:extract} Extracted DPS cross sections from template fits. Scenario 1 (Scenario 2) generates data based on correlated (uncorrelated) DPS and 
extracts the signal assuming uncorrelated (correlated) DPS. We also give the value for the effective cross section in the two scenarios 
(based on a $\sigma_{\text{eff}}$ = 15 mb used in the main analysis).}
\end{table} 

\begin{figure}
   \includegraphics[width=0.32\textwidth]{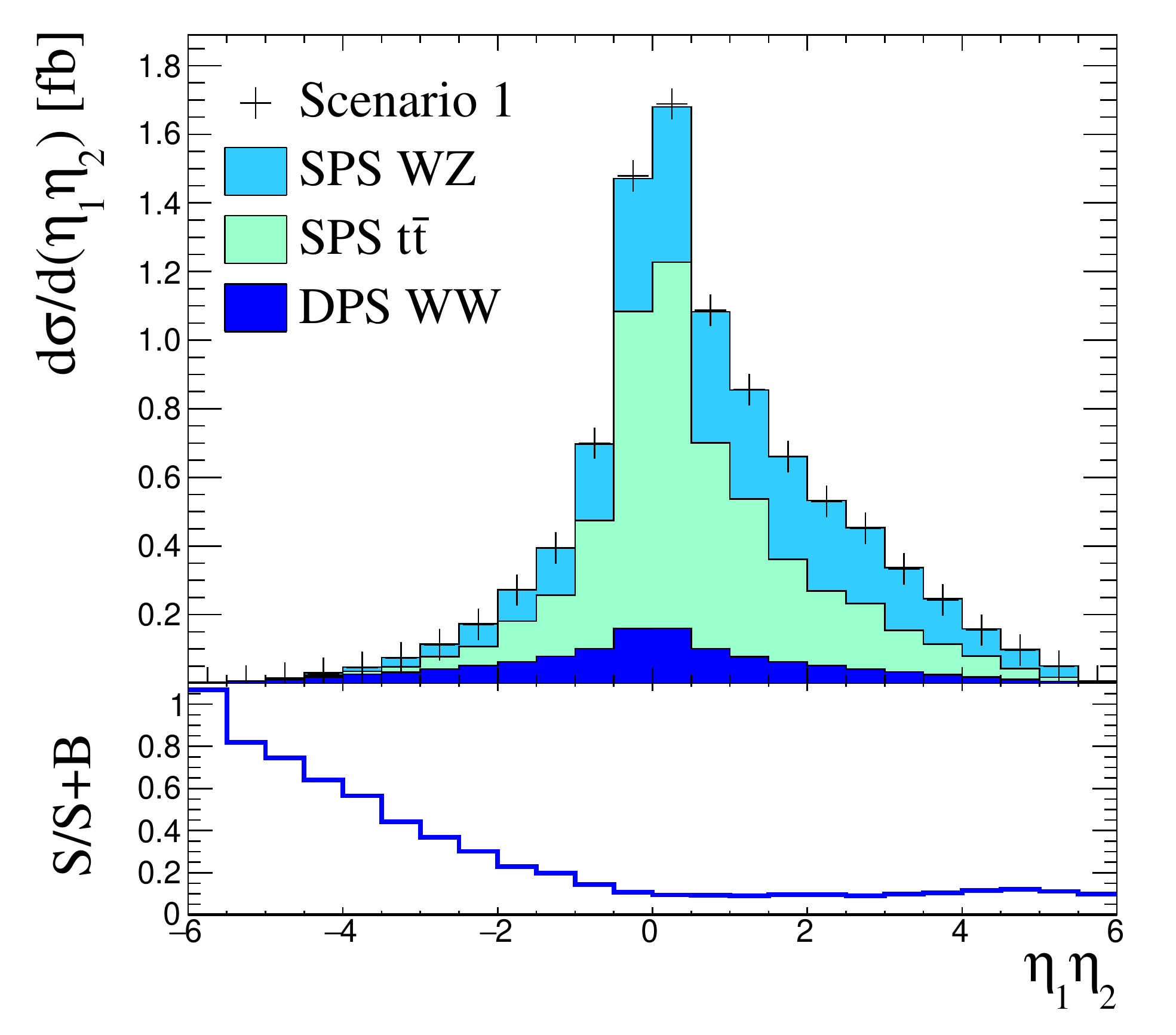}
   \includegraphics[width=0.32\textwidth]{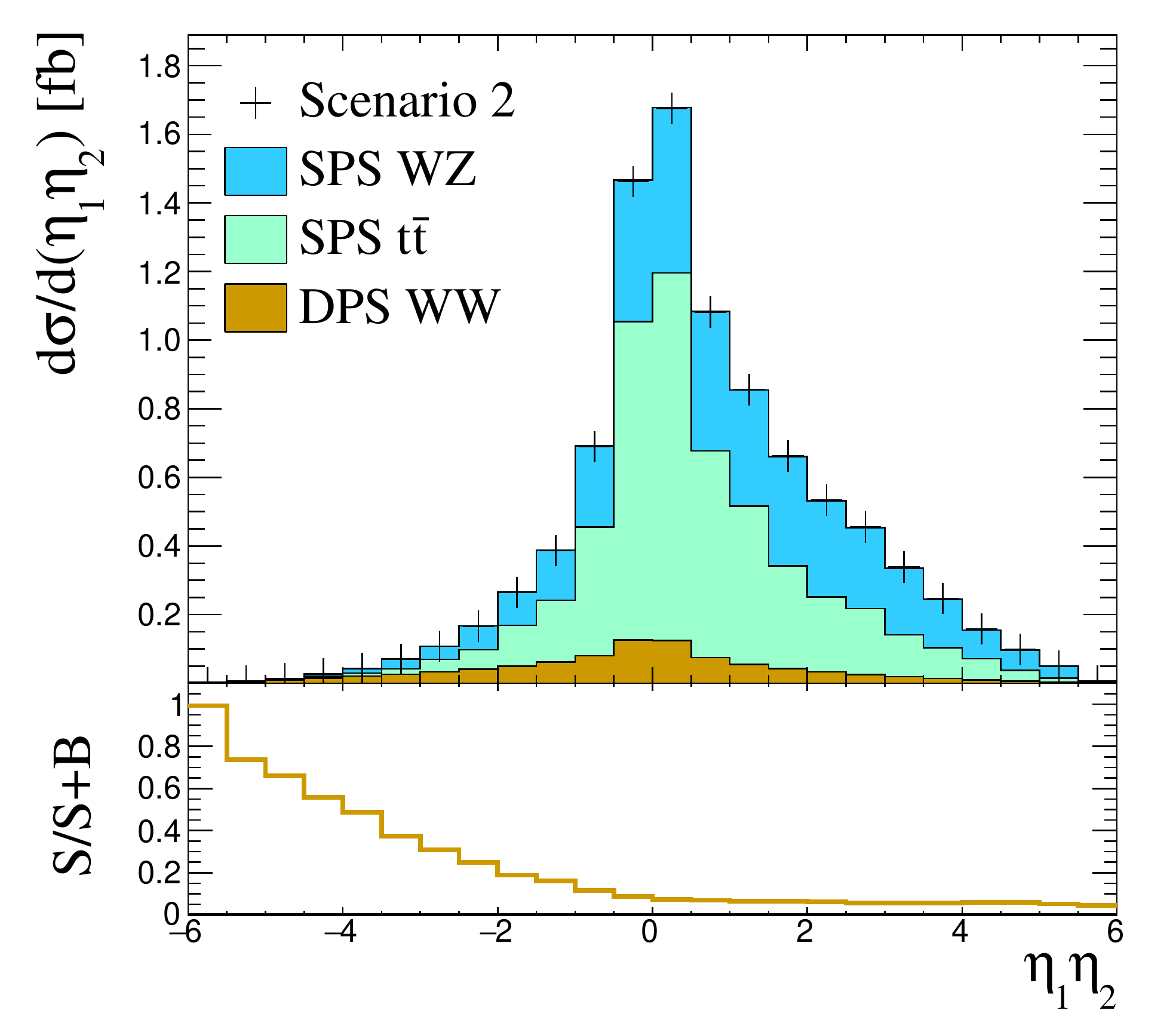}
    \includegraphics[width=0.32\textwidth]{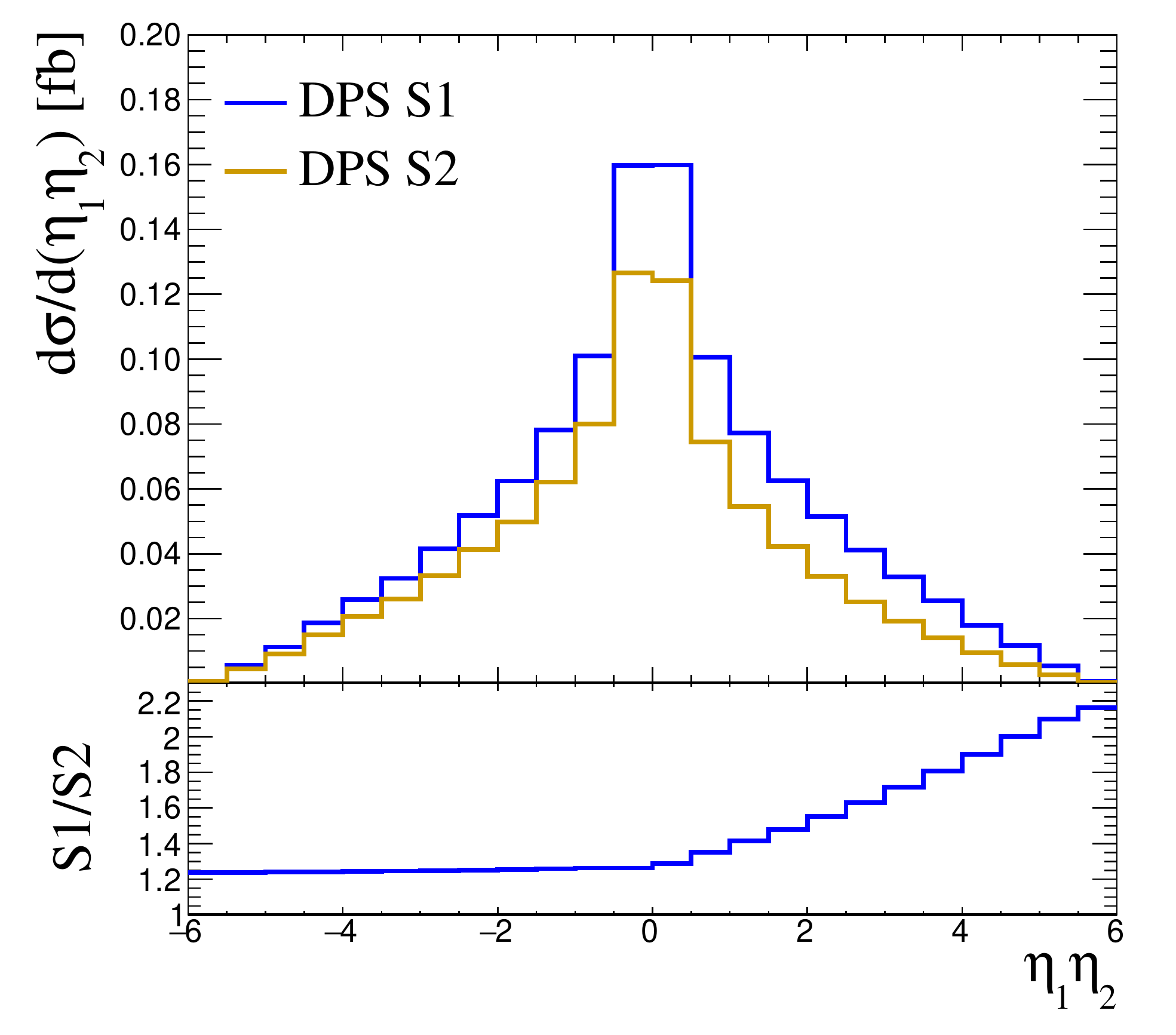}
   \caption{Results of the template fits. Left: scenario 1: min-corr model fitted to data with mix-pol DPS contribution. 
   Middle: scenario 2: mix-pol model fitted to data with min-corr DPS contribution. Right: the extracted DPS signals in the two scenarios (S1 and S2).} 
   \label{fig:temp_fit}
\end{figure}

The fitted templates for the two scenarios and the comparison of the extracted DPS distributions are shown in Fig. \ref{fig:temp_fit}.
The 30\% span of the DPS production cross section found by our simple treatment illustrates the danger in using correlation sensitive variables in the template fits. 
The difference is equivalent to the variation of $\sigma_{\text{eff}}$ by 30\%.

%

% !TEX root = WWLongPaper.tex
\clearpage
%%%%%%%%%%%%%%%%%%%%%%%%%%%%%%%%%%%%%%%%
\section{Conclusions}
\label{Conclusions}
%%%%%%%%%%%%%%%%%%%%%%%%%%%%%%%%%%%%%%%%
We have demonstrated that the LHC has the potential to discover correlations between two partons inside a single proton. 
We have shown a path towards this discovery in double same-sign W-boson production, including a detailed treatment of signal and background processes. 

Double parton scattering has undergone a staggering development in the last decade. The advances in theory, phenomenology, and experiments offer a 
realistic opportunities for measurements of quantum correlations between two proton constituents. While most past measurements of DPS have relied on the 
assumption that partonic correlations in DPS are not quantitatively impactful, the integrated luminosity collected at the LHC now starts to enable detailed tests of the properties of DPS.

We have examined the impact that correlations between the spin of two partons, and between their momenta, can have on the cross section of 
the SSW process. In particular, we have provided the analysis of four models of parton correlations, essentially extending the study on spin correlations of~\cite{Cotogno:2018mfv} and 
reinforcing the conclusions drawn therein. To isolate and measure these correlations, we have identified a handful of promising variables, some of which have a 
clear benchmark value for uncorrelated DPS. Therefore, any measured deviations from the uncorrelated values can be directly related to interparton correlations. A 
detailed study of the single parton scattering background processes has allowed us to closely examine how to maximize the purity of the signal, which is 
essential for measuring correlations, while, at the same time, keeping a large enough cross section to have sufficient statistical power. As a result, 
we have estimated the integrated luminosity necessary for experiments to start probing correlations in DPS. We have shown the dependence of the estimate on 
the absolute size of the DPS cross section as well as the exact amount of correlations. 

The asymmetry between the number of outgoing leptons from the W-boson decays which end up in the same vs opposite hemispheres is one of the most promising 
variables. Likewise, several additional variables, such as bin-by-bin ratios of cross section, 
plotted against the sum (difference) of muon rapidities, and the corresponding linear slope, show promise. The signatures of correlations in these variables have 
been demonstrated to survive after background removal and phase-space 
reductions. We have further found that, although correlations between longitudinal momenta also affect the same distributions, 
the main suspect for creating large correlations in this process is the spin of the partons. Because of the differences in angular 
momentum between different quark helicities, polarization has a direct and calculable impact on the hard partonic cross sections. 
The high-luminosity program HL-LHC will be able to deliver more precise information 
about the impact of correlations in the SSW process~\cite{Azzi:2019yne}. Nonetheless, the LHC in its current set-up would already have the potential to put restrictions 
on the models we have presented and  would be able to start discriminating between presence and absence of correlations in SSW.

First experimental measurements of correlations between two partons inside the proton are still to come. Phenomenological studies are therefore relying on
models to make predictions about the correlations, and until experimentally confirmed, their exact size is uncertain. 
It is ultimately important to realize that even a null result of a correlation measurement would be an important step towards a better understanding of DPS 
and the distributions of two quarks or gluons inside the proton.

\section*{Acknowledgements}
%%%%%%%%%%%%%%%%%%%%%%%%%%%%%%%%%
We thank Piet J. Mulders and Jonathan Gaunt for useful discussions. SC acknowledges support from he Agence Nationale de la Recherche under the Projects No. ANR-18-ERC1-0002 and No. ANR-16-CE31-0019 and the European Research Council (ERC) under the program QWORK (contract no. 320389). TK acknowledges support from the Alexander von Humboldt Foundation.
 MM acknowledges support from the grant LTC17038 of the INTER-EXCELLENCE program at the Ministry of Education, Youth and Sports of the Czech Republic.

\appendix{
% !TEX root = WWLongPaper.tex
%%%%%%%%%%%%%%%%%%%%%%%%%%%%%%%%%%%%%%%%
\section{Coupling factors}

\label{AppendixA}
%%%%%%%%%%%%%%%%%%%%%%%%%%%%%%%%%%%%%%%%
The coupling factors that enter the cross section formula of eq.~\eqref{CrosSecSab} are derived in~\cite{Kasemets:2013nma}. Since the leptons are the result of the decay of a $W^+$ boson with mass M and width $\Gamma_W$, we introduced the factors $K_{q_i \bar q_j}$ given by:
\begin{equation}
\label{Kappas}
K_{q_i \bar q_j}=\frac{\alpha^2}{4N_c}\frac{|V_{q_i q_j}|^2}{(2\sin \theta_w)^4}\frac{q_i^2}{(q_i^2-m_W^2)^2+m^2_W\Gamma^2_W},\quad (e_{q_i}-e_{q_j}=1),
\end{equation}
where $N_c=3$ is the number of colors, $V_{q_i q_j}$ a CKM matrix element, $\theta_w$ the weak mixing angle, $\alpha$ the electromagnetic fine structure constant and $e_{q_i}$ the charge of quark $q_i$~\cite{Patrignani:2016xqp}.

 }

%%%%%%%%%%%%%%%%%%%%%%%%%%%%%%%%%
%%%%%%%%%%%%%%%%%%%%%%%%%%%%%%%%%

\appendix

%%%%%%%%%%%%%%%%%%%%%%%%%%%%%%%%%
%%%%%%%%%%%%%%%%%%%%%%%%%%%%%%%%%
\bibliographystyle{JHEP}
\bibliography{Bibliography/Unique_ref}
%\bibliography{Bibliography/referencesWW}
%%%%%%%%%%%%%%%%%%%%%%%%%%%%%%%%%

\end{document}